\pdfoutput=1
\documentclass[10pt,twocolumn,a4paper]{article}

\usepackage{amsmath}
\usepackage[T1]{fontenc}
\usepackage[utf8]{inputenc}
\usepackage{newtxtext}
\usepackage{newtxmath}

\DeclareEncodingSubset{TS1}{ntxtlf}{0}
\usepackage[a4paper,top=18mm,bottom=20mm,left=13.5mm,right=13.5mm,columnsep=5.5mm]{geometry}
\usepackage{graphicx}
\graphicspath{{figures/}}
\usepackage{float}
\usepackage{booktabs,longtable,array}
\usepackage[detect-all]{siunitx}
\usepackage{authblk}
\usepackage[super,comma,sort&compress]{natbib}
\usepackage{xcolor}
\usepackage[font=small,labelfont=bf,labelsep=period,justification=justified,singlelinecheck=false]{caption}
\usepackage{microtype}
\usepackage[colorlinks=true,linkcolor=black,citecolor=black,urlcolor=blue!50!black]{hyperref}
\usepackage[capitalise,nameinlink]{cleveref}
\usepackage{dblfloatfix} %
\newcommand{\pf}{\mathrm{PF}/\tau}
\usepackage{titlesec}
\titleformat{\section}{\normalfont\large\bfseries}{\thesection}{0.6em}{}
\titleformat{\subsection}{\normalfont\normalsize\bfseries}{\thesubsection}{0.5em}{}
\titleformat{\subsubsection}{\normalfont\normalsize\bfseries\itshape}{\thesubsubsection}{0.5em}{}
\titlespacing*{\section}{0pt}{2.2ex plus 0.6ex minus 0.3ex}{1.0ex plus 0.2ex}
\titlespacing*{\subsection}{0pt}{1.6ex plus 0.5ex minus 0.3ex}{0.6ex plus 0.2ex}
\titlespacing*{\subsubsection}{0pt}{1.1ex plus 0.4ex minus 0.2ex}{0.4ex plus 0.1ex}
\crefname{section}{Sec.}{Secs.}\Crefname{section}{Section}{Sections}
\crefname{table}{Table}{Tables}\crefname{figure}{Fig.}{Figs.}\crefname{equation}{Eq.}{Eqs.}

\title{\Large\bfseries High-throughput screening of Heusler alloys for thermoelectrics combining machine-learned densities of states and first-principles transport}

\author[1]{Shivam Randive}
\author[2]{Rahul Dixit}
\author[1,*]{Himanshu Pandey}
\affil[1]{Condensed Matter \& Low-Dimensional Systems Laboratory, Department of Physics, Sardar Vallabhbhai National Institute of Technology, Surat, India}
\affil[2]{Department of Artificial Intelligence, Sardar Vallabhbhai National Institute of Technology, Surat, India}
\newcommand{\email}[1]{\href{mailto:#1}{#1}}
\date{\small $^{*}$Corresponding author: H.~Pandey (\email{hp@phy.svnit.ac.in})}

\pdfpageattr{/Group << /S /Transparency /CS /DeviceRGB /I true >>}
\begin{document}
\twocolumn[
\begin{@twocolumnfalse}
\maketitle
\vspace{-4mm}
\begin{abstract}
\noindent
Thermoelectrics turn waste heat into electricity, but finding new ones requires transport calculations too costly for whole databases.
Searches therefore choose what to calculate from the electronic density of states (DOS), increasingly predicted by machine learning, yet
such choices are rarely tested. We test them on all 106,235 Heusler structures in the DXMag database, screening 51,257 on
density-functional spectra and 54,978 on neural-network predictions, and judging both by 3,625 band-structure and transport
calculations. A DOS-derived Seebeck indicator ranks most compounds passing every transport and stability test within the top 1\% of
their population, and ten of the eleven established half-Heusler thermoelectrics calculated pass. Predicted spectra lose most band gaps,
unnoticed by spectral-similarity metrics, yet still rank passing compounds near the top. The screen certifies 80 computational
candidates, 61 without reported measurements. Computed or predicted, the DOS can choose what to calculate; only a band structure shows
whether a gap exists.
\end{abstract}


\vspace{4mm}
\end{@twocolumnfalse}
]

\section{Introduction}\label{sec:intro}\label{sec:intro-te}\label{sec:r-physics}\label{sec:intro-ml}

Thermoelectric generators convert a temperature difference directly into electrical power and could recover part of the heat that
industry and transport reject\cite{Bell2008Waste,He2017Advances}. Their efficiency is measured by the dimensionless figure of merit
$zT=S^2\sigma T/(\kappa_e+\kappa_L)$, where $S$ is the Seebeck coefficient (the thermopower), $\sigma$ the electrical conductivity, $T$ the
temperature, and $\kappa_e$ and $\kappa_L$ the electronic and lattice thermal conductivities\cite{Snyder2008}. The numerator, the power
factor $S^2\sigma$, is governed by the electronic structure and by the scattering of the carriers. The figure of merit is largest in a doped semiconductor
whose chemical potential lies near a band edge, provided that the band gap is wide enough for carriers thermally excited across it not to
reduce the thermopower\cite{Sofo1994OptimumGap}. On the electronic side, a search for new thermoelectrics is therefore a search for
semiconductors with suitable band edges.

First-principles calculations can predict the power factor. From a density-functional-theory (DFT) band structure they compute $S$ and
$\sigma/\tau$, the conductivity per relaxation time $\tau$, usually in the constant-relaxation-time approximation (CRTA), in which every
electronic state has the same $\tau$\cite{Madsen2006a,BoltzTraP2,Yang2008,Chen2016,Ricci2017}. Each such calculation needs a
self-consistent DFT run, a band structure on a dense mesh of wavevectors and an interpolation of the bands. Large searches therefore first
rank or filter compounds with a cheaper electronic-structure signal, such as a density-of-states (DOS) effective mass\cite{Gorai2017}, a gap
near the Fermi level\cite{Ricci2020GappedMetals} or features of the DOS itself\cite{Jia2022UnsupHH}. Such a signal decides which compounds
receive a transport calculation at all. Whether a DOS-based decision selects the compounds that the transport calculation would confirm,
and whether it still does so when the DOS itself comes from a machine-learning model, has rarely been tested at scale.

Heusler compounds offer a clear baseline for this test, because simple electron-counting rules already predict many of their
semiconductors. Half-Heusler compounds (composition XYZ) with eight or eighteen valence electrons per formula unit are often
semiconducting, because these counts fill the bonding states formed by hybridisation of the atoms on the different
sublattices\cite{Kandpal2006,Graf2011}. The best half-Heusler thermoelectrics have eighteen: $n$-type MNiSn and $p$-type MCoSb alloys
(M = Ti, Zr, Hf)\cite{Chen2013HHProgress,Quinn2021}, NbFeSb\cite{Fu2015}, ZrCoBi\cite{Zhu2018ZrCoBi} and TaFeSb\cite{Zhu2019} reach
figures of merit near or above one at high temperature. Full Heuslers (X$_2$YZ) follow the same pattern at 24 valence electrons:
Fe$_2$-based compounds with this count have large calculated power factors\cite{Bilc2015FullHeuslerPF}, and some full-Heusler
semiconductors are predicted to have very low lattice thermal conductivity\cite{He2016FullHeuslerKL}. Because the good compounds sit at
these closed-shell electron counts, counting electrons is already a strong screen, and a DOS screen of this family is useful only if it
does better. Heusler compounds are also numerous and structurally varied. They include regular and inverse full-Heusler and half-Heusler prototypes,
dozens of elements on each site, and cubic and tetragonal forms of each composition; for many full Heuslers the tetragonal form is the
calculated ground state\cite{Faleev2017Tetragonal}. Transport screens have so far concentrated on cubic structures, above all on
half-Heuslers\cite{Carrete2014NanograinedHH,BhattacharyaMadsen2016pHH}, so the tetragonal and metastable structures remain largely
unexamined.
\begin{figure*}[t]
\centering
\includegraphics[width=\textwidth]{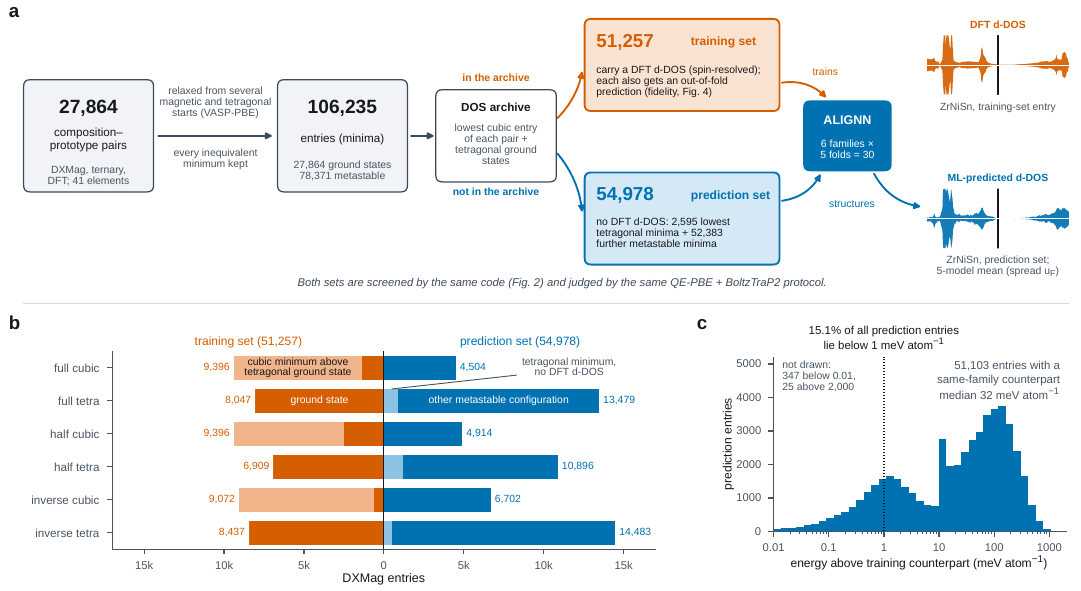}
\caption{\textbf{One database, two populations.}
\textbf{a},~The DXMag Heusler database: 27,864 composition--prototype pairs, relaxed from several magnetic and tetragonal starting
configurations, give 106,235 energy minima (entries). Its DOS archive holds the spin- and orbital-resolved DFT DOS of the lowest-energy
cubic entry of every pair and of the tetragonal entries that are ground states. These 51,257 entries form the training set (vermilion;
triangles in later figures): they train the ALIGNN model (30 fold models), and each also receives an out-of-fold prediction. The other
54,978 entries form the prediction set (blue; circles), screened on the five-model mean of the predicted d-DOS and its spread $u_F$.
Thumbnails: the DFT d-DOS of the training-set ZrNiSn entry and the predicted d-DOS of the prediction-set ZrNiSn entry
($\pm2.4$~eV; spin down below). \textbf{b},~Entries per family, set and state. \textbf{c},~Energy of each prediction entry above its
same-family, same-composition training entry (51,103 entries; logarithmic axis); dotted line, 1~meV\,atom$^{-1}$; 15.1\% of all
prediction entries lie below it, most likely relaxations that returned to the training crystal. d-DOS, d-orbital-projected density
of states; ALIGNN, atomistic line graph neural network.}
\label{fig:workflow}
\end{figure*}

The DXMag Computational HeuslerDB of Xiao and Tadano contains these structures\cite{XiaoTadano2025,XiaoTadano2026npj}
(\cref{fig:workflow}a). Like an earlier
high-throughput Heusler library\cite{Sanvito2017HeuslerMagnets}, it was built to find magnetic materials. Each of 27,864 combinations of
composition and prototype was relaxed from several magnetic and tetragonal starting configurations, and every inequivalent energy
minimum reached was kept. The result is 106,235 entries, each with its energy and its distance from the convex hull, a measure of
thermodynamic stability. A companion archive gives spin- and orbital-resolved DFT densities of states for about half of the
entries\cite{DXMagDOS}; matched to the database by identifier, these are the lowest-energy cubic entry of every composition and prototype
and the tetragonal entries that are ground states. The database contains no charge or thermoelectric transport, only anomalous Hall and
Nernst conductivities for 671 entries. Its documentation (\url{https://www.nims.go.jp/group/spintheory/database/}) also states that the
structures, energies and DOS were computed without spin--orbit coupling (SOC), which splits and shifts the band edges of heavy-element
Heuslers\cite{Chadov2010,Guo2016SOCANiB}. A thermoelectric search of DXMag therefore has the structures it needs, but it must decide from
the DOS which of them to calculate, and half of them have no DFT DOS.

\begin{figure*}[p]
\centering
\includegraphics[width=\textwidth]{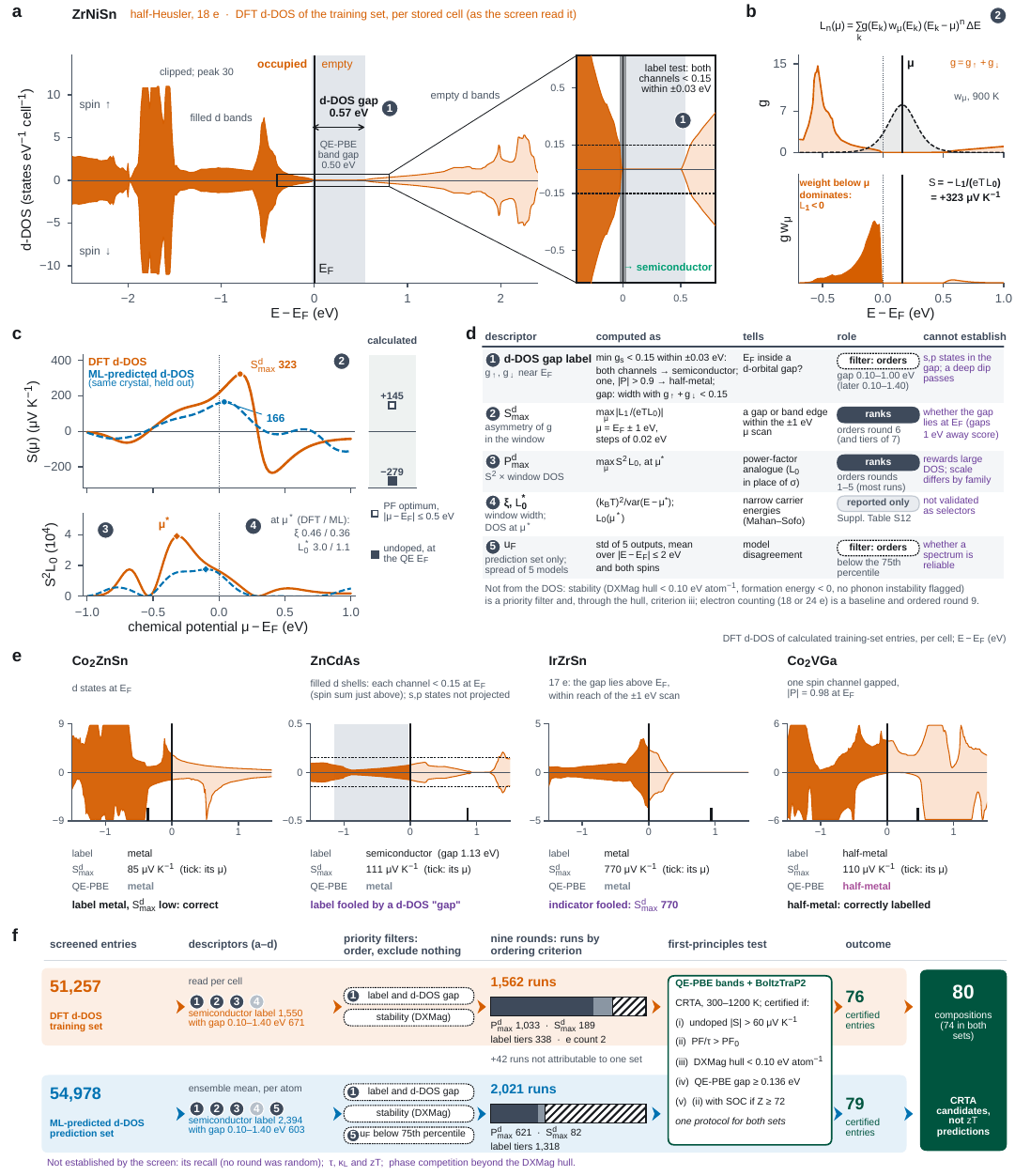}
\caption{\textbf{From a d-DOS to a screening decision.}
\textbf{a},~DFT d-DOS of ZrNiSn (training set), per stored cell as screened: spin up above, spin down below; occupied solid, empty tinted; valence d peaks between $-1.9$ and $-1.5$~eV are clipped (maximum 30). Grey band, the d-DOS gap
(spin-summed d-DOS below 0.15~states\,eV$^{-1}$; 0.57~eV, against a QE-PBE band gap of 0.50~eV). Right, the label test: both spin
channels below 0.15 (dashed) within $\pm0.03$~eV of $E_F$ (dark band). \textbf{b},~Fermi-window moments
(\cref{eq:moments,eq:seebeck}) at the $\mu$ of $S^{\mathrm d}_{\max}$: $g$ (states\,eV$^{-1}$\,cell$^{-1}$) and the scaled window
$w_\mu$ at \SI{900}{K} (top); $g\,w_\mu$ (bottom), whose larger weight below $\mu$ gives $L_1<0$ and $S>0$. \textbf{c},~$S(\mu)$ and
$S^2L_0$ ($10^4$~$\mu$V$^2$\,K$^{-2}$\,states\,eV$^{-1}$) over the $\pm1$~eV scan for the DFT and the held-out ML-predicted d-DOS of
the same crystal (dashed), both per cell; circles, $S^{\mathrm d}_{\max}$; diamonds, $P^{\mathrm d}_{\max}$ at $\mu^*$. The screen
ranks $|S|$; its sign depends on where $\mu$ lies. The strip gives the calculated thermopower at \SI{900}{K}, referred to the QE Fermi level and so kept off the $\mu$ axis. \textbf{d},~Descriptor ledger; numbered badges mark the same
descriptors in a--c and f (grey: reported only). \textbf{e},~Four calculated training-set entries (tick: the $\mu$ of
$S^{\mathrm d}_{\max}$). \textbf{f},~Selection in each set: priority filters order but exclude nothing; bars, runs by the criterion that ordered their round (Supplementary Table~S6); criteria (i)--(v) as in \cref{sec:m-cert}. d-DOS in
archive units (twice the physical values); QE-PBE, Quantum ESPRESSO with the PBE functional; CRTA, constant-relaxation-time
approximation; SOC, spin--orbit coupling.}
\label{fig:descriptors}
\end{figure*}

A screen can read a gap from the DOS in two ways (\cref{fig:descriptors}a--c). The simpler is a binary label that asks whether the d-orbital-projected DOS (d-DOS) has
a gap at the Fermi level $E_F$. The other follows from transport theory and uses more of the spectrum. The electronic transport
coefficients are moments of the transport distribution
$\Sigma(\varepsilon)\propto\sum_{n\mathbf k}v_{n\mathbf k}^2\,\tau_{n\mathbf k}\,\delta(\varepsilon-\varepsilon_{n\mathbf k})$, which
measures how much current the states at energy $\varepsilon$ can carry; $\varepsilon_{n\mathbf k}$, $v_{n\mathbf k}$ and
$\tau_{n\mathbf k}$ are the energy, velocity and relaxation time of an electron in band $n$ with wavevector $\mathbf k$. The moments are
taken over the Fermi window $-\partial f/\partial\varepsilon$, where $f$ is the Fermi--Dirac distribution; the window is a few $k_BT$ wide,
with $k_B$ the Boltzmann constant, so only states near the chemical potential $\mu$ contribute. The conductivity is the zeroth moment. The
Seebeck coefficient is the first moment divided by the zeroth, that is, it is set by the mean energy of the current-carrying states
relative to $\mu$. If $v^2\tau$ is the same for all states in the window, $\Sigma(\varepsilon)$ reduces to the DOS $g(\varepsilon)$, and
the thermopower takes the Cutler--Mott form\cite{CutlerMott1969,Jonson1980}. In this form $S(\mu)$ is large only where the window-weighted
DOS is small and asymmetric about $\mu$, that is, at a band edge or in a gap. Its largest value grows with the width of the gap, as in
the Goldsmid--Sharp estimate $E_g\approx2e|S|_{\max}T_{\max}$, where $e$ is the elementary charge and $T_{\max}$ the temperature of the
thermopower maximum\cite{GoldsmidSharp1999,Gibbs2015GapEstimate}. The second reading of the DOS, which we call the Seebeck indicator, is the largest
value of this $S(\mu)$. Unlike the label, it weighs how far states lie from $\mu$ on the scale of the thermal energy.

We read both from the d-DOS because the gaps of transition-metal Heuslers open between d-derived states formed by hybridisation of the transition-metal
sublattices\cite{Kandpal2006,Galanakis2002HH,Galanakis2002FH}; the price is that s- and p-derived states inside the d gap are not seen. The d-DOS is also resolved by spin, so that the label can tell half-metals, which are gapped in one spin
channel only, and exchange-split magnets apart from non-magnetic semiconductors; the indicator reads the spin sum.

The Cutler--Mott derivation also shows where the approach is expected to fail. A constant $v^2\tau$ is a poor assumption at band extrema, where the velocities
vanish. For a non-degenerate parabolic band with its edge at $\varepsilon_0$, $\Sigma\propto(\varepsilon-\varepsilon_0)^{3/2}$ whereas
$g\propto(\varepsilon-\varepsilon_0)^{1/2}$, and the extra power of $(\varepsilon-\varepsilon_0)$ in $\Sigma$ raises the mean energy of the current-carrying
states by $k_BT$, so the DOS form underestimates $|S|$ by $k_B/e\approx\SI{86}{\micro V.K^{-1}}$, an error
comparable with the thermopower thresholds that screens apply. The conductivity, moreover, depends on the band velocities and on the
valley degeneracy and the band mass separately, whereas the DOS combines degeneracy and mass into a single DOS effective
mass\cite{Pei2011,Gibbs2017,Xing2017}. Together with the form of
the thermopower derived above, these limits lead to two expectations that the calculations of this study can test. A DOS-based indicator
should recognise gap-like electronic structure, and so help to choose what to calculate, but it should not rank compounds by power
factor. A binary label, which ignores the thermal window, should do worse than the Seebeck indicator.

For the half of DXMag without a DFT spectrum, a machine-learned model is the practical source of one: computing the missing spectra would
take 54,978 further DFT calculations, fifteen times the number of transport calculations they are meant to prioritise. Machine-learning
models now predict the DOS of a crystal from its composition and
structure\cite{Yeo2019,Chandrasekaran2019,BenMahmoud2020,Kaundinya2022,Fung2022,DOSTransformer,PETMADDOS}; among them are atomistic line
graph neural networks (ALIGNN)\cite{ALIGNN}, which have been trained on tens of thousands of spectra\cite{Kaundinya2022}. Thermoelectric
screens assisted by machine learning have instead learned transport labels directly\cite{Choudhary2020TE,Sheng2020ActiveLearningPF}. The
closest DOS-based study evaluated a gap feature of predicted spectra on one hundred new DFT calculations, but against the DFT DOS rather
than against transport\cite{Mat2Spec}. A good average fit does not show that predicted spectra support screening decisions. A screen
depends on whether a gap exists at $E_F$, and modest spectral errors can fill a gap\cite{PETMADDOS} or shift $E_F$\cite{BenMahmoud2020,How2025}.
More generally, accurate regression does not guarantee accurate selection\cite{Borg2023}, especially near a threshold\cite{Bartel2020,Riebesell2025}. Apparent accuracy can also be inflated by random
splits\cite{Meredig2018}, by near-duplicate entries\cite{Omee2024,Li2023b} and by overlap between pretraining and test
data\cite{Kapoor2023,Magar2022}. If a screen fails on predicted spectra, it could fail for two different reasons: the DOS may be a poor signal, or the model may be a poor source of it. Only the same screen,
run on DFT spectra and assessed in the same way, can tell the two apart.

\begin{table*}[t]
\centering
\caption{\textbf{The two populations and how each is screened.} DXMag release 2025-09-09. Composition strings keep the DXMag site order, so they distinguish half-Heusler site permutations; reduced formulas do not. Label thresholds are converted to physical units (archive factor of two removed); the training-set screen reads the d-DOS per stored primitive cell of three or four atoms. QE-PBE: Quantum ESPRESSO PBE calculations (\cref{sec:m-dft}); certified: passing all criteria of \cref{sec:m-cert}; $u_F$: ensemble spread (\cref{sec:m-model}); UUID: identifier of a DXMag entry.  Per-family values are given in Supplementary Table~S1.}
\label{tab:populations}
\footnotesize
\begin{tabular}{@{}>{\raggedright}p{0.29\textwidth} >{\raggedright}p{0.32\textwidth} >{\raggedright\arraybackslash}p{0.32\textwidth}@{}}
\toprule
 & training set & prediction set \\
\midrule
\multicolumn{3}{@{}l}{\emph{Population}}\\
entries (DXMag UUIDs) & 51,257 & 54,978 \\
what the entries are & lowest-energy cubic entry of every composition and prototype; lowest tetragonal entry where it is the ground state & 2,595 lowest tetragonal minima without a spectrum; 52,383 further metastable minima \\
compositions (strings / reduced formulas) & 18,792 / 14,580 & 17,149 / 14,291 \\
energy above the training entry of the same family and composition & 0 (by construction) & median 32~meV\,atom$^{-1}$; 15.1\% within 1~meV\,atom$^{-1}$ \\
DXMag hull distance $<\SI{0.10}{eV.atom^{-1}}$ & 6.0\% & 3.3\% \\
electron-count rule: 18 (half) / 24 (full, inverse) & 2.5\% / 3.1\% & 2.0\% / 3.6\% \\
DXMag phonon results & 8,198 entries & none \\
\multicolumn{3}{@{}l}{\emph{Screen}}\\
d-DOS screened & DFT d-DOS of the archive, per stored cell & ensemble-mean prediction, per atom \\
label threshold, physical units & 0.025 (half) or 0.019 (full, inverse) states\,eV$^{-1}$\,atom$^{-1}$ per spin & 0.075 states\,eV$^{-1}$\,atom$^{-1}$ per spin \\
d-DOS labels: semiconductor / half-metal / metal & 1,550 / 962 / 48,745 & 2,394 / 473 / 52,111 \\
ensemble-spread priority filter & none & $u_F$ below its 75th percentile \\
\multicolumn{3}{@{}l}{\emph{First-principles assessment}}\\
runs (QE-PBE + BoltzTraP2) & 1,562 & 2,021 \\
certified entries / compositions & 76 / 75 & 79 / 79 \\
\bottomrule
\end{tabular}

\end{table*}
DXMag makes this separation possible, because its DOS archive divides it into two sets of similar size. We call the 51,257 entries with a
DFT d-DOS the \emph{training set}: they can be screened on their own DFT spectra, and they train the model. The other 54,978 entries form
the \emph{prediction set}, which can be screened only on spectra predicted by an ALIGNN model trained on the training set. Every training
entry also has an out-of-fold predicted spectrum, made by the cross-validation model for which it was a test entry, so the fidelity of
predicted spectra can be measured on crystals whose DFT spectra are known. The two sets share their compositions but, apart from near-duplicates,
not their structures: the prediction set holds the other minima of the same compositions, almost all of them higher in energy. Both are screened
by the same code and judged by the same first-principles protocol (\cref{fig:descriptors}f). We screen each set separately, select entries
for calculation in score-ordered rounds, and assess them in 3,625 first-principles runs, each a Quantum ESPRESSO band structure followed
by a BoltzTraP2 transport calculation, recomputing runs of heavy-element compounds with SOC. Because the screen itself chose what to calculate,
with only some rounds sampling entries that it would reject, the calculations measure how well the screen concentrates the compounds that
pass; they cannot measure how many it misses. Four questions organise the results. (Q1)~Does a gap in the d-DOS at $E_F$, read from DFT or
from predicted spectra, identify compounds that are gapped in a band-structure calculation? (Q2)~Does the Seebeck indicator, on either
kind of spectrum, concentrate the entries that pass the first-principles criteria; does it add to electron counting; and does it rank
them? (Q3)~How do SOC and the choice among structural variants change which entries pass? (Q4)~Which compositions pass, and how do they relate
to known and less-explored thermoelectric chemistry? The first three questions test the screen. The fourth uses it, and its answer is a
list of computational candidates, not of measured thermoelectrics.

\section{Data and methods}\label{sec:methods}

The study applies one screen to two kinds of spectra. A difference in outcome can be attributed to the source of the spectra only if both
populations are screened by the same code, the predicted spectra come from a documented model, and one recorded first-principles protocol selects and assesses the entries of both. \Cref{sec:m-data}
describes the database and the two populations, \cref{sec:m-model} the model, \cref{sec:m-descriptors} the screening indicators,
\cref{sec:m-queues} the selection of entries, \cref{sec:m-dft,sec:m-transport} the first-principles calculations and the certification
criteria, and \cref{sec:ml-eval} the evaluation; \cref{tab:populations} summarises where the two populations nonetheless differ.

\subsection{Two density-of-states populations from one database}\label{sec:m-data}
\emph{The database.} All structures and reference data come from the ternary, DFT-computed part of DXMag\cite{XiaoTadano2025,XiaoTadano2026npj},
metadata release 2025-09-09 (CC BY 4.0; \url{https://www.nims.go.jp/group/spintheory/database/}); its machine-learned quaternary and
all-d entries\cite{XiaoTadano2026npj} are not used. The 27,864 combinations of composition and prototype draw on 41 elements (27 d-block
metals, La, Lu and twelve elements of groups 13--15). They comprise 9,396 half-Heusler XYZ, 9,072 regular and 9,072 inverse X$_2$YZ, and
324 X$_3$Z compounds, which we group with the full Heuslers. Each combination was relaxed with the Vienna Ab initio Simulation Package
(VASP), projector-augmented-wave (PAW) potentials and the Perdew--Burke--Ernzerhof (PBE) functional, collinearly spin-polarised and
without SOC or a Hubbard correction (as documented on the database page). The relaxations started from several magnetic configurations
and a series of tetragonal strains, and every inequivalent minimum, distinguished by energy and magnetic moments, was kept. This gives
106,235 entries: 27,864 ground states and 78,371 metastable states. Each entry lists its cell, positions, energy, formation energy,
magnetic moments and distance from a convex hull built with the competing phases of the Open Quantum Materials Database (OQMD, version
1.6)\cite{XiaoTadano2025,OQMD2013,OQMD2015}. For a subset, harmonic phonons are given at the commensurate $q$-points of a
$2\sqrt2\times2\sqrt2\times2$ supercell. We group the entries into six families: the full, half and inverse prototypes, each in a cubic
($Fm\bar3m$, $F\bar43m$) and a tetragonal ($I4/mmm$, $I\bar4m2$) form.

\emph{Two populations.} Spin- and orbital-resolved DOS are distributed separately, as the HeuslerDB electronic-DOS dataset\cite{DXMagDOS}
(VASP-PBE, primitive cell\cite{XiaoTadano2025}, pymatgen format\cite{pymatgen}). Its archives contain 51,257 spectra, all of which we used,
although the record description quotes 51,373. Matched to the database by identifier, the files follow a simple rule: every composition
and prototype has the spectrum of its lowest-energy cubic entry, and the lowest-energy tetragonal entry has one where it is the ground
state. DOS availability therefore splits the database into two populations of similar size (\cref{tab:populations}). The \emph{training
set}, 51,257 entries with a DFT spectrum, trains the DOS model and is screened on its own DFT spectra. The \emph{prediction set}, the other
54,978 entries (2,595 lowest tetragonal minima without a spectrum and 52,383 further metastable minima), is screened on predicted spectra
(Supplementary Fig.~S1). Every prediction entry shares its composition with a training entry. Of the 51,103 prediction entries that also have a training
entry of the same family, none lies below it, and the median energy difference is 32~meV\,atom$^{-1}$; 15.1\% of all prediction entries
lie within 1~meV\,atom$^{-1}$ of their training counterpart, most likely relaxations that returned to the same crystal. The prediction set thus samples alternative configurations of known compositions, not new
chemistry. The two sets are screened and evaluated separately throughout and meet only in the final list of compositions.

\emph{Structures, units and counting.} The crystal graphs that the model reads (\cref{sec:m-model}) are built from the structure stored with
each spectrum (training set) or from the DXMag cell and positions (prediction set), without relaxation. The archive spectra carry a
uniform factor of two: integrating the archive's spin-resolved total DOS up to $E_F$ gives twice the DXMag magnetisation (median ratio 2.00
over 150 magnetic entries; Supplementary Note~1, Fig.~S2 and Table~S2). Archive spectra, model targets and predictions therefore all carry
this factor relative to physical values; ratios are unaffected, and absolute thresholds are converted where they matter. We count in three units. An \emph{entry} is one DXMag record, identified by its universally unique identifier (UUID). A \emph{run}
is one first-principles calculation; of 3,625 runs, 1,562 belong to the training set, 2,021 to the prediction set and 42, run under a
composition name without a UUID, to neither. A \emph{composition} merges entries with the same DXMag composition string across families and sets;
because the string keeps the site order, half-Heusler site permutations count as different compositions. Prediction entries that share
composition, family and space group with a training entry and differ from it by less than 5~meV\,atom$^{-1}$ in DXMag energy and 1\% in
volume are \emph{near-duplicates} (Supplementary Note~2). The \emph{reference crystal} of a composition is its calculated entry of lowest
DXMag energy, and a \emph{structural variant} is any other calculated entry at least 5~meV\,atom$^{-1}$ above it. Compounds are named
conventionally in the text and figures (ZrNiSn for the DXMag string NiZrSn, whose first element occupies the tetrahedral site);
Supplementary tables list DXMag strings, and Supplementary Table~S3 gives both. Throughout, $E_F$ in a d-DOS (model target, label and
indicators) is the Fermi level of the archive spectrum; in our first-principles calculations it is their own Fermi level
(\cref{sec:m-dft}).

\subsection{The spin-resolved d-DOS model}\label{sec:m-model}
\begin{figure*}[t]
\centering
\includegraphics[width=\textwidth]{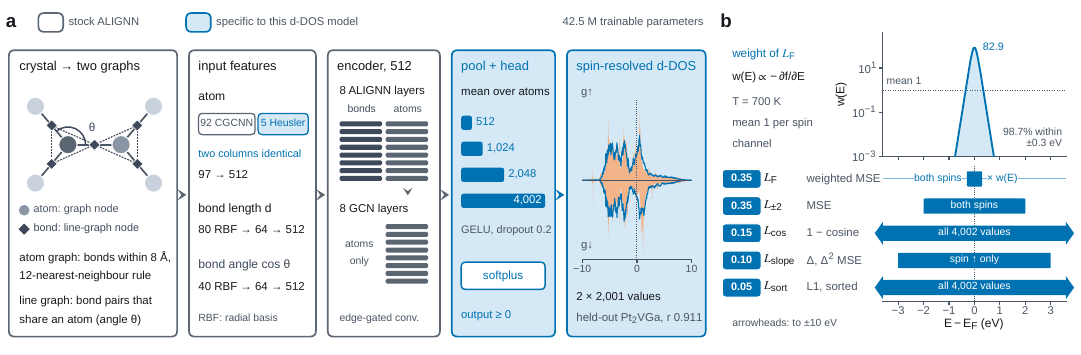}
\caption{\textbf{The d-DOS model and its Fermi-focused loss, as implemented.} \textbf{a},~Each crystal becomes an atom graph (nodes,
atoms; edges, bonds to nearest neighbours) and a line graph (nodes, bonds; edges, pairs of bonds that share an atom, carrying the bond
angle). The stock encoder of the atomistic line graph neural network (ALIGNN; grey outline) is followed by the parts specific to this model (blue): a mean over atoms, a three-layer head drawn
with bar length proportional to layer width (from the 512 pooled features), and a softplus that keeps the spin-resolved d-DOS non-negative; ``two columns identical'': the d-electron count enters twice. Inset energy axis, $E-E_F$ (eV). The inset is the held-out
training-set spectrum at the median $r$ (DFT d-DOS vermilion, prediction blue). \textbf{b},~The weight $w(E)$ of $\mathcal L_\mathrm F$ as
implemented (logarithmic axis), and the energy range, spin channels and weight of each of the five terms of \cref{eq:loss}. Settings are
those of all 30 fold models (\cref{sec:m-model}).}
\label{fig:model}
\end{figure*}
\emph{Target.} For each training entry the archive provides the d-projected DOS $N^{\mathrm d}_s(E)$ of spin channel $s$, summed over sites
and orbitals, for a structure of $N$ atoms. The model learns the per-atom d-DOS on a fixed grid of energies measured from $E_F$:
\begin{equation}
g_s(E_k)=\max\Bigl[0,\ \tfrac1N\,N^{\mathrm d}_s(E_F+E_k)\Bigr],\qquad E_k=-10+0.01k~\mathrm{eV}, \label{eq:target}
\end{equation}
with $k=0,\dots,2000$; the archive spectra are interpolated linearly onto this grid, without smoothing. The two spin channels together form
a vector of 4,002 values (\cref{fig:model}a, right). The reasons for the d projection and the spin resolution are given in \cref{sec:intro}.

\emph{Representation and architecture.} The model is an atomistic line graph neural network\cite{ALIGNN}, which encodes each crystal as two
graphs (\cref{fig:model}a). In the atom graph, edges join each atom to its nearest neighbours (12 nearest neighbours with a
neighbour-search radius of \SI{8}{\angstrom}, the settings of the ALIGNN graph builder) and carry radial-basis expansions of the bond
length. In the line graph, the nodes are bonds, and the edges join bonds that share an atom and carry the cosine of the angle between
them, so that three-body geometry enters the atom features. The line graph matters here because the cubic and tetragonal forms of a
Heusler prototype differ in bond angles as well as bond lengths. Each atom starts from 97 features. Ninety-two are the one-hot encoding
of nine elemental properties introduced with the crystal graph convolutional neural network (CGCNN)\cite{CGCNN} and used by default in
ALIGNN. The other five are Heusler-motivated: the d-electron count (included twice), the largest magnitude of a common oxidation state, the
distance of the d count from half or full filling, and the period of the d shell. These columns are per-atom proxies for the
valence-electron counts on which the electron-counting rules of Heuslers rest\cite{Kandpal2006,Graf2011}. The model receives no magnetic
moments, so it must infer magnetic order and exchange splitting from the structure alone. The ALIGNN encoder is used without modification
apart from its size: eight ALIGNN layers, each an edge-gated convolution on the line graph followed by one on the atom graph, and eight
further atom-graph convolutions, all 512 features wide, where the published configuration uses four, four and 256\cite{ALIGNN}. The
task-specific parts are a three-layer head (512, 1,024, 2,048 and 4,002 units, Gaussian error linear unit (GELU) activations, dropout 0.2)
acting on the atom-averaged features, and a softplus output, $\ln(1+e^x)$, that keeps the predicted DOS non-negative.

\emph{A loss focused on the Fermi level.} Thermoelectric screening reads the DOS within a few $k_BT$ of $E_F$, so the loss weights that
region most strongly (\cref{fig:model}b):
\begin{equation}
\mathcal{L}=0.35\,\mathcal{L}_\mathrm{F}+0.35\,\mathcal{L}_{\pm2}+0.15\,\mathcal{L}_\mathrm{cos}+0.10\,\mathcal{L}_\mathrm{slope}+0.05\,\mathcal{L}_\mathrm{sort}. \label{eq:loss}
\end{equation}
The first two terms are mean-squared errors near $E_F$. $\mathcal L_\mathrm F$ is weighted by the Fermi window $w(E)=-\partial f/\partial E$
at \SI{700}{K}, normalised to unit mean over the grid, so that the weight at $E_F$ is 83 times the average and 98.7\% of it lies within
$\pm\SI{0.3}{eV}$; $\mathcal L_{\pm2}$ is the mean-squared error within \SI{2}{eV} of $E_F$. The other three terms compare the spectra as
a whole: $\mathcal L_\mathrm{cos}$ is the cosine distance between predicted and DFT spectra; $\mathcal L_\mathrm{slope}$ compares first
and (with weight 0.1) second finite differences within \SI{3}{eV} of $E_F$, for the spin-up channel only; and $\mathcal L_\mathrm{sort}$ is
the L1 distance between the sorted values of the two spectra, taken over both spins and all energies. Equations are given in
Supplementary Note~3. No term rewards a DOS of exactly zero, so a small residual DOS inside a gap costs little (\cref{sec:r-model}). The
training temperature (\SI{700}{K}) is lower than the screening temperature (\SI{900}{K}), so the window used in training is slightly
narrower than the one the indicators use.

\emph{Training and prediction.} Each family was split by shuffled five-fold cross-validation (seed 42): in each fold, 72\% of the family
trained a model, 8\% served for validation and 20\% tested it, giving 30 models. The family models were to be fine-tuned from an
all-family model pretrained on every training spectrum; the run records confirm this for two of the 30 models and leave the parent of the
others unconfirmed (Supplementary Table~S4 and Fig.~S3). All were trained with the AdamW optimiser, a reduce-on-plateau learning-rate
schedule and stochastic weight averaging (SWA)\cite{SWA} over the last fifth of the epochs (settings in Supplementary Table~S5): 300
epochs with batch size 128 (cubic full Heuslers: 150 or 200 epochs, batch size 64), learning rate $10^{-3}$ and weight decay
$5\times10^{-5}$. Each fold model kept two sets of weights: a best-validation checkpoint, which gives all held-out metrics, and the SWA
weights, which give all predictions. Each prediction-set structure was passed through the five SWA models of its family, and the predicted d-DOS is the
mean of their outputs. The population standard deviation of the five outputs, $u_s(E)$, measures their spread; its average over
$|E-E_F|\le\SI{2}{eV}$ and both spin channels is $u_F$. Every training crystal was a test entry for one of the five models but a training
or validation entry for the other four, so a prediction entry that nearly duplicates a training crystal receives a spectrum close to
that crystal's fitted, in-sample spectrum. Because the pretraining
saw every spectrum that later served as a test entry, the held-out metrics of any fine-tuned model are upper bounds. The provenance
records are given in Supplementary Note~3, the model and its deployment in Supplementary Figs.~S4 and S5, and the implementation details that earlier internal descriptions of the workflow stated differently in Supplementary Note~4.

\subsection{DOS-derived screening indicators}\label{sec:m-descriptors}
\Cref{fig:descriptors} shows each descriptor on a real spectrum and how it is used; panel d summarises what each one reads, its role
and what it cannot establish. \emph{Indicators.} As outlined in \cref{sec:intro}, replacing the transport distribution\cite{BoltzTraP2} by the spin-summed d-DOS
$g=g_\uparrow+g_\downarrow$, that is, assuming the same $v^2\tau$ for all states in the window, gives the Cutler--Mott form of the
thermopower\cite{CutlerMott1969,Jonson1980}. With the Fermi window $w_\mu(E)=-\partial f/\partial E$ at $T=\SI{900}{K}$, the temperature of
interest for half-Heusler power generation,
\begin{gather}
L_n(\mu)=\sum_k g(E_k)\,w_\mu(E_k)\,(E_k-\mu)^n\,\Delta E , \label{eq:moments}\\
S(\mu)=-\frac{1}{eT}\,\frac{L_1(\mu)}{L_0(\mu)} , \label{eq:seebeck}
\end{gather}
with $e$ the elementary charge, $\Delta E=\SI{0.01}{eV}$, and $\mu$ scanned over $E_F\pm\SI{1}{eV}$ in steps of \SI{0.02}{eV}. $L_0$ is the
DOS averaged over the thermal window around $\mu$, $L_1$ measures how unevenly the states in the window lie above and below $\mu$, and
$L_2$ how widely they are spread about $\mu$; $S(\mu)$ is thus the mean energy of the window-weighted states relative to $\mu$, divided by $-eT$. Four indicators
are formed:
\begin{equation}
\begin{gathered}
S^{\mathrm d}_{\max}=\max_\mu|S(\mu)|,\qquad P^{\mathrm d}_{\max}=\max_\mu S^2L_0,\\
\xi=\frac{(k_BT)^2}{L_2/L_0-(L_1/L_0)^2}\Big|_{\mu^*},\qquad L_0^{*}=L_0(\mu^*),
\end{gathered}\label{eq:ad}
\end{equation}
where $\mu^*$ maximises $S^2L_0$. The \emph{d-DOS Seebeck indicator} $S^{\mathrm d}_{\max}$ is the largest thermopower that a shift of $\mu$
within \SI{1}{eV}, that is, doping, could reach. Because $S(\mu)$ is large only where the window-weighted DOS is small and asymmetric, next to
a band edge or a deep minimum, and its maximum grows with the width of a gap\cite{GoldsmidSharp1999,Gibbs2015GapEstimate},
$S^{\mathrm d}_{\max}$ acts as a gap detector smoothed over the thermal window. The other three indicators are $P^{\mathrm d}_{\max}$, the
power-factor analogue, in which $L_0$ takes the place of the conductivity; $\xi$, an inverse energy variance of the transport window,
motivated by the argument of Mahan and Sofo that a narrow distribution of carrier energies favours a large figure of
merit\cite{MahanSofo1996}; and $L_0^*$, the window-weighted DOS at the optimum. $S^{\mathrm d}_{\max}$ and $\xi$ do not depend on the
overall scale of $g$, whereas $P^{\mathrm d}_{\max}$ and $L_0^*$ do. For 13\% of entries in each set the optimum $\mu^*$ lies at the edge of
the scan, where $P^{\mathrm d}_{\max}$ and $L_0^*$ are lower bounds. All four indicators are
evaluated in Supplementary Table~S12; the main text uses $S^{\mathrm d}_{\max}$ and $P^{\mathrm d}_{\max}$.

\emph{The d-DOS gap label.} An entry is labelled a \emph{semiconductor} if both spin channels fall below 0.15~states\,eV$^{-1}$ (archive
units) somewhere within \SI{0.03}{eV} of $E_F$, a \emph{half-metal} if only one does and the spin polarisation at $E_F$ exceeds 0.9 in
magnitude, and a \emph{metal} otherwise. The \emph{d-DOS gap} is the width of the contiguous region around $E_F$ in which $g$ stays below
the same threshold. Because the label tests each channel separately but the gap is measured on their sum, a semiconductor label can carry
a gap of zero width (541 training and 1,156 prediction entries). The same code screens both sets, but its input is normalised differently.
The prediction set is screened on the per-atom ensemble mean, whereas the training-set screen reads the archive d-DOS per stored primitive cell, without dividing by its three or four atoms. In physical units the label threshold is therefore 0.075~states\,eV$^{-1}$\,atom$^{-1}$ per
spin in the prediction set and 0.025 or 0.019 in the training set, three to four times stricter (\cref{tab:populations}). $S^{\mathrm d}_{\max}$
and $\xi$, which do not depend on the scale of $g$, are unaffected, and raw indicator values and label rates are compared only within a
population. Two baselines are evaluated alongside the indicators: an earlier composite of four d-DOS features at $E_F$, computed for the
prediction set only, which ordered
part of the first selection round (Supplementary Note~5), and the electron-count rule, which flags half-Heuslers with 18 and full and
inverse Heuslers with 24 valence electrons.

\subsection{Selection of entries for first-principles assessment}\label{sec:m-queues}
Entries were calculated in nine successive rounds (\cref{fig:descriptors}f). Three \emph{priority filters} were recorded for each entry: a semiconductor label with a
d-DOS gap of 0.10--\SI{1.00}{eV}; stability (DXMag hull distance below \SI{0.10}{eV.atom^{-1}}, negative formation energy and no phonon instability, that is, DXMag neither marks the entry unstable nor
lists a frequency below $-5$~cm$^{-1}$; an entry lacking a value passed); and, for the prediction set, a model spread $u_F$ below its 75th percentile over
all families. The filters ordered the selection and excluded nothing. Later rounds, and the analyses below, use a wider \emph{gap range}, a
d-DOS gap of 0.10--\SI{1.40}{eV}, chosen after the first results. Within each round, entries were grouped into tiers of d-DOS gap or of model confidence ($u_F$) and ordered within a
tier by $P^{\mathrm d}_{\max}$ or $S^{\mathrm d}_{\max}$; some rounds deliberately sampled other labels (Supplementary Note~2, Table~S6 and
Fig.~S6). Because the rounds drew on the two sets in different proportions, the sets were sampled differently. Rounds 7 and 8, ordered by label tiers, supplied 1,318 of the 2,021 prediction-set runs (65\%) but 338
of the 1,562 training-set runs (22\%), and no certified entry (certification is defined in \cref{sec:m-cert}). No round was random, so
outcome rates are conditional on the selection, and the screen's recall, the fraction of all passing compounds that it would find, is not
measured. $S^{\mathrm d}_{\max}$ was identified as the best prioritiser only retrospectively. All indicator values and labels analysed here
are those of the final screen.

\subsection{First-principles electronic structure and spin--orbit coupling}\label{sec:m-dft}
Self-consistent field (SCF) and non-self-consistent (NSCF) calculations used Quantum ESPRESSO (QE)\cite{QE2009,QE2017} with the PBE
functional\cite{PBE1996}, the efficiency set of the Standard Solid-State Pseudopotentials (SSSP)\cite{SSSP2018}, Marzari--Vanderbilt
smearing of 0.01~Ry\cite{MV1999} and Monkhorst--Pack meshes\cite{MP1976}. Wavefunction cutoffs were set to the largest SSSP value among the
species (30--90~Ry; for SOC, 1.15 times the value of the fully relativistic potentials), and the SCF threshold was $10^{-8}$~Ry. The meshes
had $n_i=\max(6,\operatorname{round}(K/|\mathbf a_i|))$ points along the $i$-th reciprocal-lattice direction, where $\mathbf a_i$ is the
$i$-th lattice vector, with $K=\SI{40}{\angstrom}$ for the shifted SCF mesh and \SI{55}{\angstrom} for the unshifted NSCF mesh. Every
calculation used the DXMag structure, symmetrised and reduced to its primitive cell with spglib at a tolerance of $10^{-3}$ and not relaxed
further (Supplementary Table~S7). Compounds containing V, Cr, Mn, Fe, Co or Ni were calculated spin-polarised from ferromagnetic starting
moments, with one antiparallel retry on failure; the others were calculated without spin polarisation. Each run's electronic class follows
from its NSCF eigenvalues: \emph{metallic} if every spin channel has a band crossing $E_F$, \emph{half-metallic} if some but not all do, and
otherwise \emph{gapped} if the smallest band gap over channels is at least the smearing width, \SI{0.136}{eV}, and \emph{narrow-gap} below
it. We refer to these calculations as QE-PBE, to the electronic class as the QE-PBE class and to
this band gap as the QE-PBE gap. It is read on the NSCF mesh, which for 1,846 runs (98 of the 155 certified entries) is an odd,
unshifted mesh that does not contain the X and L points of the face-centred-cubic Brillouin zone; for half-Heuslers whose conduction-band
minimum lies at X these gaps are upper bounds.

\label{sec:m-soc}Runs of compounds containing an element with atomic number $Z\ge72$ (Hf, Ta, W, Re, Os, Ir, Pt, Au, Hg, Tl, Pb or Bi;
1,956 runs) were eligible for a non-collinear recalculation with SOC. It was attempted for 330 of the 337 gapped and 55 of the 68
narrow-gap runs, for 245 of the 254 metallic runs that met the Seebeck and power-factor criteria, (i) and (ii) of
\cref{sec:m-cert}, and for 63 other metallic runs. Of
these 693 attempts, 607 finished and 86 failed and were not retried (Supplementary Table~S8). Each SOC calculation reused the cell,
positions and $k$-meshes of its scalar-relativistic partner (verified for all pairs) and converged to a non-magnetic state, as did all its
scalar partners. It used fully relativistic PSlibrary PAW potentials\cite{DalCorso2014} (version 1.0.0; 0.3.1 for Co and Mn, 0.2.3 for
Ir), cutoffs set per compound, enough spinor bands to match the scalar band window, the same smearing and the same transport extraction,
so that paired quantities are defined identically. The pairs differ in the pseudopotential family as well as in the treatment of
relativity, and a same-potential scalar control was not performed, so paired differences are those of the fully relativistic
recalculation as a whole.

\subsection{Transport and certification}\label{sec:m-transport}
Transport coefficients were computed with BoltzTraP2\cite{BoltzTraP2} from the NSCF eigenvalues (interpolation multiplier 15) in the
constant-relaxation-time approximation (CRTA):
\begin{gather}
\Sigma_{\alpha\beta}(\varepsilon)=\frac{e^2\tau}{N_k}\sum_{n,\mathbf k}v_{n\mathbf k,\alpha}v_{n\mathbf k,\beta}\,
\delta(\varepsilon-\varepsilon_{n\mathbf k}), \label{eq:sigmaE}\\
\Lambda^{(m)}_{\alpha\beta}(\mu,T)=\frac{1}{V}\int\Sigma_{\alpha\beta}(\varepsilon)\,(\varepsilon-\mu)^m
\left[-\frac{\partial f}{\partial\varepsilon}\right]d\varepsilon , \label{eq:onsager}
\end{gather}
with $\sigma=\Lambda^{(0)}$ and $S=-(eT)^{-1}[\Lambda^{(0)}]^{-1}\Lambda^{(1)}$, where $N_k$ is the number of $k$-points, $V$ the cell volume
and $\alpha,\beta$ Cartesian directions. Unlike \cref{eq:moments}, these expressions include the band velocities of all bands. Spin
channels were merged with a common $\tau$, tensors were reduced to a third of their trace, and the power factor is reported per
relaxation time, $\pf=S^2\sigma/\tau$, from 300 to \SI{1200}{K}. Coefficients were evaluated at the grid point nearest the QE Fermi level,
the \emph{undoped} point, and with $\mu$ optimised for $\pf$ within $|\mu-E_F|\le\SI{0.5}{eV}$, which represents doping. For gapped runs the
QE Fermi level is fixed by the smearing and lies anywhere in the gap (6--94\% of the gap above the valence-band maximum in a sample of 80
runs), so the undoped point is a convention rather than an intrinsic chemical potential. For ranking, the window was widened to
\SI{1.0}{eV} and the carrier concentration $n$, counted from the undoped point, was capped at $|n|\le10^{20}$ or $10^{21}$\,cm$^{-3}$,
separately for hole (p-type) and electron (n-type) doping; points whose Seebeck sign contradicts the carrier type were excluded. Runs
completed before the correction of a driver error, which had interpolated the self-consistent instead of the NSCF eigenvalues, were recomputed; we call this the mesh correction (Supplementary Fig.~S7).

\label{sec:m-cert}A run, and the entry it belongs to, is \emph{certified} if it passes five criteria in sequence: (i) the \emph{Seebeck
criterion}, $|S|>\SI{60}{\micro V.K^{-1}}$ at the undoped point at some temperature; (ii) the \emph{power-factor criterion}, a best $\pf$, over temperature and over the undoped and the
optimised $\mu$, above $\pf_0=\SI{1.09e12}{W.m^{-1}.K^{-2}.s^{-1}}$; (iii) the \emph{stability criterion}, a
DXMag hull distance below \SI{0.10}{eV.atom^{-1}}; (iv) a scalar-relativistic QE-PBE gap of at least \SI{0.136}{eV}; and (v) for compounds
with $Z\ge72$, criterion (ii) re-evaluated with SOC. Runs passing (i)--(iii) are \emph{qualified}; qualified, gapped runs with $Z\ge72$ are
the \emph{SOC candidates}; heavy-element runs without a finished SOC calculation are not certified. The threshold $\pf_0$ is the median
\SI{900}{K} value of the first 1,401 completed runs; it was set during the campaign and then held fixed. The same median over mesh-corrected
runs, $\SI{1.15e12}{W.m^{-1}.K^{-2}.s^{-1}}$, would certify 77 compositions (Supplementary Table~S9), and the development of the criteria
is recorded in Supplementary Note~6, Table~S10 and Fig.~S8. The stability threshold is stricter than the \SI{0.22}{eV.atom^{-1}} that Xiao
and Tadano calibrated on experimentally known Heuslers\cite{XiaoTadano2025}. Because the best $\pf$ is taken over 300--\SI{1200}{K} while
$\pf_0$ is a \SI{900}{K} median, the power-factor criterion favours high temperature; counts at \SI{900}{K} are therefore reported alongside,
and the sensitivity of the certified count to every threshold is given in Supplementary Table~S9. For ranking, each composition is represented by its certified entry with the highest $\pf$, and compositions are compared on
$|S|$ and $\pf$ at \SI{900}{K} and $|n|\le\SI{e21}{cm^{-3}}$, with $\mu$ in the widened window, for p- and n-type doping separately and with
SOC values for heavy compounds; we call this the ranking convention. Pareto layers follow from non-dominated sorting: the first layer holds the compositions
that no other composition exceeds in both $|S|$ and $\pf$ for the same carrier type. Certification is a computational status under the CRTA, not a figure of merit:
$\tau$ and $\kappa_L$ are not computed.

\begin{figure*}[t]
\centering
\includegraphics[width=\textwidth]{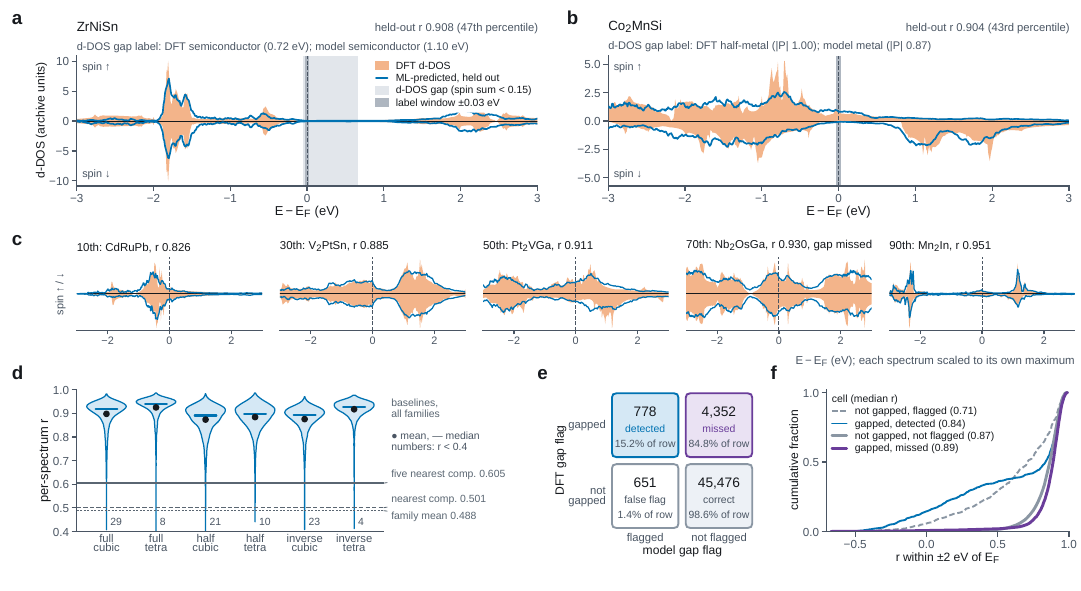}
\caption{\textbf{Predicted spectra reproduce the shape of the d-DOS but miss most gaps.} All panels use the 51,257 held-out training-set
spectra, each predicted by the best-validation checkpoint of the fold that held it out; DFT d-DOS as vermilion fill, ML prediction as
blue line, spin down below. \textbf{a,b},~ZrNiSn (half-Heusler, 18 valence electrons) and half-metallic Co$_2$MnSi with their d-DOS gap labels ($|P|$, spin polarisation at $E_F$); shading,
the d-DOS gap of the label (spin-summed DFT d-DOS below 0.15 around $E_F$: \SI{0.72}{eV} for ZrNiSn, none for Co$_2$MnSi, whose label
rests on one spin channel). Labels in a,b are read per atom, the prediction-set convention, so the ZrNiSn d-DOS gap differs from the per-cell reading of
Fig.~\ref{fig:descriptors}a (0.57~eV). \textbf{c},~Entries nearest the 10th to 90th percentiles of per-spectrum Pearson $r$; ``gap missed'' refers to the flag of e. \textbf{d},~$r$ by
family (violins above $r=0.4$; numbers, spectra below it) and the all-family mean $r$ of three baselines (grey lines).
\textbf{e},~Held-out gap flag, a test separate from the screening label (spin-summed d-DOS below 0.05 anywhere within \SI{1}{eV} of $E_F$; DFT as truth); missed gaps in purple. \textbf{f},~Distributions of $r$
within \SI{2}{eV} of $E_F$ for the cells of \textbf{e}. DOS values are in archive units, a factor of two above physical values
(\cref{sec:m-data}); the metrics are upper bounds, because pretraining, where used, saw the test spectra (\cref{sec:m-model}). d-DOS, d-orbital-projected density of states.}
\label{fig:fidelity}
\end{figure*}
\subsection{Evaluation}\label{sec:ml-eval}
\label{sec:m-stats}\emph{Held-out spectra.} Each training entry was predicted once, by the best-validation checkpoint of the fold in which it
was a test entry. Pearson's correlation coefficient $r$ between predicted and DFT spectra was computed per spectrum, over all 4,002 values
and within \SI{2}{eV} of $E_F$, and compared with three baselines that involve no trained model: the family-mean DOS, the DOS of the nearest entry,
and the mean DOS of the five nearest entries in element-fraction space. A spectrum is flagged as gapped when $g$ falls below
0.05~states\,eV$^{-1}$\,atom$^{-1}$ (archive units) anywhere within \SI{1}{eV} of $E_F$. With the DFT flag as truth, gapped spectra that the
model flags are true positives (TP) and those it misses false negatives (FN); non-gapped spectra are true negatives (TN) or false positives
(FP). The recall is $\mathrm{TP}/(\mathrm{TP}+\mathrm{FN})$, the fraction of gaps that the model detects, and the balanced accuracy (BA) and
Matthews correlation coefficient (MCC) are
\begin{gather}
\mathrm{BA}=\tfrac12\left[\frac{\mathrm{TP}}{\mathrm{TP}+\mathrm{FN}}+\frac{\mathrm{TN}}{\mathrm{TN}+\mathrm{FP}}\right], \label{eq:ba}\\
\mathrm{MCC}=\frac{\mathrm{TP}\,\mathrm{TN}-\mathrm{FP}\,\mathrm{FN}}
{\sqrt{(\mathrm{TP}{+}\mathrm{FP})(\mathrm{TP}{+}\mathrm{FN})(\mathrm{TN}{+}\mathrm{FP})(\mathrm{TN}{+}\mathrm{FN})}}. \label{eq:mcc}
\end{gather}

\emph{Screening performance.} Indicators were evaluated within each population on the calculated entries (1,562 training-set and 2,016
prediction-set entries; five prediction entries were calculated twice), with certified entries as positives. We report the area under the
receiver-operating-characteristic curve (ROC-AUC), computed as the Mann--Whitney statistic, that is, the probability that a randomly
chosen certified entry outranks a randomly chosen uncertified one; the area under the precision--recall curve (precision--recall AUC);
and the \emph{recovered fraction} $R(q)$, the fraction of all certified entries that lie within the top fraction $q$ of calculated entries
ranked by the indicator, for which random ordering gives $q$. ROC-AUC differences were assessed by a paired bootstrap over entries (2,000
resamples), which captures sampling variation within the calculated set but not the effect of the selection. Rank correlations are
Spearman's $\rho$.

\emph{Paired and literature comparisons.} SOC effects are summarised over the 607 paired runs by the change of electronic class, the gap
difference for runs gapped in both calculations, and the ratio of best $\pf$ values; structural variants are compared with the reference
crystal of their composition in the same way. Each certified composition was searched in the literature under its conventional name and
assigned the strongest thermoelectric evidence found for the same composition in the same structure type: \emph{established} (a benchmark
in reviews), \emph{measured} (thermoelectric measurements reported), \emph{computed only}, or \emph{no study found}. Compositions whose
experimentally reported phase is not the Heusler structure, or for which DFT places another structure below the half-Heusler, were
flagged. Every source was checked against its Crossref record (Supplementary Table~S3). The audit is targeted rather than exhaustive, and
the known-compound check involves no numerical comparison with published coefficients.

\section{Results}\label{sec:results}

\subsection{The screening task}\label{sec:r-pop}
\begin{figure*}[t]
\centering
\includegraphics[width=\textwidth]{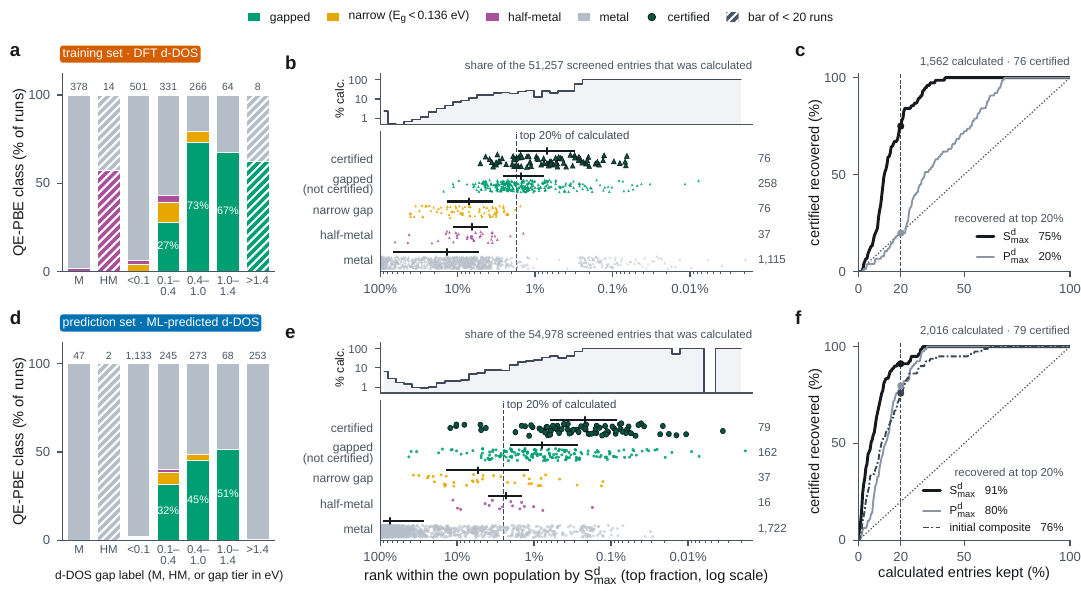}
\caption{\textbf{The gap label passes mostly metals; the Seebeck indicator places the certified entries at the top.}
Top, training set (DFT d-DOS); bottom, prediction set (predicted d-DOS). \textbf{a,d},~QE-PBE class of the
calculated runs in each d-DOS gap label (M, metal; HM, half-metal; otherwise gap tier in eV); numbers above the bars, runs; white
numbers, gapped share; hatching, labels with fewer than 20 runs. \textbf{b,e},~Calculated entries placed by the rank of
$S^{\mathrm d}_{\max}$ within their own population (top fraction, logarithmic, higher rank to the right) and grouped by first-principles
outcome (exclusive lanes, so ``gapped'' means gapped but not certified; entries per lane on the right); bars above each lane, interquartile range and median; upper strip, share of each rank bin that was calculated; certified entries dark green with a dark edge; triangles, training set; circles, prediction set. \textbf{c,f},~Share of certified entries kept when only the highest-ranked fraction of
calculated entries is retained, for each screening score (dotted, random order; initial composite, the four-feature d-DOS score that ordered part of the first round, Supplementary Note~5); the key gives the value at 20\%. Panels b,c,e,f count entries
(1,562 and 2,016; five prediction-set entries were calculated twice), a,d runs. d-DOS, d-orbital-projected density of states; QE-PBE, Quantum ESPRESSO calculation with the PBE functional.}
\label{fig:screening}
\end{figure*}
In this study a candidate must be gapped and near-stable (criteria iii and iv of \cref{sec:m-cert}), and both properties are rare in
DXMag. Judged by their
own d-DOS, both populations are overwhelmingly metallic: only 1,550 training entries (3.0\%) and 2,394 prediction entries (4.4\%) carry a
semiconductor label, and 671 and 603 of these have a d-DOS gap within the 0.10--\SI{1.40}{eV} gap range (\cref{fig:descriptors}f). Near-stable
entries are similarly scarce. Only 6.0\% of the training set and 3.3\% of the prediction set lie within \SI{0.10}{eV.atom^{-1}} of the
hull, and no prediction entry lies on it (\cref{tab:populations}). By its own labels, fewer than one entry in a hundred is both gapped and
near-stable, so the screen has to find a small minority among metals. The prediction set also lies higher in energy than the training
set: its entries lie a median 32~meV\,atom$^{-1}$ above their training counterpart, consistent with its smaller
near-stable share. The distribution is bimodal (\cref{fig:workflow}c),
with 15.1\% of the set within 1~meV\,atom$^{-1}$, most likely relaxations that returned to the training crystal.

\subsection{Predicted spectra reproduce shape but not gaps}\label{sec:r-model}
\begin{figure}[t]
\centering
\includegraphics[width=\columnwidth]{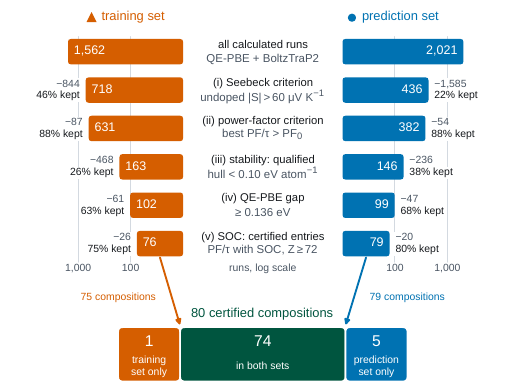}
\caption{\textbf{Each criterion removes runs; the certified entries of both sets map to 80 compositions.}
Calculated runs passing the criteria of \cref{sec:m-cert} in sequence; training set left (vermilion), prediction set right (blue). Bar length is logarithmic in the number of runs (bars start at 10 runs); rows (i)--(v) as in \cref{sec:m-cert} and \cref{fig:descriptors}f; white numbers give runs that pass, grey numbers runs removed, and ``kept'' the percentage of
the runs entering a criterion that pass it. The spin--orbit-coupling (SOC) step also removes heavy-element runs without a finished SOC calculation (five
training-set, one prediction-set). The tiles give certified compositions by set.}
\label{fig:cascade}
\end{figure}
Every prediction-set indicator inherits the errors of its predicted spectrum, so we first ask how good these spectra are. On held-out
training spectra the mean per-spectrum Pearson correlation with DFT is 0.895 (0.874--0.925 by family; Supplementary Figs.~S9--S11 and
Table~S11). The baselines reach 0.488 for the family-mean spectrum, 0.501 for the nearest composition and 0.605 for the mean of the five
nearest (\cref{fig:fidelity}d); they involve no pretraining, so the comparison favours the model. Across the range of quality, from the
tenth to the ninetieth percentile of $r$ (\cref{fig:fidelity}c), peak positions and band widths are reproduced. The one systematic
exception, the filled Zn 3d band near $-\SI{7}{eV}$ of twelve Zn-rich prediction entries, lies outside the screening window
(Supplementary Note~3).

Gaps are not reproduced. Of the 5,130 held-out spectra whose DFT d-DOS is gapped within \SI{1}{eV} of $E_F$, the model flags 778, a recall
of 0.15; the balanced accuracy (\cref{eq:ba}) is 0.57 and the Matthews correlation coefficient (\cref{eq:mcc}) 0.25 (\cref{fig:fidelity}e).
The recall varies about eightfold between families, from 0.04 (inverse cubic) to 0.33 (full tetragonal), and is 0.21 for the cubic
half-Heuslers that dominate the candidates (Supplementary Table~S11). The spectral metric does not register the failure, because a gap is
a narrow interval of near-zero weight: filling it barely changes the correlation over the whole spectrum. Spectra with a missed gap have a
median $r$ of 0.909, against 0.912 for correctly non-gapped spectra, and within \SI{2}{eV} of $E_F$ they score higher, 0.889 against 0.866.
Spectra in which the model does detect a gap are, by contrast, less faithful (median $r$ within \SI{2}{eV} of $E_F$, 0.84; \cref{fig:fidelity}f). The misses are systematic, a smoothing of
the spectrum across the gap (\cref{fig:fidelity}a--c).

Several features of the model favour this error. A squared error is small for small values: a residual 0.1~states\,eV$^{-1}$ inside
a gap costs one hundredth of a unit error at a d-band peak. The Fermi weight raises the cost by up to 83 near $E_F$, but it falls off within about \SI{0.3}{eV}, so in gaps several tenths of an electronvolt wide much of the residual lies where the
weight is small. If the Fermi level of the archive
spectra falls at varying positions inside a gap, as it does in our QE runs (\cref{sec:m-transport}), the band edges relative to $E_F$ vary
between similar compounds, and a least-squares fit averages them into the gap. Finally, the softplus output approaches zero only
asymptotically, and the five-model mean used for the prediction set can only add weight where one member places an edge. Two familiar
compounds show the two forms that the error takes (\cref{fig:fidelity}a,b). For ZrNiSn the predicted band edges lie too far apart, so the
predicted d-DOS gap is wider than the DFT one, yet residual weight inside it halves the thermopower that a screen would read from the
prediction (\cref{sec:r-indicators}). For the half-metal Co$_2$MnSi the spin-down gap is partly filled, and the label read from the prediction is metal.
A further limitation, unrelated to gaps, is that the model receives no magnetic information. It infers the magnetic state, and with it the exchange splitting, from the geometry
alone, so distinct magnetic minima at nearly the same geometry receive nearly the same spectrum.

These metrics come from best-validation checkpoints and, for fine-tuned models, from spectra that the pretraining had already seen, so
they are upper bounds; the deployed five-model mean was not evaluated on held-out spectra. Because these conditions favour the model, the
true gap recall on unseen crystals is unlikely to exceed 0.15. A predicted spectrum can thus be trusted for its shape but not for whether it has a gap near $E_F$, which is what a screen reads. Similar failures have been reported for other machine-learned DOS models: gaps lost to
residual in-gap weight\cite{PETMADDOS} or a misplaced $E_F$\cite{How2025}, as well as spurious predicted gaps\cite{Mat2Spec}.

\subsection{A binary gap label fails as a classifier even on DFT spectra}\label{sec:r-labels}
\begin{figure*}[t]
\centering
\includegraphics[width=\textwidth]{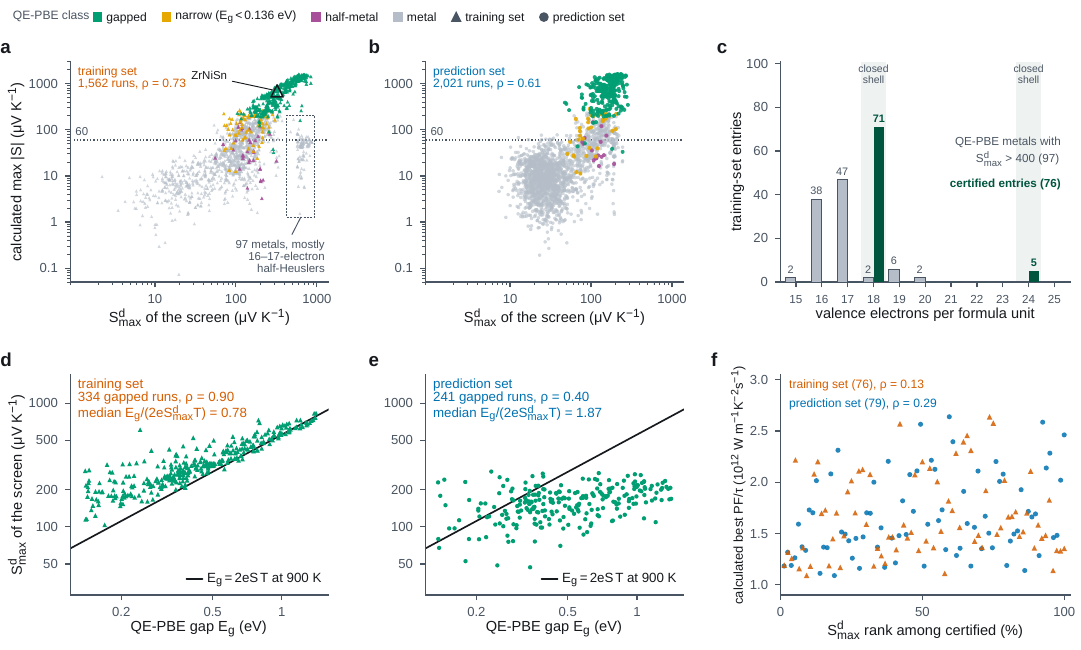}
\caption{\textbf{The Seebeck indicator tracks computed thermopower but does not rank power factors.}
\textbf{a,b},~$S^{\mathrm d}_{\max}$ (\cref{eq:ad}) of each run against its QE-PBE + BoltzTraP2 maximum undoped $|S|$ (at the QE Fermi level)
over 300--\SI{1200}{K}, for the 1,562 training-set (\textbf{a}, DFT d-DOS) and 2,021 prediction-set runs (\textbf{b}, predicted
d-DOS), coloured by QE-PBE class; dotted line, the \SI{60}{\micro V.K^{-1}} criterion. The dashed box in a marks the 97 metals with
$S^{\mathrm d}_{\max}>\SI{400}{\micro V.K^{-1}}$, 85 of them 16- or 17-electron half-Heuslers (open triangle, ZrNiSn of \cref{fig:descriptors}). \textbf{c},~Valence electrons per formula unit of these 97 metals (grey) and of the 76 certified training-set entries (dark
green); shaded, the closed-shell counts 18 and 24.
\textbf{d,e},~$S^{\mathrm d}_{\max}$ against QE-PBE gap for the gapped runs of each set; line, $E_g=2eST$ at \SI{900}{K}, the fixed-temperature form of the Goldsmid--Sharp estimate used here. \textbf{f},~Percentile rank of $S^{\mathrm d}_{\max}$ within the certified entries of each set (76 and 79)
against best $\pf$. $\rho$, Spearman's rank correlation. Sets share a panel only in f (within-set ranks); 42 runs not
attributable to one set are omitted; transport in the constant-relaxation-time approximation. d-DOS, d-orbital-projected density of states; QE-PBE, Quantum ESPRESSO calculation with the PBE functional.}
\label{fig:indicator}
\end{figure*}
Question Q1 asks whether the d-DOS gap label identifies gapped compounds. On predicted spectra the label could fail simply because
the spectra are wrong, so the training set provides the decisive test: there the label reads DFT spectra, and it can be compared with
the QE-PBE class of each calculated run (\cref{fig:screening}a,d). Of the 1,170
calculated semiconductor-labelled runs of the training set, 738 (63.1\%) are QE-PBE metals and 23 half-metals, so 761 lack a gap in at
least one spin channel. In the prediction set, where the labels come from predicted spectra, 1,678 of 1,972 (85.1\%) are metals. The
failures concentrate outside the gap range. Runs with a near-zero d-DOS gap (below \SI{0.10}{eV}) are 94\% (training) and 98\% (prediction)
metallic, many of them carrying a zero-width gap, and wide d-DOS gaps above \SI{1.40}{eV}, common only in the prediction set, are almost all
metallic there. Within the gap range, by contrast, 27--73\% of training-set runs and 32--51\% of prediction-set runs are gapped, depending on the d-DOS-gap tier
(\cref{fig:screening}a,d), against at most 0.4\% of runs labelled metallic, half-metallic or near-zero-gap. The label thus narrows the search by two orders of magnitude, yet
most entries it passes are still metals.

On DFT spectra the failure cannot come from a model, and two features of the label can explain it. The first is the d projection: the
d-DOS gap of a gapped entry lies on or above its QE-PBE gap, as expected when s and p states inside a d-orbital gap are ignored (Supplementary Fig.~S12); ZnCdAs, a metal
with a 1.13~eV d-DOS gap, is an extreme case (\cref{fig:descriptors}e). The second is the fixed threshold: a semiconductor label needs only that both spin channels dip below
0.15~states\,eV$^{-1}$ somewhere within \SI{0.03}{eV} of $E_F$, so a spectrum with a deep minimum near $E_F$ passes, and 541
semiconductor-labelled training entries have a d-DOS gap of zero width (\cref{sec:m-descriptors}). The selection rounds show what this costs in practice (Supplementary Table~S6). The two rounds ordered by label tier (one of them
by $S^{\mathrm d}_{\max}$ within its tiers), 1,656 runs drawn from semiconductor-labelled entries of lower priority, produced 26 gapped and
no certified runs, whereas the round ordered by $S^{\mathrm d}_{\max}$ alone produced 119 gapped and 47 certified runs of 272. Later rounds
drew on what earlier ones had left, so this contrast overstates the difference, but the label by itself did not find the compounds that
transport calculations certify.

A gap is required for certification, but its width also matters. Among gapped runs, the share that passes the Seebeck, power-factor and stability
criteria rises with the QE-PBE gap and is highest for gaps of 0.6--\SI{1.2}{eV} (53--61\% in the training set, 58--61\% in the prediction
set; Supplementary Fig.~S12). The rise is consistent with bipolar suppression of the thermopower in narrow-gap compounds at high
temperature, where carriers excited across the gap reduce $|S|$\cite{Sofo1994OptimumGap}. The decline beyond \SI{1.2}{eV}, by contrast, is
not physical: the $\pm\SI{0.5}{eV}$ chemical-potential window of the certification, measured from a Fermi level that the smearing places
somewhere inside the gap, cannot reach the band edges of wide-gap entries.

\subsection{What the transport calculations certify}\label{sec:r-decisions}
The certification criteria (\cref{sec:m-cert}) define the entries that an indicator should find.
Applying the criteria in sequence shows which of them shape the final list (\cref{fig:cascade}). The Seebeck and power-factor criteria
keep 40\% of the training-set runs and 19\% of the prediction-set runs. Stability then removes three-quarters (training) and three-fifths
(prediction) of what remains, and the gap requirement removes 61 and 47 qualified runs, almost all metals (57 and 45). These metals pass
the power-factor criterion only when the chemical potential is shifted by several tenths of an electronvolt, a shift that a metal, with
its large DOS at $E_F$, cannot sustain. SOC then removes 21 and 19 heavy-element runs that fall below $\pf_0$, and five and one more remain
unresolved without a finished SOC calculation (\cref{sec:r-soc}). The result is 76 certified training-set entries (75
compositions) and 79 certified prediction-set entries (79 compositions). Their union contains 80 compositions: 74 certified in both sets,
five only in the prediction set (Rh-containing half-Heuslers with Hf, Ta or Bi whose training entries were left unresolved by failed SOC
calculations) and one, ScPdAs, only in the training set.

The thresholds that bind are the hull cut and the power factor, not the thermopower. The certified count is unchanged for Seebeck
thresholds between 40 and \SI{100}{\micro V.K^{-1}}, but it moves from 49 to 104 compositions for hull cuts of 0.025 to
\SI{0.15}{eV.atom^{-1}} and from 105 to 65 for power-factor thresholds of 0.8 to 1.2\,$\pf_0$ (Supplementary Table~S9). The criteria were
set during the campaign, so the cascade describes the final protocol, not a prospective test.

\subsection{The Seebeck indicator prioritises but does not rank}\label{sec:r-indicators}
Question Q2 asks whether the Seebeck indicator concentrates these entries. Ranked by $S^{\mathrm d}_{\max}$, the
certified entries sit at the top of each population, with median ranks in the top 0.7\% (training) and 0.2\% (prediction), whereas metals
spread across the entire calculated range (\cref{fig:screening}b,e). The prediction set illustrates this. Its calculated entries
reach deep into the population (their median lies 64\% of the way down the ranking), yet 72 of the 79 certified entries are among the
1,000 highest-ranked of 54,978. Keeping only the fifth of the calculated entries with the highest $S^{\mathrm d}_{\max}$
retains 75\% of the certified training-set entries and 91\% of the certified prediction-set entries, against 20\% for random ordering; in
the training set this would cut the calculations needed per certified entry from about 21 to about 5.5 (\cref{fig:screening}c,f).

The ROC-AUC of $S^{\mathrm d}_{\max}$ is 0.869 (training) and 0.938 (prediction), against 0.649 and 0.877 for the
power-factor analogue $P^{\mathrm d}_{\max}$ (paired-bootstrap differences $+0.221$ [0.171, 0.267] and $+0.061$ [0.047, 0.076];
Supplementary Tables~S12 and~S13), and its precision--recall AUC, 0.158 and 0.340, is three to nine times the base rate. The higher
prediction-set values do not show that predicted spectra rank better. They can be explained by the selection: 65\% of the calculated prediction-set
runs, against 22\% of training-set runs, come from the low-priority label rounds, which supplied many low-ranked negatives and no
certified entry (\cref{sec:m-queues}). $P^{\mathrm d}_{\max}$ does worse because $S^2L_0$ rewards a large DOS, which metals have, and because
its scale differs between families. The initial four-feature composite, computed for the prediction set only, also ranks the calculated
entries less well (ROC-AUC 0.869, $0.068$ [0.038, 0.101] below $S^{\mathrm d}_{\max}$), and none of the 100 entries it ranked highest in
the whole prediction set was certified (Supplementary Note~5).

These ROC-AUCs come from a score-ordered sample. The hull distance, moreover, enters both a priority filter and criterion (iii), and a
feature shared by selection and certification could inflate the apparent performance. On entries that meet the stability criterion alone, however, the ROC-AUCs
are 0.873 (training) and 0.927 (prediction), so the shared feature does not drive the result. In the prediction set, all 79 certified entries, and 475 calculated
entries in total, are near-duplicates of training entries whose DFT spectra were training targets. This follows mainly from how the database and the model are built: the prediction set holds other minima of the
training compositions, and the model, seeing geometry only, gives near-duplicate geometries near-identical spectra. The prediction-set figures therefore
test whether the deployed model reproduces decisions for crystals close to its training data, not for new crystals.

Electron counting, the baseline set out in \cref{sec:intro}, needs no spectrum at all, so the indicator is worth computing only if it
does better. Counting is
statistically indistinguishable from the indicator in the training set (ROC-AUC 0.858; difference $+0.011$ [$-0.011$, 0.035]) and slightly
weaker in the prediction set (0.891). The indicator nevertheless adds information that the rule does not have. Among the 18- and
24-electron entries alone, where the rule has no discriminating power, $S^{\mathrm d}_{\max}$ still reaches ROC-AUCs of 0.805 and 0.808 (training and prediction), and 0.829 and 0.836 within the gap range. Among the entries that electron counting already favours, the d-DOS thus separates those that are certified from those that are not.

The indicator works where the label does not because it reads the DOS differently. $S^{\mathrm d}_{\max}$ integrates the asymmetry
of the DOS over a thermal window several tenths of an electronvolt wide, so it responds to the depth and width of a low-DOS region
rather than to a single value below a threshold; a deep minimum raises it even when the minimum is not empty (\cref{fig:indicator}). Across
all calculated runs it tracks the thermopower of the transport calculation, with Spearman correlations of 0.73 (training) and 0.61
(prediction) against the calculated undoped $|S|$, maximised over temperature (\cref{fig:indicator}a,b). For gapped runs on DFT spectra it
follows the gap closely, as the Goldsmid--Sharp estimate of \cref{sec:intro} suggests: Spearman 0.90 against the QE-PBE gap, with
$E_g/(2eS^{\mathrm d}_{\max}T)$ a median 0.78 (\cref{fig:indicator}d). This ratio is our fixed-temperature analogue of the estimate,
evaluated at \SI{900}{K} with $S$ maximised over $\mu$ rather than at the temperature of the thermopower maximum. The estimate itself
becomes unreliable for gaps well below $10k_BT$, about \SI{0.8}{eV} at \SI{900}{K}\cite{Gibbs2015GapEstimate}, and the ratio lies below
one because the d-DOS gap exceeds the QE-PBE gap. On predicted spectra the relation weakens to a Spearman 0.40, and the median ratio rises
to 1.87 (\cref{fig:indicator}e). Residual weight inside a gap lowers the thermopower that the gap implies: for ZrNiSn the predicted d-DOS
gap is wider than the DFT one, but $S^{\mathrm d}_{\max}$ falls from 323 to \SI{166}{\micro V.K^{-1}} (\cref{fig:descriptors}c; a certified prediction-set entry is shown in Supplementary Fig.~S13). The two readings of the DOS respond differently to this error. Residual weight
can flip a threshold label, whereas it only lowers $S^{\mathrm d}_{\max}$, and for these entries it did not remove the certified ones from
the top of the ranking.

In the training set, the indicator's highest-scoring false positives are mostly 16- and 17-electron half-Heuslers
(\cref{fig:indicator}c; IrZrSn in \cref{fig:descriptors}e). Of the 1,115 calculated training-set metals,
97 have an $S^{\mathrm d}_{\max}$ above \SI{400}{\micro V.K^{-1}}, higher than most certified entries, and 85 of them are half-Heuslers with
16 or 17 valence electrons. With one or two electrons fewer than eighteen, these compounds have the 18-electron gap in their d-DOS a few
tenths of an electronvolt from $E_F$, so the $\pm\SI{1}{eV}$ scan of $\mu$ finds it; the undoped compound, however, is a metal, and only 20
of the 97 reach the \SI{60}{\micro V.K^{-1}} Seebeck criterion. Electron counting removes these compounds, which is why it matches the
indicator on DFT spectra. Predicted spectra produce no such metals above \SI{400}{\micro V.K^{-1}}, because their residual weight caps the
indicator.

\begin{figure*}[t]
\centering
\includegraphics[width=\textwidth]{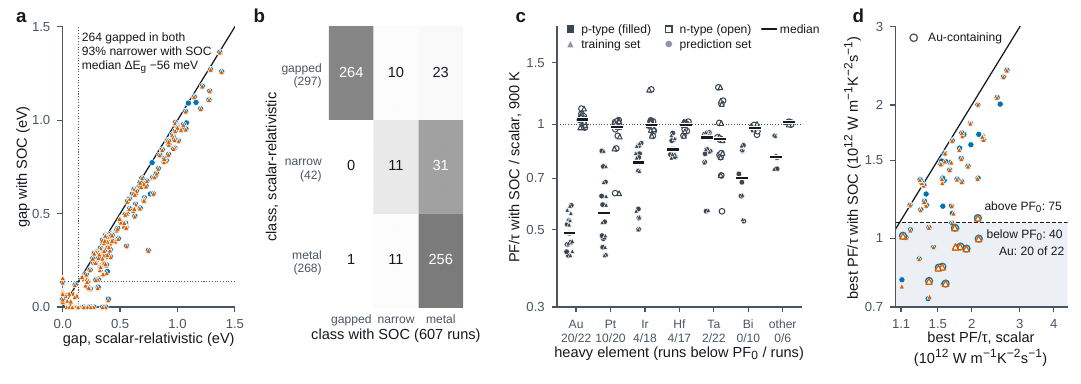}
\caption{\textbf{Spin--orbit coupling suppresses p-type power factors of heavy-element candidates.}
Paired scalar-relativistic and spin--orbit-coupling (SOC) calculations of the same structure and $k$-meshes (\cref{sec:m-soc}); triangles, training set;
circles, prediction set (near-duplicate pairs overlap as a triangle on a circle). \textbf{a},~QE-PBE gap without and with SOC for the 351 runs gapped or
narrow-gap in at least one calculation; metallic results at zero; solid line, $y=x$; dotted, the \SI{0.136}{eV} gapped/narrow boundary.
\textbf{b},~Electronic class of all 607 paired runs, including six not attributable to one set, without (rows, totals in parentheses)
and with SOC (columns); shading, row fraction; no half-metals occur because all paired runs converged non-magnetic. \textbf{c},~SOC-to-scalar $\pf$ ratio at \SI{900}{K} and $|n|\le\SI{e21}{cm^{-3}}$ of
the 115 heavy-element runs gapped and qualified (criteria i--iii) without SOC, for p-type (filled markers) and n-type doping (open markers), each run
under the first of its heavy elements in the order shown; bars, medians. \textbf{d},~Best $\pf$ (300--\SI{1200}{K}) without and with
SOC for the same 115 runs; open markers, Au-containing; shaded, below $\pf_0$. The paired calculations also differ in pseudopotential
family. $\pf$, power factor per relaxation time; $\pf_0$, the power-factor threshold (\cref{sec:m-cert}); QE-PBE, Quantum ESPRESSO
calculation with the PBE functional.}
\label{fig:soc}
\end{figure*}
The indicator does not, however, rank the candidates. Among certified entries its rank is only weakly
correlated with the calculated power factor (Spearman $\rho=0.13$ and 0.29; \cref{fig:indicator}f). The certified entries might span too
narrow a range for a correlation to appear, but over all gapped runs the correlation is no stronger (0.09 and 0.23), so range restriction
does not explain it. This is the behaviour anticipated in \cref{sec:intro}: the DOS carries neither band velocities nor valley degeneracy
separately from band mass\cite{Pei2011,Gibbs2017,Xing2017}, and a constant $v^2\tau$ fails at band extrema. For ZrNiSn the calculated thermopower at \SI{900}{K} is $-279$~\si{\micro V.K^{-1}} at the undoped point and $+145$~\si{\micro V.K^{-1}} at
the power-factor optimum (\cref{fig:descriptors}c); neither value can be read off the d-DOS curve, in part probably because the QE
Fermi level lies higher in the gap than the archive $E_F$, which sits at the valence-side edge of the d-DOS gap. Candidates are therefore ranked only on computed quantities.

\subsection{Spin--orbit coupling suppresses p-type power factors of heavy-element candidates}\label{sec:r-soc}\label{sec:r-robust}
Question Q3 asks how SOC and the choice among structural variants change which entries pass. DXMag was built, and our candidates were first
qualified, in the scalar-relativistic approximation, without SOC; a heavy-element candidate is certified only if its power factor
survives when SOC is included. Across the 607 paired calculations (\cref{fig:soc}), SOC narrows
gaps: of 264 runs gapped in both calculations, 93\% have a smaller gap with SOC, by a median of \SI{56}{meV} (11\% of the scalar gap). It
changes the class of 76 runs, 64 of them towards metallic behaviour (\cref{fig:soc}a,b). The gap reduction is largest for W-, Pt- and
Au-containing compounds (median 148, 101 and \SI{97}{meV}) and smallest for Pb and Bi (29 and \SI{32}{meV}), although Pb and Bi are the
heaviest of these elements. What matters, then, appears to be how much heavy-atom character the band edges carry. This is consistent with the character
of the edges in 18-electron half-Heuslers: both have strong transition-metal d character, and the valence band is mixed with p
states of the main-group atom\cite{Kandpal2006}, so a heavy transition metal (W, Pt, Au) contributes to both edges, whereas a heavy
main-group atom (Pb, Bi) contributes mainly to the valence band.

\begin{figure}[t]
\centering
\includegraphics[width=\columnwidth]{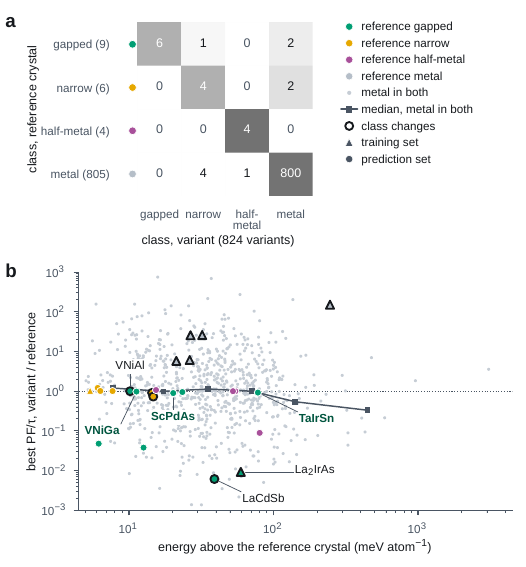}
\caption{\textbf{Higher-energy structural variants rarely change the electronic class.}
A variant is a calculated entry at least 5~meV\,atom$^{-1}$ above the reference crystal, the lowest-energy calculated entry of its
composition in either set (824 variants of 491 compositions; 661 prediction-set, 163 training-set). \textbf{a},~QE-PBE class of the reference
crystal (rows, totals in parentheses) against that of the variant (columns); shading, row fraction; dots, the colours of b. \textbf{b},~Best $\pf$ (300--\SI{1200}{K}) of each variant relative to its reference, against its energy above it.
Grey points are the 800 pairs metallic in both crystals, and the line their median in energy bins up to 1,000~meV\,atom$^{-1}$ (one
metallic variant lies at 3,080~meV\,atom$^{-1}$); coloured markers are the 24 other pairs, coloured by the class of the reference crystal
and outlined where the class changes. Triangles, training set; circles, prediction set. QE-PBE class, electronic class from the Quantum ESPRESSO calculation with the PBE functional. TaIrSn, VNiGa and ScPdAs are the certified
compositions with a calculated variant; the other labels name the gapped references whose variant changes class.}
\label{fig:variants}
\end{figure}
For the SOC candidates, the heavy-element runs that are gapped and qualified in the scalar calculation, the effect falls on the power
factor rather than on the gap. Of the 115 with a finished SOC calculation, 40 fall below $\pf_0$ with SOC, removing 20 of 60 candidate
compositions, yet none of the 40 loses its gap (\cref{fig:soc}d). A gap check, with or without SOC, would not have caught them. Separating the
carrier types shows where the loss occurs (\cref{fig:soc}c). At \SI{900}{K} and $|n|\le\SI{e21}{cm^{-3}}$ the p-type power factor of the
candidates falls to a median 0.76 of its scalar value, whereas the n-type power factor is unchanged (median 1.00; Supplementary
Table~S8). Both carrier types are computed with the same potentials, so the asymmetry is unlikely to come from the change of
pseudopotential. It is what is expected when SOC splits degenerate valence-band states near the band edge, lifting part of their degeneracy,
and affects the conduction-band edge much less, as earlier calculations on ANiB half-Heuslers and on ZrNiPb
found\cite{Guo2016SOCANiB,Guo2016ZrNiPb}. This asymmetry explains why the effect depends on the heavy element. Ten of the eleven
Au-containing candidate compositions are lost (20 of their 22 runs): their p-type power factor halves (median ratio 0.49), and all 22 runs
had relied on p-type doping; only HfAuAl remains certified. Bi-containing candidates are weakened on the p-type side too (median 0.76),
but none of their 12 runs (seven compositions) is lost, because with SOC nine of them reach their best power factor with n-type doping.
For ZrCoBi, this weakening of the p-type side agrees with the slight SOC splitting of its valence bands at L\cite{Zhu2018ZrCoBi}.

The change of pseudopotential cannot be excluded entirely. The 27 gapped pairs whose gap changes by less than \SI{10}{meV}
still have a median $\pf$ ratio of 0.89, close to the median over all 607 pairs, 0.87 (Supplementary Fig.~S14);
part of the average reduction may therefore come from the change of potential. HfNiSn, an established thermoelectric that falls 4\% below
$\pf_0$ with SOC, lies within this uncertainty. Two certified entries, a near-duplicate pair of NbOsSb, have a scalar gap of \SI{0.26}{eV}
that closes with SOC. They remain certified because criterion (iv) uses the scalar gap, but they should be regarded as marginal.

\subsection{Structural variants rarely change the electronic class}\label{sec:r-variants}
The tetragonal and metastable structures of DXMag were one reason to screen the whole database (\cref{sec:intro}). Because the prediction
set consists of alternative configurations of known compositions, its calculated members show whether a different structure can turn a
composition into a candidate. The calculations include 824 structural variants of 491 compositions, entries at least
5~meV\,atom$^{-1}$ above the reference crystal of their composition (median 33~meV\,atom$^{-1}$, within the range of known metastable
crystals\cite{Sun2016Metastability,Aykol2018Metastable}), most of them from the prediction set. Their electronic class rarely changes
(\cref{fig:variants}a). Eight hundred pairs are metallic in both crystals. Of the nine variants whose reference crystal is gapped, six remain
gapped, two become metallic and one narrow-gap, with power factors from below 1\% to about 100\% of the reference value. Four variants of
metallic references open a narrow gap; their 6 to 26 times larger power factors reflect how small the metallic reference value is. No
calculated variant is gapped where the reference crystal of its composition is not.

Three certified
compositions have a calculated variant, and none of the variants improves on its reference crystal (\cref{fig:variants}b). A tetragonal
variant of TaIrSn, 78~meV\,atom$^{-1}$ above the certified cubic crystal, remains gapped (\SI{0.79}{eV} against \SI{0.96}{eV}) and meets
the Seebeck, power-factor and stability criteria with 92\% of the reference power factor, but lacks an SOC calculation. A tetragonal variant
of VNiGa, \SI{11}{meV.atom^{-1}} higher, keeps its gap and 98\% of the power factor but lies just beyond the hull cut, and a tetragonal
variant of ScPdAs keeps a smaller gap (\SI{0.18}{eV}) and fails the hull cut.

This sample cannot establish a general rule. It contains only nine variants of gapped references; it was selected mostly on predicted spectra, which detect few gaps (held-out recall 0.15); and the QE calculations started from ferromagnetic moments or without spin polarisation, so the magnetic minima
that define much of the prediction set were recalculated without their own magnetic configurations.

\subsection{The certified compositions in context}\label{sec:r-survive}
\begin{figure*}[t]
\centering
\includegraphics[width=\textwidth]{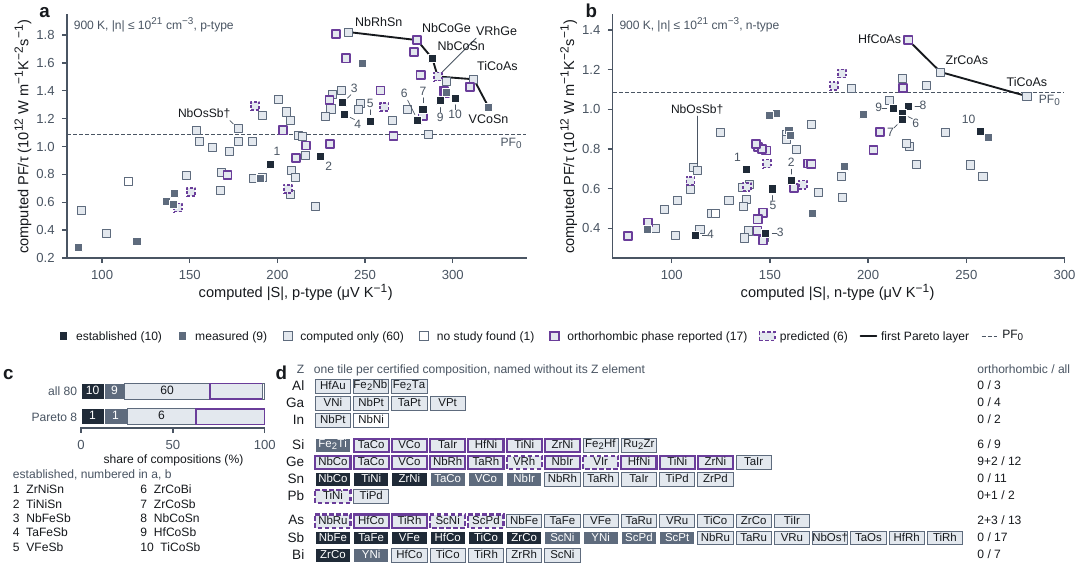}
\caption{\textbf{Certification recovers the established half-Heuslers among mostly unmeasured compositions.}
\textbf{a,b},~Computed $|S|$ and $\pf$ of the 80 certified compositions (155 certified entries) at \SI{900}{K} and
$|n|\le\SI{e21}{cm^{-3}}$ for p-type (a) and n-type (b) doping, one square per composition, with spin--orbit-coupling (SOC) values for atomic number $Z\ge72$. The black line
joins the first Pareto layer (named; \cref{tab:pareto}), numbers mark the ten certified established thermoelectrics (key in c), and the dashed line is the power-factor threshold $\pf_0$; squares below it pass at another temperature, doping or carrier type. \textdagger,~NbOsSb,
certified on a scalar-relativistic gap (\SI{0.26}{eV}) that closes with SOC. \textbf{c},~Prior thermoelectric evidence (Supplementary
Table~S3) for all 80 compositions and the eight on the Pareto layer; purple outline, share with an orthorhombic phase. \textbf{d},~The 80 compositions by Z-site element, tiles named
without the Z element; right, compositions with a reported (+ predicted) orthorhombic phase. Fill gives
the evidence class; a purple outline marks an orthorhombic phase, reported (solid) or predicted (dashed). Fill, from dark to light: established, measured, computed only (outlined), no study found (white, outlined). Fe$_2$ and Ru$_2$ tiles are
24-electron full Heuslers, all others 18-electron compounds. Values in the constant-relaxation-time approximation.}
\label{fig:candidates}
\end{figure*}
\begin{table*}[t]
\centering
\caption{\textbf{First Pareto layer of the certified compositions.} Non-dominated compositions in computed $|S|$ and $\pf$ at \SI{900}{K} and $|n|\le\SI{e21}{cm^{-3}}$, for p- or n-type doping, among the 80 certified compositions; all are cubic half-Heuslers. Set: certified in the training (T) and prediction (P) sets. $E_g$: QE-PBE gap (eV), from Quantum ESPRESSO with the PBE functional. $E_\mathrm{hull}$: DXMag hull distance (eV\,atom$^{-1}$). SOC: spin--orbit-coupling values used ($Z\ge72$). $S$ in \si{\micro V.K^{-1}}, $\pf$ in \SI{e12}{W.m^{-1}.K^{-2}.s^{-1}} ($\pf_0=1.09$), mostly at the carrier cap (\cref{sec:r-survive}); layer: Pareto layer within each carrier type. Constant-relaxation-time values, not $zT$ predictions. Lit.: prior thermoelectric evidence (Supplementary Table~S3): E, established; M, measured; C, computed only; $^{*}$orthorhombic phase reported experimentally; $^\dagger$orthorhombic ground state predicted.}
\label{tab:pareto}
\footnotesize
\begin{tabular}{@{}l c cc c rrc rrc c@{}}
\toprule
 & & & & & \multicolumn{3}{c}{p-type} & \multicolumn{3}{c}{n-type} & \\
\cmidrule(lr){6-8}\cmidrule(l){9-11}
compound & set & $E_g$ & $E_\mathrm{hull}$ & SOC & $S$ & PF/$\tau$ & layer & $S$ & PF/$\tau$ & layer & lit. \\
\midrule
TiCoAs & T{+}P & 1.32 & 0.031 & no & 312 & 1.48 & 1 & $-$281 & 1.07 & 1 & C \\
NbCoGe & T{+}P & 1.09 & 0.000 & no & 280 & 1.76 & 1 & $-$218 & 1.11 & 3 & C$^{*}$ \\
NbCoSn & T{+}P & 1.04 & 0.039 & no & 288 & 1.63 & 1 & $-$220 & 1.02 & 3 & E \\
VCoSn & T{+}P & 0.66 & 0.036 & no & 320 & 1.28 & 1 & $-$172 & 0.47 & 11 & M \\
VRhGe & T{+}P & 0.45 & 0.050 & no & 291 & 1.50 & 1 & $-$138 & 0.61 & 12 & C$^\dagger$ \\
NbRhSn & T{+}P & 0.76 & 0.044 & no & 241 & 1.82 & 1 & $-$139 & 0.39 & 16 & C \\
HfCoAs & T{+}P & 1.29 & 0.003 & yes & 296 & 1.39 & 4 & $-$221 & 1.35 & 1 & C$^{*}$ \\
ZrCoAs & T{+}P & 1.23 & 0.006 & no & 274 & 1.26 & 7 & $-$237 & 1.19 & 1 & C \\
\bottomrule
\end{tabular}

\end{table*}
The remaining question (Q4) concerns the certified compositions themselves. Because every certified prediction entry is a near-duplicate of a training entry, the 155 certified entries correspond to about 80 distinct crystals. Supplementary Table~S14 lists all 80 compositions. All but six
are certified only as cubic half-Heuslers; five are cubic full Heuslers, and one, ScPdAs, is certified in a cubic and a tetragonal entry
that are the same crystal within numerical precision. Of the 155 entries, 145 have 18 valence electrons and ten have 24. The 24-electron
group is the Fe$_2$VAl-type full Heuslers Fe$_2$NbAl, Fe$_2$TaAl, Fe$_2$TiSi, Fe$_2$HfSi and Ru$_2$ZrSi, with QE-PBE gaps of
0.20--\SI{0.37}{eV} and n-type power factors close to $\pf_0$, the family in which large calculated power factors were reported
before\cite{Bilc2015FullHeuslerPF}. For each of the 80 compositions DXMag gives phonons for a training-set crystal (release 2025-09-09; prediction entries have none),
and all 80 are dynamically stable at the commensurate $q$-points, whereas 102 of the 530 calculated training runs with phonon results
are unstable. Because phonon stability was one of the priority filters that ordered the selection, this agreement is not an
independent test.

Most of the compounds already known to be good thermoelectrics are on the list. Ten of the eleven established
half-Heusler thermoelectrics that were calculated (Supplementary Table~S15) are certified, in both populations: TiNiSn, ZrNiSn, TiCoSb,
ZrCoSb, HfCoSb, VFeSb, NbFeSb\cite{Fu2015}, TaFeSb\cite{Zhu2019}, NbCoSn and ZrCoBi\cite{Zhu2018ZrCoBi}, reviewed together in
refs.~\citenum{Quinn2021} and~\citenum{Zeier2016}. The eleventh, HfNiSn, is the SOC-limited case of \cref{sec:r-soc}. On the calculated
$|S|$--$\pf$ maps the established compounds lie among the other certified compositions rather than above them: NbCoSn is on the p-type
Pareto layer, whereas the established n-type MNiSn sit below $\pf_0$ in the n-type map at the ranking convention
(\cref{fig:candidates}a,b). The CRTA power factor per relaxation time is therefore not a merit order. A targeted literature audit places all
80 compositions on a scale of prior evidence (\cref{fig:candidates}c; Supplementary Table~S3). Besides the ten established compositions,
nine have reported thermoelectric measurements, among them ScPdSb, ScPtSb and YNiSb\cite{Oestreich2003ScMSb}, YNiBi\cite{Li2015JAP},
TaCoSn\cite{Li2019TaCoSn} and NbIrSn\cite{Hohl1998NbIrSn}; 60 have only computational predictions, at least eleven of them only within
high-throughput screens\cite{Carrete2014,Pak2025}; and for NbNiIn no thermoelectric study was found. Sixty-one of the 80 compositions
therefore have no reported thermoelectric measurement of the Heusler phase, including six of the eight compositions on the first Pareto layer, which
are listed in \cref{tab:pareto}.

The audit also flags compositions whose experimentally reported phase is not the half-Heusler, and these flags follow the Z-site element.
For 17 certified compositions (six of them only through secondary sources) the reported phase is an orthorhombic TiNiSi- or Co$_2$Si-type
structure, for example HfCoAs\cite{Kleinke1998MMAs} and TaRhGe and NbRhGe\cite{Dinges2010TaRhGe} (sources for all 17 in Supplementary
Table~S3), and for six more DFT places an orthorhombic structure below the half-Heusler\cite{Gautier2015ABX,Bilinska2023HHPhases}. ZrCoAs
belongs to the series of arsenides in which orthorhombic HfCoAs was reported\cite{Kleinke1998MMAs}, but its structure could not be
confirmed, and it is not counted. The reported phases cover all six certified half-Heuslers with Si, nine of the twelve compositions with
Ge and two of the thirteen with As, but none of the 35 with Sn, Sb or Bi (\cref{fig:candidates}d). Three of the eight first-Pareto-layer
compositions carry such a flag, NbCoGe and HfCoAs (reported) and VRhGe (predicted), and ZrCoAs may be a fourth. A convex hull built
from the phases of a general-purpose database need not contain the competing ternary phase, so a small DXMag hull distance does not
establish the half-Heusler structure, and, conversely, published data for the orthorhombic phase say nothing about the half-Heusler one. For the flagged compositions on the first Pareto layer, phase
stability has to be settled before transport.

\begin{table*}[t]
\centering
\caption{\textbf{What each headline result establishes.} T, training set; P, prediction set. Near-duplicates are prediction entries that nearly duplicate a training entry (defined in \cref{sec:m-data}). CRTA, constant-relaxation-time approximation; SOC, spin--orbit coupling.}
\label{tab:evidence}
\footnotesize
\begin{tabular}{@{}>{\raggedright}p{0.25\textwidth} >{\raggedright}p{0.17\textwidth} >{\raggedright}p{0.23\textwidth} >{\raggedright\arraybackslash}p{0.27\textwidth}@{}}
\toprule
result & rests on & establishes & main qualification \\
\midrule
held-out $\bar r=0.895$, gap recall 0.15 & held-out training spectra; best-validation models & shape reproduced, gaps mostly missed & upper bound: pretraining, where used, saw the test spectra \\
63.1\% (T) and 85.1\% (P) of labelled semiconductors are metals & calculated runs; DFT (T) or predicted (P) d-DOS & the binary gap label fails as a classifier, even on DFT spectra & score-ordered sample; thresholds differ between sets \\
top 20\% by $S^{\mathrm d}_{\max}$ hold 75\% (T) and 91\% (P) of certified entries & calculated entries of each set & the indicator prioritises & score-ordered sample; sets sampled differently (65\% of P, 22\% of T runs from low-priority label rounds); hull shared with a criterion; P certified entries are near-duplicates of T crystals \\
ROC-AUC of $S^{\mathrm d}_{\max}$ 0.805 (T), 0.808 (P) among 18- and 24-electron entries & calculated entries of each set & the indicator adds to electron counting & score-ordered sample; rule is family-specific (18 half, 24 full and inverse) \\
$\rho(S^{\mathrm d}_{\max},\pf)=0.13$ (T), 0.29 (P); 0.09, 0.23 over gapped runs & certified entries; all gapped runs & the indicator does not rank & CRTA power factor per $\tau$ \\
80 certified compositions (47 at \SI{900}{K}, $|n|\le\SI{e21}{cm^{-3}}$) & QE-PBE + BoltzTraP2, both sets & a CRTA list under declared criteria & $\approx$80 distinct crystals; criteria set during the campaign; no $\tau$, $\kappa_L$; wide gaps under-represented \\
SOC: p-type $\pf$ ratio 0.76, n-type 1.00; 40 of 115 SOC-calculated candidates fall below $\pf_0$ & 607 paired heavy-element runs & the SOC penalty is carrier-selective (p-type) & selected subset; pseudopotentials also change \\
no variant gapped where its reference crystal is not & 824 calculated variants & within the calculated sample, variants add no gapped crystal & 9 gapped-reference pairs; chosen on predicted labels; magnetic states not reproduced \\
10 of 11 established half-Heuslers certified & known compositions calculated & criteria recover known chemistry & tests the criteria, not the ranking \\
\bottomrule
\end{tabular}

\end{table*}
Any ranking of the certified compositions also depends on the transport conditions at which it is read. All 155 entries reach their best
CRTA power factor at \SI{1200}{K}, the top of the temperature scan, and 94 have their optimum at the edge of the chemical-potential window.
At \SI{900}{K} and $|n|\le\SI{e21}{cm^{-3}}$, the convention used for ranking, 91 entries (47 compositions) remain above $\pf_0$, and at
$|n|\le\SI{e20}{cm^{-3}}$ none does. The first Pareto layer at that convention comprises NbCoGe, NbCoSn, TiCoAs, VCoSn, NbRhSn and VRhGe for
p-type and HfCoAs, ZrCoAs and TiCoAs for n-type doping (\cref{tab:pareto}), all cubic half-Heuslers certified in both sets. Against the best
point over all conditions (the earlier best-over-all-conditions set is listed in Supplementary Table~S16, and the robustness of the
certified set is shown in Supplementary Fig.~S15), the ranks agree moderately for p-type doping (Spearman 0.67 for $|S|$, 0.68 for $\pf$)
but hardly for n-type (0.07 and $-0.23$). At this convention most optima sit at the carrier cap (72 of 80 compositions for p-type, 70 for
n-type), so the compositions are compared at nearly fixed carrier concentration. There $|S|$ follows the DOS effective mass, whereas the
power factor also depends on the conductivity mass. The certified compositions are therefore best read as a set of computational
candidates, and the Pareto layer as one convenient view of that set.

\section{Discussion}\label{sec:discussion}

A DOS-based screen can ask one descriptor three questions: whether a compound is gapped, whether it deserves a transport calculation, and
which compound is best. The calculations give each question a different answer (\cref{tab:evidence}). A binary label read from the d-DOS
at $E_F$ answers the first poorly, even on DFT spectra, because the d projection ignores s and p states inside a d-orbital gap and a fixed threshold accepts a deep minimum as a gap; most entries it passes are metals. The Seebeck indicator answers the second. It reads
the DOS as the thermopower does, as an asymmetry about $\mu$ weighted over a thermal window, so it responds to the depth and width of a low-DOS region rather than to a single value below a
threshold. The same property makes it tolerant of the main error of predicted spectra: residual weight in a gap lowers $S^{\mathrm d}_{\max}$ but, for the entries calculated here, did not remove the certified ones from
the top of the ranking. The certified entries had median ranks within the top 1\% of each population, and the indicator still separated
them from other entries among the 18- and 24-electron compounds that electron counting already favours. The third question lies beyond
the DOS, as the derivation in \cref{sec:intro} predicts. The power factor depends on band velocities and valley structure that the DOS does
not contain, and the indicator was only weakly correlated with the calculated power factor, among the certified entries and over all
gapped runs alike.

These answers suggest an order of steps for screens of this kind. Stability comes first, because certification requires it. Electron counting comes with it, because it costs nothing and removes the indicator's typical high-scoring false positive: a 16- or 17-electron
half-Heusler whose 18-electron gap lies within the scanned range of $\mu$ but not at $E_F$, which describes 85 of the 97 calculated
training-set metals with $S^{\mathrm d}_{\max}$ above \SI{400}{\micro V.K^{-1}}. $S^{\mathrm d}_{\max}$, from a DFT or a predicted DOS, then
orders what remains; in the training set, calculating only the top fifth would have cut the calculations per certified entry from about 21
to about 5.5, at the cost of a quarter of the certified entries. A first-principles band structure then decides whether a gap exists,
because neither the label nor a predicted spectrum can. Heavy-element compounds need a fully relativistic recalculation before they are
ranked: SOC moved 40 of the 115 gapped, qualified heavy-element runs below $\pf_0$, mostly through the p-type power factor (part of the reduction
may come from the change of pseudopotential), and none of them lost its gap. Only explicit transport, ideally with scattering, should
rank the certified compounds.

This order leaves phase competition open. A hull built from a general-purpose database need not contain the TiNiSi- and Co$_2$Si-type
phases reported or predicted for 23 of the 80 certified compositions, all but one of them with Si, Ge or As on the Z site
(\cref{sec:r-survive}). For this chemistry a small hull distance is therefore a weaker guarantee of the Heusler structure than its value
suggests\cite{Bartel2022StabilityReview}. Adding these orthorhombic structures to the reference hull would make the stability step of the
order meaningful where it is now weakest.

For the machine-learned spectra, the model reproduced the shape of the d-DOS but missed most gaps, and the
correlation coefficient by which DOS models are usually judged did not register the failure. On predicted spectra the indicator still
placed the certified entries near the top, and the form of the error is consistent with this. Residual weight inside a gap, which an
$E_F$-weighted least-squares objective makes cheap (\cref{sec:r-model}), turns a threshold label into a wrong answer but only compresses the
thermopower indicator: on predicted spectra $S^{\mathrm d}_{\max}$ follows the QE-PBE gap with a Spearman correlation of 0.40 instead of
0.90, yet 72 of the 79 certified prediction entries ranked among the 1,000 highest of 54,978. For screening, the useful measure of a DOS
model is therefore the fidelity of the decisions made from it: whether a gap is detected, and whether the ranking of entries survives.

The predicted spectra did not, however, extend the list of candidates, and neither did the structural variants. Every certified prediction entry is a near-duplicate of a training entry, including those of the five compositions certified
only in the prediction set, whose training entries were left unresolved by failed SOC calculations; the prediction-set figures
therefore describe spectra close to the training data. Within the calculated
sample, moreover, no variant was gapped where its reference crystal was not: the tetragonal and metastable minima that motivated the
screen added structural variety but no candidate. The model sees geometry only, without magnetic moments, and so assigns near-duplicate
geometries near-identical spectra; resolving the two spin channels helps the label separate half-metals, but the indicator reads the spin sum, so spin resolution does
not change it. The ensemble spread $u_F$ ordered the selection, but whether it marks the reliable predictions was not assessed. Three
changes would make such a model more useful to a screen, each following from a result here. Gap recall and the yields of screening decisions (for example, the fraction of selected entries that
are certified) should be reported alongside spectral metrics, because the correlation missed the failure. Evaluation should use splits that hold out whole
compositions in every training stage, because pretraining here saw the test spectra. And the loss should include a term on the weight
inside DFT gaps, because the present objective leaves that weight cheap.

The strongest results of this study involve no machine learning: on DFT spectra the gap label fails as a classifier, the Seebeck indicator
concentrates the certifiable entries, and transport calculations are needed to rank them. These results rest on a selection ordered by the
screen, with criteria set during the campaign, so the screen's recall is unmeasured and the certified count depends on the thresholds; the
list is a prioritised sample, not a census. It is also biased against wide gaps, because the $\pm\SI{0.5}{eV}$ window of the certification,
measured from a Fermi level placed by the smearing, cannot reach the band edges of many entries with gaps above about \SI{1.2}{eV}.

The electronic classes also carry the biases of the calculations, two of which act on the gap in opposite directions. PBE underestimates
gaps\cite{PerdewLevy1983,Borlido2019GapBenchmark}, whereas the unshifted NSCF meshes, which miss the X point in many runs, overestimate them
where the conduction-band minimum lies there. The QE runs started from ferromagnetic moments or without spin polarisation, so they do not
reproduce the magnetic configuration of each DXMag entry, and the magnetic minima of the prediction set were tested as structures only.
The undoped point of the Seebeck criterion is set by the smearing, and optima at the edges of the temperature and chemical-potential scans
make the reported power factors bounds within the scanned conditions.

The candidate list rests on CRTA transport at the PBE level, without $\tau$, $\kappa_L$, convergence studies or relaxation in QE.
Calculations that go beyond the CRTA show how much scattering can change such conclusions. For half-Heuslers, energy-dependent electron--phonon scattering can
reorder power factors relative to the CRTA\cite{Graziosi2020} and accounts for their unusually large values\cite{Zhou2018NonBonding},
and the assumed scattering model changes the predicted benefit of band alignment\cite{Kumarasinghe2019CoHH}. In other materials,
scattering-resolved calculations have ranked carrier mobilities across a range of semiconductors better than the CRTA\cite{Ganose2021},
improved high-throughput transport predictions for chalcogenides\cite{Xi2018} and removed the expected benefit of band convergence in
Zintl and full-Heusler compounds\cite{Park2021}. The lattice term, which is not computed here, also varies widely: $\kappa_L$ spans more than two orders of magnitude
across mechanically stable half-Heuslers\cite{Carrete2014}. A computational prediction can also be hard to test: an ab initio estimate
of $zT>2$ for n-type LiZnSb\cite{Madsen2006a} could not be tested, because the samples made were p-type, with $zT$ of 0.02--0.08 at
\SI{523}{K}\cite{Toberer2009}.

The model evaluation, finally, is favourable by construction: pretraining overlapped the test spectra, the
splits were random within families, the parent of 28 of the 30 models is unconfirmed, and the deployed five-model mean was not evaluated
on held-out spectra. These qualifications bear on the prediction-set figures; the results on DFT spectra do not depend on the model.

The 80 certified compositions are therefore candidates for calculations beyond the constant-relaxation-time approximation, not predicted
thermoelectrics. The next steps follow from the limitations above, in order: phase competition against TiNiSi- and Co$_2$Si-type
structures, first for the 23 compositions with a reported or predicted orthorhombic phase; scattering-limited transport and lattice
thermal conductivity for those that remain; and synthesis, beginning with the compositions whose half-Heusler phase has already been made.

\section*{Data availability}
The DXMag structures and metadata (release \mbox{2025-09-09}) and the HeuslerDB electronic-DOS dataset (doi:10.48505/nims.5794) are available
from NIMS under CC BY 4.0 (\url{https://www.nims.go.jp/group/spintheory/database/})\cite{DXMagDOS}; the DFT spectra used for training are not redistributed. The machine-learned
d-DOS of the 54,978 prediction-set entries (five-model mean and spread), the held-out predictions of the best-validation checkpoints of the same fold models for the 51,257
training-set entries, the prediction-set screening table, the input structures, an index of DXMag identifiers and metadata, and the 30
trained model weights with their configurations are deposited at Zenodo under CC BY 4.0 (doi:\href{https://doi.org/10.5281/zenodo.23329059}{10.5281/zenodo.23329059}). The per-run
first-principles results of all 3,625 runs (electronic class, QE-PBE gap, transport summaries, certification status and screening
indicators), the 607 paired scalar and SOC results, the structural-variant comparison, the rankings of the 80 certified compositions and the
literature-status table are included in the code repository.

\section*{Code availability}
The code used to train and apply the d-DOS model, compute the screening indicators, run and aggregate the Quantum ESPRESSO and BoltzTraP2
calculations, and produce the figures and tables is available at \url{https://github.com/yednaph/heusler-dos-te-screening} (release v1.0.0). It is released under the MIT licence, except the first-principles workflow, which is
released under GPL-3.0-or-later because it derives from BoltzTraP2. Inputs that are not redistributed are listed in the repository.

\section*{Acknowledgements}
We thank E.~Xiao and T.~Tadano for making the DXMag Computational HeuslerDB and its electronic-DOS dataset openly available. The d-DOS models were trained and applied on Aurora, the high-performance computing cluster (NVIDIA H100 GPU server) at Sardar Vallabhbhai National Institute of Technology, Surat, whose computing time we acknowledge; the first-principles calculations were run on local workstations.

\section*{Author contributions}
S.R.: Methodology, Software, Investigation, Formal analysis, Data curation, Visualization, Writing -- original draft. H.P.: Methodology, Software, Validation, Supervision, Resources, Project administration, Writing -- review \& editing. R.D.: Visualization, Writing -- review \& editing. All authors reviewed the manuscript.

\section*{Competing interests}
The authors declare no competing interests.

\bibliographystyle{unsrtnat}
\bibliography{references}

\clearpage
\newgeometry{margin=20mm,onecolumn}
\onecolumn
\setcounter{page}{1}\renewcommand{\thepage}{S\arabic{page}}
\setcounter{figure}{0}\setcounter{table}{0}\setcounter{equation}{0}
\renewcommand{\thefigure}{S\arabic{figure}}\renewcommand{\theHfigure}{S\arabic{figure}}
\renewcommand{\thetable}{S\arabic{table}}\renewcommand{\theHtable}{S\arabic{table}}
\renewcommand{\theequation}{S\arabic{equation}}\renewcommand{\theHequation}{S\arabic{equation}}
\begin{center}
{\LARGE\bfseries Supplementary Information for\\[2pt] ``High-throughput screening of Heusler alloys for thermoelectrics combining
machine-learned densities of states and first-principles transport''\par}
\vskip 1.5em
{\large Shivam Randive, Ayana Ghosh, Rahul Dixit, Himanshu Pandey\par}
\end{center}
\vskip 1.5em
\noindent All figures and tables are generated from stored results by the scripts of the code repository (main text, Code availability).
Throughout, ``training set'' denotes entries screened on their DFT d-DOS and ``prediction set'' entries screened on
ML-predicted d-DOS. The two sets are reported separately, and their raw screening scores are not compared.

\section*{Supplementary Note 1: normalisation of the DOS archive}

\noindent For a sample of magnetic training entries ($|M|\ge1\,\mu_B$ per formula unit in DXMag; 25 per family, fixed random seed), the
archive's spin-resolved total DOS was integrated up to $E_F$, $m=\int^{E_F}[D_\uparrow(E)-D_\downarrow(E)]\,dE$, and divided by the number
of formula units of the structure stored with the spectrum. The ratio of this moment to the DXMag total magnetisation per formula unit is
2.00 (median; interquartile range 1.97--2.03) and lies between 1.99 and 2.01 in every family (Supplementary Fig.~S2). The archive DOS
therefore carries a factor of two relative to its stored structure, and so do the per-atom d-DOS targets of the main text (Eq.~1) and the
model's predictions. The d-projected channel is assumed to carry the same factor; this was not checked
separately. An earlier audit of 340 files, which integrated the total DOS to $E_F$ and compared it with the valence-electron count,
found the same factor. Ratios of DOS values, and hence the d-DOS Seebeck indicator $S^{\mathrm d}_{\max}$ and the Mahan--Sofo
sharpness $\xi$, are unaffected. Absolute thresholds are affected: the \SI{0.15}{states.eV^{-1}} classifier threshold, the
\SI{0.05}{states.eV^{-1}.atom^{-1}} gap flag and the $g(E_F)=2$ reference of the initial composite score (Supplementary Note~5) all
correspond to half these values in physical units. The training-set screen, which works per stored cell, differs further by the number of
atoms in that cell (three or four). Supplementary Table~S2 lists the consequences for each indicator and threshold. Script: \texttt{scripts/figures/dos\_normalisation\_check.py}.

\section*{Supplementary Note 2: selection rounds and counting units}

\noindent\textbf{Selection rounds.} DFT calculations were not drawn from a single ranked list. Nine selection rounds (rounds 1--9,
Supplementary Table~S6) were assembled in succession between July and October 2026, each ordered by the screening scores current when it
was built (main text, Methods). Most rounds followed the d-DOS power-factor indicator $P^{\mathrm d}_{\max}$; round~6 followed the d-DOS
Seebeck indicator $S^{\mathrm d}_{\max}$ without priority filters; rounds 7 and 8 took semiconductor-labelled entries by gap tier; and
round~9 took entries obeying the electron-count rule (18 or 24 valence electrons). Later rounds were therefore shaped by what earlier
rounds had shown, and outcome rates differ strongly between rounds (Supplementary Table~S6). No round was a random or stratified sample, so
every rate reported in the paper is conditional on this selection and the recall of the screen is not measured.

\noindent\textbf{Counting units.} \emph{Runs} are individual QE-PBE/BoltzTraP2 calculations: 3,625 in total, of which 1,562 are from
the training set, 2,021 from the prediction set and 42 could not be attributed unambiguously to either set; these 42 are excluded from all per-set statistics. \emph{Entries} are distinct DXMag records (UUIDs); all matching is by
UUID, never by formula, because different entries can share a formula. \emph{Compositions} merge entries with the same DXMag composition string
across structural variants and across sets; the string keeps the site order, so it distinguishes half-Heusler site permutations, which a
reduced formula does not. The certified counts are 76 training-set and 79 prediction-set entries, corresponding to 75
and 79 compositions; their union is 80 compositions (74 certified in both sets, 5 in the prediction set only, 1 in the training set only).

\noindent\textbf{Near-duplicate entries.} Because every prediction-set entry shares its composition with a training-set entry, some
computed entries in the two sets are near-duplicates of the same crystal: a prediction-set entry is a near-duplicate of a training-set
entry when the two share composition, family and space group and differ by less than 5~meV\,atom$^{-1}$ in DXMag energy and 1\% in
volume. Among all computed entries, 475 prediction-set entries have such a
near-duplicate training-set entry (largest differences 4.78~meV\,atom$^{-1}$ and 0.86\% in volume). Among the 155 certified entries, 79
prediction-set entries have a near-duplicate training-set entry (largest differences: \SI{1.73}{meV.atom^{-1}} in DXMag energy and 0.20\% in volume), so the 155 entries correspond to about 81 distinct
crystals, or 80 when the cubic and tetragonal PdScAs entries, which are the same crystal within numerical precision, are counted
once, as in the main text. This is why certified results are reported per set and merged only at the level of compositions, with provenance.

\section*{Supplementary Note 3: d-DOS model and training details}

\noindent The main text summarises the d-DOS model; this note gives the full specification as read from the training and inference code,
the per-fold configuration files, the tensor shapes of the deployed checkpoints and the surviving run records (settings collected in
Supplementary Table~S5).

\noindent\textbf{Representation.} ALIGNN represents a crystal by its atom graph $G$, with atoms as nodes and neighbour pairs (including
periodic images) as edges, and by the line graph $L(G)$, whose nodes are the bonds of $G$ and whose edges connect pairs of bonds that share
an atom and carry the bond angle $\theta$. Graphs use the 12-nearest-neighbour rule of the package with an \SI{8}{\angstrom} cutoff, and are built from the
structure stored with each training spectrum or from the listed DXMag cell of a prediction-set entry, without relaxation. Bond lengths are
expanded in 80 radial basis functions and bond-angle cosines in 40, with 64-dimensional embeddings. Each atom carries 97 input features:
the 92 CGCNN descriptors plus five Heusler-specific features, namely the d-electron count, the largest magnitude of a common oxidation
state, the distance of the d count from a half-filled or filled shell, the period of the d shell (its principal quantum number plus one)
and a duplicate of the d-electron count. The d count is read from the electron configuration, so p-block elements with a filled d shell
(Ga to Bi) carry a value of ten.

\noindent\textbf{Edge-gated convolution.} For node features $\mathbf h_i$ and edge features $\mathbf e_{ij}$, each convolution computes
\begin{gather}
\mathbf m_{ij}=\mathbf A\mathbf h_i+\mathbf B\mathbf h_j+\mathbf C\mathbf e_{ij},\quad
\boldsymbol{\eta}_{ij}=\frac{\mathrm{sigm}(\mathbf m_{ij})}{\sum_{k}\mathrm{sigm}(\mathbf m_{ik})+\epsilon},\label{eq:egc1}\\
\mathbf h_i'=\mathbf h_i+\mathrm{SiLU}\Bigl[\mathrm{BN}\Bigl(\mathbf U\mathbf h_i+\sum_j\boldsymbol{\eta}_{ij}\odot\mathbf V\mathbf h_j\Bigr)\Bigr],\quad
\mathbf e_{ij}'=\mathbf e_{ij}+\mathrm{SiLU}[\mathrm{BN}(\mathbf m_{ij})],\label{eq:egc2}
\end{gather}
where $\mathbf m_{ij}$ is the edge message, $\boldsymbol{\eta}_{ij}$ the normalised edge gate, $\mathbf A,\mathbf B,\mathbf C,\mathbf U,\mathbf V$
are learned matrices, sigm is the logistic function, $\epsilon$ a small constant that prevents division by zero, BN batch normalisation,
$\mathrm{SiLU}(x)=x\,\mathrm{sigm}(x)$, $\odot$ the element-wise product, and the sums over $j$ and $k$ run over the neighbours of node $i$.
An ALIGNN layer applies this update first on $L(G)$, where the nodes are bonds and the edges bond pairs carrying the angle expansion, and
then on $G$, using the updated bond features as edge features; angular information thereby reaches the atom features. The encoder has eight
ALIGNN layers followed by eight graph-convolution layers acting on $G$ alone, all with 512 features (package default: four and four layers
of 256 features). The atom features are then averaged over the crystal into one 512-value vector. No layer of the encoder was modified.

\noindent\textbf{Output head.} The single linear output layer of the package was replaced by a head $512\rightarrow1024\rightarrow2048
\rightarrow4002$ with GELU activations and dropout 0.2 after the first two layers, followed by $\mathrm{softplus}(z)=\ln(1+e^{z})$, so that
the predicted spin-resolved d-DOS $\hat{\mathbf y}\in\mathbb R^{4002}$ ($2\times2{,}001$ points, \SI{-10}{} to \SI{10}{eV} in steps of
\SI{0.01}{eV}) is non-negative. The model has 42,520,354 trainable parameters, $1.08\times10^{7}$ of them in the head.

\noindent\textbf{Loss.} The loss, evaluated on the softplus output and averaged over a batch, is
\begin{equation}
\mathcal{L}=0.35\,\mathcal{L}_\mathrm{F}+0.35\,\mathcal{L}_{\pm2}+0.15\,\mathcal{L}_\mathrm{cos}+0.10\,\mathcal{L}_\mathrm{slope}
+0.05\,\mathcal{L}_\mathrm{sort}.\label{eq:si-loss}
\end{equation}
$\mathcal L_\mathrm F=\frac1{4002}\sum_k\tilde w_k(\hat y_k-y_k)^2$ is a mean-squared error weighted by the Fermi window
$w(E)=-\partial f/\partial E$ at \SI{700}{K}, normalised to unit mean ($\tilde w_k$). $\mathcal L_{\pm2}$ is the mean-squared error within
$|E_k|\le\SI{2}{eV}$, averaged over the two spin channels. $\mathcal L_\mathrm{cos}=1-\hat{\mathbf y}\cdot\mathbf y/(\lVert\hat{\mathbf
y}\rVert\lVert\mathbf y\rVert+\epsilon)$ is the cosine distance. $\mathcal L_\mathrm{slope}$ is the mean-squared error of first differences
plus 0.1 times that of second differences over $|E_k|\le\SI{3}{eV}$; as implemented it acts on the spin-up channel only.
$\mathcal L_\mathrm{sort}$ is the L1 distance between the sorted values of prediction and target over the whole 4,002-value vector, so it
pools both spin channels and ignores energy positions. Two further terms in the code had zero weight. At \SI{700}{K} the normalised Fermi
weight has a full width at half maximum of \SI{0.21}{eV} and a peak of 83 at $E_F$, and 99\% of it lies within $\pm\SI{0.3}{eV}$; per energy
bin near $E_F$, $\mathcal L_\mathrm F$ therefore outweighs $\mathcal L_{\pm2}$ by a factor of about 17.

\noindent\textbf{Training.} Each family was divided by shuffled five-fold cross-validation (seed 42), with 10\% of the non-test entries held
out for validation, so that each model trained on 72\% of its family, was selected on 8\% and tested on 20\%. Optimisation used AdamW
(learning rate $10^{-3}$, weight decay $5\times10^{-5}$, gradient clipping at 1.0) with a reduce-on-plateau schedule (factor 0.5; patience
25, or 15 for cubic full-Heusler folds 0--2). The schedule monitors the validation loss. Stochastic weight averaging (SWA) averaged the weights over the last 20\% of epochs, during
which a single annealed learning rate of $5\times10^{-5}$ applied to all parameters; the batch-normalisation statistics were then
re-estimated in one pass with momentum 0.1. Runs lasted 300 epochs with batch size 128, except cubic full Heusler (150 or
200 epochs, batch 64). The best-validation checkpoint, chosen on the Pearson $r$ over the concatenated validation spectra before SWA
starts, gives all held-out metrics; the SWA weights give all predictions. Data augmentation, early stopping and learning-rate warm-up were
configured but not executed. Seeds were fixed and cuDNN was set to deterministic mode; bitwise reproducibility was not tested.

\noindent\textbf{Pretraining and fine-tuning.} An all-family (foundation) model of the same architecture was first trained on all 51,257
training-set spectra for 100 epochs (SWA from epoch 85). Its validation subset was part of its own training data, so every later test entry
was seen during pretraining. Fine-tuning loads all weights of the parent's SWA checkpoint, including the head, and trains the encoder at the learning rate and weight
decay scaled by $s$ and the head at the full rate. A variance-aware initialisation of the last output layer was used only for the
foundation run.

\noindent\textbf{Provenance record.} For cubic full-Heusler folds 0 and 1, the run metadata and optimiser states show initialisation from
the pretrained model with $s=0.1$. For the other 28 models, four project sources give different parents (Supplementary Table~S4), the
checkpoints do not record them, the launch script was edited after training, and the code silently falls back to random initialisation if
a parent path is missing; their parent, learning-rate scaling and weight loading are therefore unconfirmed. The only surviving training
record, that of cubic full-Heusler fold~1 (Supplementary Fig.~S3), shows the validation $r$ peaking at epoch~2 (0.827) and declining to
0.786 by epoch~149, while the validation loss rose from 0.050 to 0.058. The best-validation checkpoints are those of epoch~2 (fold~1) and
epoch~0 (fold~0); the SWA weights average epochs 120--149.

\noindent\textbf{Inference.} All six screening scripts specify the five SWA fold models of the entry's family; the mean of their softplus
outputs is the predicted d-DOS and their population standard deviation $u_s(E)$ its spread, whose mean over $|E|\le\SI{2}{eV}$ and both
channels is $u_F$. The run summaries for cubic full and inverse Heuslers list five checkpoints (4,504 entries in \SI{219}{s}; 6,702 in
\SI{316}{s}), and the stored arrays confirm five members in every family (Note~4). Per-member predictions were not saved. The final re-screen was completed on 23 August 2026;
its command-line arguments were not logged. The predicted d-DOS feeds the indicators $S^{\mathrm d}_{\max}$, $P^{\mathrm d}_{\max}$, $\xi$
and $L_0^{*}$ of the main text. One systematic artefact lies outside the screening window: for 42 held-out and 12 prediction-set
spectra, all of Zn-rich compounds, the filled Zn 3d band near \SI{-7}{eV} is predicted too high, so that the occupied d-states per atom
(physical units) exceed ten, which no DFT target does; it does not enter any indicator, which uses states within about \SI{2}{eV} of $E_F$.

\noindent\textbf{Software and hardware.} alignn 2026.4.1, PyTorch 2.2.2 (CUDA 12.1), DGL 1.1.3 and Python 3.11 (versions at export), on
NVIDIA H100 GPUs; spglib 2.0.2, BoltzTraP2 24.7.2 and Quantum ESPRESSO 7.1 for the DFT part (Supplementary Table~S7).

\section*{Supplementary Note 4: implementation record}

\noindent This note records implementation facts that earlier descriptions of the workflow stated differently, and how each was
established. The main text follows the implementation throughout.
\begin{enumerate}\setlength{\itemsep}{1pt}
\item \emph{Ensemble size.} All six screening scripts specify the five fold models of the family, and the two surviving inference
summaries list five checkpoints. The stored arrays confirm this for every family: for $n$ members the ratio of the population standard
deviation to the mean cannot exceed $\sqrt{n-1}$, and its maximum is exactly 2.000 ($n=5$) in all six families.
\item \emph{Transfer lineage.} Initialisation from the all-family model is recorded in the run files for two of the 30 models (cubic full
Heusler, folds 0 and 1); for the others the parent is not recorded (Supplementary Table~S4 and Note~3).
\item \emph{Loss.} Five terms are active, with the weights of Eq.~(\ref{eq:si-loss}) in all 30 configuration files; the code defaults (seven
terms) were overridden. Of the two zero-weight terms, the asymmetric peak term indexes the spin-down window one bin early; it has no
effect on any model. The sorted-value term, named after the Wasserstein distance in the code, pools both spin channels and ignores
energy positions, and the slope term acts on the spin-up channel only.
\item \emph{Interpolation and features.} Targets are linearly interpolated. The five Heusler-specific features are those of Note~3.
\item \emph{Normalisation layers.} The encoder uses batch normalisation; after SWA its statistics were re-estimated in one pass over the
training data with the default momentum of 0.1, that is as an exponential rather than a cumulative average.
\item \emph{Configured but not executed.} Data augmentation, early stopping and learning-rate warm-up appear in the configuration files but
were not executed.
\item \emph{Evaluation.} All held-out metrics were computed with the best-validation checkpoints, and the pretrained model had seen every
test entry. The SWA window is the last 20\% of each run: epochs 240--299 for the 300-epoch runs and 120--149 or 160--199 for the cubic
full-Heusler runs of 150 and 200 epochs.
\item \emph{Determinism and versions.} The code fixes the seeds and sets cuDNN to deterministic mode; bitwise reproducibility was not tested.
The training script was modified after training, and no version-control commit was recorded; the released code is that of the
repository cited in the main text.
\item \emph{Structures.} The inference code reads the DXMag cell and positions unchanged; no structure was relaxed.
\item \emph{Grouped splits.} Splits grouped by the Z-site element were prepared but never evaluated.
\item \emph{Parameter count.} The deployed checkpoints have 42,520,354 trainable parameters (from the tensor shapes); counts of about
$4.257\times10^{7}$ include the batch-normalisation statistics.
\item \emph{Screening statements.} The power-factor threshold was derived from the first 1,401 runs of the campaign and then held fixed;
metallic shares are reported per set (63.1\% and 85.1\%); and the priority filters ordered the selection rounds without excluding entries.
\end{enumerate}

\section*{Supplementary Note 5: the initial composite score}

\noindent In the initial screen (July 2026) the prediction set was ranked by a composite score computed from the ensemble-mean d-DOS
(called D34 in the project code),
\begin{equation}
Q=s_\mathrm{M}\,\phi\,\frac{1}{1+|g(E_F)-2|}\,\exp\!\Bigl[-\min_{\pm1\,\mathrm{eV}}g_\uparrow\,\min_{\pm1\,\mathrm{eV}}g_\downarrow\Bigr],
\label{eq:initial}
\end{equation}
where $s_\mathrm{M}=\max_s|\pi^2k_B^2T\,g_s'(E_F)/3e\,g_s(E_F)|$ is the larger over the two spin channels $s$ of the
Mott-formula Seebeck coefficient at $T=\SI{700}{K}$, and $\phi$ is the larger over the channels of the peak height within \SI{2}{eV} of $E_F$ divided by
its full width at half maximum. Values are in archive units, so the reference $g(E_F)=2$ corresponds to 1 per atom physically
(Supplementary Note~1). The score exists for the prediction set only. Taking the larger single-channel Mott term ignores that the two spin
channels conduct in parallel, $S=(\sigma_\uparrow S_\uparrow+\sigma_\downarrow S_\downarrow)/(\sigma_\uparrow+\sigma_\downarrow)$, so that
a large term in a weakly conducting channel is outweighed by the other, more conducting channel. This is one reason why the score rewarded
metals. Of its 100 best prediction-set entries, 83 carried a metal label, 98 lay at least
\SI{0.1}{eV.atom^{-1}} above the hull, 7 were computed and none was certified (Supplementary Fig.~S8a).

\section*{Supplementary Note 6: development of the workflow}
\noindent The certification criteria and the indicators of the main text are the final protocol. They were reached while the 3,625
calculations proceeded, and each step was motivated by a first-principles observation (Table~S10): an earlier composite score was replaced by
the Cutler--Mott indicators, stability, label and ensemble spread became priority filters, the electronic class was taken from the QE-PBE band
structure, a single power-factor threshold was fixed from the first 1,401 runs, a QE-PBE gap became a requirement, heavy-element runs were
re-checked with SOC, and ranking moved to computed transport. Because these choices were made after part of the results had been seen, the
cascade of the main text is not a prospective test; the sensitivity of the certified count to each threshold is given in Table~S9.

\section*{Supplementary Tables}

\begin{table}[H]\centering\scriptsize
\caption{\textbf{Characterisation of the two sets by family} (T, training set; P, prediction set). Compositions are site-ordered DXMag
composition strings; atoms per listed cell are medians; hull distances in eV\,atom$^{-1}$; $|M|$ in $\mu_B$ per formula unit, as a share of
entries with a known moment; VEC is the valence-electron count per formula unit.}
\begin{tabular}{@{}llrrrrrrrr@{}}
\toprule
family & set & entries & compositions & atoms/cell & hull$<$0.10 (\%) & median hull & $|M|>0.1$ (\%) & 18 e$^-$ (\%) & 24 e$^-$ (\%) \\
\midrule
full cubic & T & 9,396 & 9,396 & 4 & 9.5 & 0.37 & 67.0 & 3.7 & 4.8 \\
 & P & 4,504 & 3,572 & 4 & 7.3 & 0.41 & 60.0 & 3.2 & 6.6 \\
full tetra & T & 8,047 & 8,047 & 8 & 13.5 & 0.28 & 25.0 & 3.1 & 3.6 \\
 & P & 13,479 & 7,816 & 8 & 5.5 & 0.37 & 36.6 & 2.2 & 4.0 \\
half cubic & T & 9,396 & 9,396 & 3 & 3.7 & 0.65 & 66.5 & 9.4 & 4.8 \\
 & P & 4,914 & 3,907 & 3 & 4.6 & 0.68 & 62.4 & 11.6 & 3.8 \\
half tetra & T & 6,909 & 6,909 & 6 & 1.5 & 0.58 & 27.6 & 6.1 & 5.6 \\
 & P & 10,896 & 6,256 & 6 & 0.7 & 0.71 & 40.2 & 4.7 & 4.5 \\
inverse cubic & T & 9,072 & 9,072 & 4 & 2.2 & 0.45 & 69.4 & 3.7 & 4.9 \\
 & P & 6,702 & 4,453 & 4 & 1.9 & 0.50 & 60.7 & 2.8 & 6.2 \\
inverse tetra & T & 8,437 & 8,437 & 8 & 5.3 & 0.31 & 24.6 & 2.9 & 4.7 \\
 & P & 14,483 & 8,076 & 8 & 2.2 & 0.41 & 36.6 & 2.3 & 5.1 \\
\bottomrule
\end{tabular}

\end{table}

\begin{table}[H]\centering\scriptsize
\caption{\textbf{Consequences of the DOS normalisation} for each indicator and threshold (Supplementary Note~1).}
\begin{tabular}{@{}>{\raggedright}p{0.24\textwidth} >{\raggedright}p{0.22\textwidth} >{\raggedright}p{0.22\textwidth} >{\raggedright\arraybackslash}p{0.24\textwidth}@{}}
\toprule
quantity & prediction set (per atom, $\times2$) & training-set screen (per cell, $\times2N$) & consequence \\
\midrule
$S^{\mathrm d}_{\max}$, sharpness $\xi$ & invariant & invariant & none \\
$P^{\mathrm d}_{\max}$, $L_0^{*}$ (scale-dependent) & $\times2$, uniform & $\times6$ (half) or $\times8$ (full, inverse) & rankings within the prediction set unchanged; in the training set $P^{\mathrm d}_{\max}$ favours full and inverse over half Heuslers by 4/3, which affects the $P^{\mathrm d}_{\max}$-ordered rounds and its training-set AUC \\
classifier threshold 0.15 & physical 0.075 per atom and spin & physical 0.025 (half) or 0.019 (full, inverse) & training-set label 3--4$\times$ stricter, by family; d-DOS gap widths and the 0.10--1.40~eV window select different populations in the two sets \\
held-out gap flag 0.05 (spin-summed) & physical 0.025 per atom & not used & defines the confusion matrix of main-text Fig.~4e \\
initial composite score: $1/(1+|g(E_F)-2|)$ and $\exp(-\min g_\uparrow\min g_\downarrow)$ & peak at physical 1; exponent $\times4$ & not computed & evaluated as implemented; its failure stands \\
loss terms & MSE terms $\times4$, cosine invariant & not used & relative term weights shift uniformly \\
\bottomrule
\end{tabular}

\end{table}

{\scriptsize\setlength{\tabcolsep}{3pt}
\begin{longtable}{@{}l l l >{\raggedright\arraybackslash}p{2.4cm} >{\raggedright\arraybackslash}p{5.4cm} >{\raggedright\arraybackslash}p{3.2cm}@{}}
\caption{\textbf{Prior thermoelectric evidence for the 80 certified compositions} (targeted literature audit, Methods). DXMag site-ordered
composition and conventional name; class by the strongest evidence for the same composition in the same structure type; polymorph: the
composition is reported experimentally in an orthorhombic structure (exp.) or such a ground state is predicted (pred.); screen-level predictions count as computational. Absence of a
report is not evidence that none exists. Every DOI was checked against its Crossref record.}\label{tab:si-literature}\\
\toprule
composition & conventional & class & poly. & evidence & principal source (DOI) \\
\midrule
\endfirsthead
\toprule
composition & conventional & class & poly. & evidence & principal source (DOI) \\
\midrule
\endhead
\bottomrule
\endfoot
CoHfSb & HfCoSb & established &  & Experimental: pristine and Pt/Sn-doped HfCoSb synthesised and Seebeck/resistivity measured; parent of the widely studied p-type (Hf,Zr,Ti)CoSb(Sn) half-Heusler family. & Xia 2000, J. Appl. Phys. 88, 1952 (10.\allowbreak{}1063/\allowbreak{}1.\allowbreak{}1305829) \\
CoNbSn & NbCoSn & established &  & Experimental: phase-pure/directionally solidified NbCoSn synthesised with measured Seebeck, resistivity and PF; Pt-substituted n-type NbCoSn with measured zT. & Kimura 2008, Appl. Phys. Lett. 92, 012105 (10.\allowbreak{}1063/\allowbreak{}1.\allowbreak{}2828713) \\
CoTiSb & TiCoSb & established &  & Experimental: pristine TiCoSb and (Ti,Zr,Hf)CoSb compounds synthesised with measured resistivity, Seebeck and thermal conductivity; long-standing p/n-type half-Heusler TE family. & Sekimoto 2005, Mater. Trans. 46, 1481 (10.\allowbreak{}2320/\allowbreak{}matertrans.\allowbreak{}46.\allowbreak{}1481) \\
CoZrBi & ZrCoBi & established &  & Experimental: p-type ZrCoBi-based half-Heuslers synthesised with measured zT and device conversion efficiency. & Zhu 2018, Nat. Commun. 9, 2497 (10.\allowbreak{}1038/\allowbreak{}s41467-\allowbreak{}018-\allowbreak{}04958-\allowbreak{}3) \\
CoZrSb & ZrCoSb & established &  & Experimental: pristine and doped ZrCoSb synthesised with measured Seebeck/resistivity; core member of p-type MCoSb half-Heusler family. & Xia 2000, J. Appl. Phys. 88, 1952 (10.\allowbreak{}1063/\allowbreak{}1.\allowbreak{}1305829) \\
FeNbSb & NbFeSb & established &  & Experimental: p-type NbFeSb-based half-Heuslers (Ti/Hf/Zr-doped) synthesised with measured zT > 1; widely recognised TE family. & Fu 2015, Nat. Commun. 6, 8144 (10.\allowbreak{}1038/\allowbreak{}ncomms9144) \\
FeTaSb & TaFeSb & established &  & Experimental: TaFeSb-based half-Heuslers (Ti-substituted Ta1-xTixFeSb) synthesised by inverse design with measured high zT; also highlighted computationally. & Zhu 2019, Nat. Commun. 10, 270 (10.\allowbreak{}1038/\allowbreak{}s41467-\allowbreak{}018-\allowbreak{}08223-\allowbreak{}5) \\
FeVSb & VFeSb & established &  & Experimental: pristine and doped FeVSb synthesised and Seebeck/resistivity/thermal conductivity measured in several independent studies; p-type FeVSb-based alloys are a recognised HH TE family. & Young 2000, Journal of Applied Physics 87, 317-321 (10.\allowbreak{}1063/\allowbreak{}1.\allowbreak{}371863) \\
NiTiSn & TiNiSn & established &  & Experimental: canonical n-type MNiSn HH thermoelectric; pristine and Sb-doped TiNiSn synthesised and S/PF/zT measured in many studies. & Bhattacharya 2000, Applied Physics Letters 77, 2476-2478 (10.\allowbreak{}1063/\allowbreak{}1.\allowbreak{}1318237) \\
NiZrSn & ZrNiSn & established &  & Experimental: canonical n-type HH thermoelectric; pristine and doped ZrNiSn synthesised and S/PF/zT measured extensively. & Uher 1999, Physical Review B 59, 8615-8621 (10.\allowbreak{}1103/\allowbreak{}physrevb.\allowbreak{}59.\allowbreak{}8615) \\
CoTaSn & TaCoSn & exp. investig. &  & Experimental: half-Heusler TaCoSn first synthesised and TE properties measured for pristine and Nb/Sb-alloyed n-type samples (zT reported); earlier computational prediction. & Li 2019, ACS Appl. Mater. Interfaces 11, 41321 (10.\allowbreak{}1021/\allowbreak{}acsami.\allowbreak{}9b13603) \\
CoVSn & VCoSn & exp. investig. &  & One experimental study (powder metallurgy + SPS) gave multiphase non-stoichiometric CoVSn with measured, very low zT; preceded by computational prediction of high p-type performance. & Hooshmand Zaferani 2020, Energies 13, 1459 (10.\allowbreak{}3390/\allowbreak{}en13061459) \\
Fe2TiSi & Fe2TiSi & exp. investig. &  & Experimental but limited: L2\_1 Fe2TiSi realised only as epitaxial thin films (sputtering, MBE); one film study measured room-temperature Seebeck and thermal conductivity; computational TE prediction (Bilc 2015). & Shimanuki 2020, J. Appl. Phys. 127, 055106 (10.\allowbreak{}1063/\allowbreak{}1.\allowbreak{}5141949) \\
IrNbSn & NbIrSn & exp. investig. &  & Half-Heusler NbIrSn synthesised; room-temperature Seebeck coefficient and resistivity measured. Computational only: dedicated first-principles TE study (electron and phonon transport, zT) of pristine HH NbIrSn; also included in a high-throughput PF/zT screen. & Hohl 1998, J. Phys.: Condens. Matter 10, 7843 (10.\allowbreak{}1088/\allowbreak{}0953-\allowbreak{}8984/\allowbreak{}10/\allowbreak{}35/\allowbreak{}016) \\
NiScSb & ScNiSb & exp. investig. &  & Experimental: pristine ScNiSb synthesised and S, rho, kappa, zT measured in several independent studies; doped variants also measured. & Synoradzki 2019, Materials 12, 1723 (10.\allowbreak{}3390/\allowbreak{}ma12101723) \\
NiYBi & YNiBi & exp. investig. &  & One dedicated experimental study (synthesis and measured S, sigma, kappa of pristine YNiBi), plus several ab initio PF studies. & Li 2015, Journal of Applied Physics 117, 205101 (10.\allowbreak{}1063/\allowbreak{}1.\allowbreak{}4921811) \\
NiYSb & YNiSb & exp. investig. &  & Experimental thermopower/transport of pristine YNiSb measured in a comparative ScMSb/YMSb study; ab initio work predicts a high p-type PF. & Oestreich 2003, Journal of Physics: Condensed Matter 15, 635-640 (10.\allowbreak{}1088/\allowbreak{}0953-\allowbreak{}8984/\allowbreak{}15/\allowbreak{}4/\allowbreak{}304) \\
PdScSb & ScPdSb & exp. investig. &  & Experimental: pristine polycrystalline ScPdSb was synthesised and its thermoelectric properties (Seebeck, resistivity, ZT near 300 K) measured in one study; later a DFT+BoltzTraP study computed p-type transport. & Oestreich 2003, J. Phys.: Condens. Matter 15, 635 (10.\allowbreak{}1088/\allowbreak{}0953-\allowbreak{}8984/\allowbreak{}15/\allowbreak{}4/\allowbreak{}304) \\
PtScSb & ScPtSb & exp. investig. &  & Experimental: pristine polycrystalline ScPtSb was synthesised and thermoelectric properties measured (one study, small ZT near 300 K); later DFT+BoltzTraP predicted it to have the largest p-type power factor of ScMSb. & Oestreich 2003, J. Phys.: Condens. Matter 15, 635 (10.\allowbreak{}1088/\allowbreak{}0953-\allowbreak{}8984/\allowbreak{}15/\allowbreak{}4/\allowbreak{}304) \\
AuHfAl & HfAuAl & comp. investig. &  & High-throughput half-Heusler transport/quality-factor screen only. Computational only: listed among AFLOW thermodynamically stable half-Heuslers with a high-throughput lattice-thermal-conductivity estimate and in an ML/DFT kappa\_L dataset; no dedicated TE transport study, synthesis or measurement found. & Pak 2025, J. Mater. Chem. A 13, 19455 (10.\allowbreak{}1039/\allowbreak{}d5ta00397k) \\
CoHfAs & HfCoAs & comp. investig. & exp. (Kleinke 1998; ICSD 406953, Pnma) & Dedicated DFT thermoelectric study; also in high-throughput screens. Computational screening only: kappa\_L estimate in AFLOW-based high-throughput screen and an n-type conductivity-effective-mass descriptor in a 69-compound half-Heusler band-engineering table; no dedicated TE study or measurement found. & Yang 2022, J. Solid State Chem. 314, 123386 (10.\allowbreak{}1016/\allowbreak{}j.\allowbreak{}jssc.\allowbreak{}2022.\allowbreak{}123386) \\
CoHfBi & HfCoBi & comp. investig. &  & Computational TE predictions only (DFT/Boltzmann transport of Bi-based 18-electron half-Heuslers XYBi incl. HfCoBi); no synthesis or measured TE data for pristine HfCoBi found. & Paul 2024, ACS Appl. Energy Mater. 7, 9595 (10.\allowbreak{}1021/\allowbreak{}acsaem.\allowbreak{}4c01652) \\
CoNbGe & NbCoGe & comp. investig. & exp. (Jeitschko 1969, via Singh 2019 (sec.)) & Computational only: first-principles cubic-vs-orthorhombic stability and Boltzmann TE transport for cubic CoNbGe; kappa\_L estimate in AFLOW high-throughput screen. & Singh 2019, J. Phys.: Condens. Matter 32, 055505 (10.\allowbreak{}1088/\allowbreak{}1361-\allowbreak{}648x/\allowbreak{}ab4e71) \\
CoTaGe & TaCoGe & comp. investig. & exp. (Jeitschko 1969, via Singh 2019 (sec.)) & Computational only: first-principles phase-stability and Boltzmann TE transport for cubic CoTaGe; kappa\_L estimate in high-throughput screen. & Singh 2019, J. Phys.: Condens. Matter 32, 055505 (10.\allowbreak{}1088/\allowbreak{}1361-\allowbreak{}648x/\allowbreak{}ab4e71) \\
CoTaSi & TaCoSi & comp. investig. & exp. (Landolt-Bornstein, via patent (sec.)) & Computational only: first-principles TE transport for hypothetical cubic CoTaSi, whose reported ground state is orthorhombic; kappa\_L estimate in high-throughput screen. & Singh 2019, J. Phys.: Condens. Matter 32, 055505 (10.\allowbreak{}1088/\allowbreak{}1361-\allowbreak{}648x/\allowbreak{}ab4e71) \\
CoTiAs & TiCoAs & comp. investig. &  & High-throughput half-Heusler transport/quality-factor screen only. Known only computationally: kappa\_L estimate in AFLOW high-throughput screen and a DFT optoelectronic/elastic study of CoTiX (X = P, As, Sb); no TE transport study or measurement found. & Pak 2025, J. Mater. Chem. A 13, 19455 (10.\allowbreak{}1039/\allowbreak{}d5ta00397k) \\
CoTiBi & TiCoBi & comp. investig. &  & Computational TE predictions only (DFT elastic/phonon/Boltzmann TE of ZrCoBi and TiCoBi; Bi-based XYBi screening); no synthesis or measurement found. & Chen 2023, Eur. Phys. J. Plus 138, 966 (10.\allowbreak{}1140/\allowbreak{}epjp/\allowbreak{}s13360-\allowbreak{}023-\allowbreak{}04592-\allowbreak{}z) \\
CoVGe & VCoGe & comp. investig. & exp. (Jeitschko 1969, via Singh 2019 (sec.)) & Computational only: first-principles cubic stability and Boltzmann TE transport for cubic CoVGe; kappa\_L estimate in high-throughput screen. & Singh 2019, J. Phys.: Condens. Matter 32, 055505 (10.\allowbreak{}1088/\allowbreak{}1361-\allowbreak{}648x/\allowbreak{}ab4e71) \\
CoVSi & VCoSi & comp. investig. & exp. (Landolt-Bornstein, via patent (sec.)) & Computational only: first-principles TE transport for hypothetical cubic CoVSi (ground state reported orthorhombic). & Singh 2019, J. Phys.: Condens. Matter 32, 055505 (10.\allowbreak{}1088/\allowbreak{}1361-\allowbreak{}648x/\allowbreak{}ab4e71) \\
CoZrAs & ZrCoAs & comp. investig. &  & High-throughput half-Heusler transport/quality-factor screen only. Known only computationally: kappa\_L estimate in high-throughput screen and a DFT elastic/electronic/optical (photovoltaic-oriented) study; TE only mentioned as a possible application. & Pak 2025, J. Mater. Chem. A 13, 19455 (10.\allowbreak{}1039/\allowbreak{}d5ta00397k) \\
Fe2HfSi & Fe2HfSi & comp. investig. &  & Computational only: DFT (FP-LAPW) structural, optoelectronic and TE properties of L2\_1 Fe2HfSi under pressure; also in a DFT Fe2YZ band-engineering TE screen. & Azam 2023, RSC Adv. 13, 15437 (10.\allowbreak{}1039/\allowbreak{}d3ra00362k) \\
Fe2NbAl & Fe2NbAl & comp. investig. &  & Computational only: DFT prediction of large power factor (Fe2YZ band engineering) and DFT lattice-dynamics/TE study of Fe2MAl (M = V, Nb, Ta). & Ye 2024, npj Quantum Mater. 9, 60 (10.\allowbreak{}1038/\allowbreak{}s41535-\allowbreak{}024-\allowbreak{}00671-\allowbreak{}1) \\
Fe2TaAl & Fe2TaAl & comp. investig. &  & Computational only: DFT electronic structure and TE (BoltzTraP) of Fe2TaZ (Z = Al, Ga, In) and Fe2YZ band-engineering TE prediction. & Khandy 2018, RSC Adv. 8, 40996 (10.\allowbreak{}1039/\allowbreak{}c8ra04433c) \\
FeNbAs & NbFeAs & comp. investig. &  & High-throughput half-Heusler transport/quality-factor screen only. Computational screening only: kappa\_L estimate in AFLOW high-throughput screen and an n-type effective-mass descriptor in a half-Heusler band-engineering table; no dedicated TE study. & Pak 2025, J. Mater. Chem. A 13, 19455 (10.\allowbreak{}1039/\allowbreak{}d5ta00397k) \\
FeTaAs & TaFeAs & comp. investig. &  & Computational only: high-throughput DFT (deformation-potential PF, Slack kappa) identifies TaFeAs as a novel stable half-Heusler and the best TE candidate of its set; kappa\_L in AFLOW screen. & Bilinska 2023, Crystals 13, 1378 (10.\allowbreak{}3390/\allowbreak{}cryst13091378) \\
FeVAs & VFeAs & comp. investig. &  & Computational only: high-throughput DFT TE (PF/ZT) study names VFeAs among novel stable half-Heuslers with TE potential. & Bilinska 2023, Crystals 13, 1378 (10.\allowbreak{}3390/\allowbreak{}cryst13091378) \\
IrNbGe & NbIrGe & comp. investig. & exp. (Mishra 2001 (Z. Naturforsch.)) & Computational only: DFT/BoltzTraP2 power factor and zT (deformation-potential relaxation time, Slack kappa\_L) for pristine cubic HH NbIrGe in a high-throughput study; no HH synthesis found. & Bilińska 2023, Crystals 13, 1378 (10.\allowbreak{}3390/\allowbreak{}cryst13091378) \\
IrTaGe & TaIrGe & comp. investig. &  & HH TaIrGe has been synthesised (as a p-type transparent conductor, not a TE study); the TE evidence is computational (DFT/BoltzTraP zT under strain), on pristine material. & Nisha 2021, Journal of Solid State Chemistry 303, 122524 (10.\allowbreak{}1016/\allowbreak{}j.\allowbreak{}jssc.\allowbreak{}2021.\allowbreak{}122524) \\
IrTaSi & TaIrSi & comp. investig. & exp. (orthorhombic TaIrSi literature (row notes)) & High-throughput half-Heusler transport/quality-factor screen only. No TE study of cubic HH TaIrSi found; the only literature concerns the orthorhombic polymorph (structure/magnetism, superconductivity). & Pak 2025, J. Mater. Chem. A 13, 19455 (10.\allowbreak{}1039/\allowbreak{}d5ta00397k) \\
IrTaSn & TaIrSn & comp. investig. &  & Computational only: DFT/BoltzTraP TE study of pristine HH TaIrSn reporting large predicted zT; other DFT papers cover its optical/thermodynamic properties. & Kaur 2017, Physics Letters A 381, 3760-3765 (10.\allowbreak{}1016/\allowbreak{}j.\allowbreak{}physleta.\allowbreak{}2017.\allowbreak{}09.\allowbreak{}043) \\
IrTiAs & TiIrAs & comp. investig. &  & Computational only: first-principles TE study (S, sigma, kappa, zT) of pristine HH TiIrAs and TiIrSb. & Kaur 2017, Journal of Alloys and Compounds 727, 1171-1177 (10.\allowbreak{}1016/\allowbreak{}j.\allowbreak{}jallcom.\allowbreak{}2017.\allowbreak{}08.\allowbreak{}216) \\
IrVGe & VIrGe & comp. investig. & pred. (Gautier 2015, predicted) & Computational only: DFT study of structural, vibrational and TE properties of pristine HH VYGe (Y = Rh, Co, Ir); also in a high-throughput PF/zT screen. & Solola 2023, Computational Condensed Matter 36, e00827 (10.\allowbreak{}1016/\allowbreak{}j.\allowbreak{}cocom.\allowbreak{}2023.\allowbreak{}e00827) \\
NiHfGe & HfNiGe & comp. investig. & exp. (Mallick 2025 (row notes)) & Computational only for the HH polymorph: DFT/BoltzTraP zT prediction for pristine HH NiHfGe, plus DFT stability/elastic work (Osafile 2021, Mol. Phys. 119, e1936249, doi 10.1080/00268976.2021.1936249, verified). & Bendahma 2019, Computational Condensed Matter 21, e00407 (10.\allowbreak{}1016/\allowbreak{}j.\allowbreak{}cocom.\allowbreak{}2019.\allowbreak{}e00407) \\
NiHfSi & HfNiSi & comp. investig. & exp. (Saurabh 2021) & Computational only for the HH polymorph: TB-mBJ DFT study of NiHfZ (Z = Si, Ge) that includes transport/TE quantities; DFT stability study (Osafile 2021). & Nazir 2026, Silicon 18, 3035-3051 (10.\allowbreak{}1007/\allowbreak{}s12633-\allowbreak{}026-\allowbreak{}03704-\allowbreak{}5) \\
NiScAs & ScNiAs & comp. investig. & pred. (Gautier 2015, predicted) & Computational only: several DFT+Boltzmann TE studies of pristine HH ScNiAs (NiTZ family; XNiAs X = Sc, Y; Ni-based HH high-throughput). & Jaishi 2021, AIP Advances 11, 025304 (10.\allowbreak{}1063/\allowbreak{}5.\allowbreak{}0031512) \\
NiScBi & ScNiBi & comp. investig. &  & Computational only: ab initio power-factor and TE studies of p-type ScNiBi (with YNiBi, LuNiBi). & Winiarski 2020, Acta Physica Polonica A 138, 533-538 (10.\allowbreak{}12693/\allowbreak{}aphyspola.\allowbreak{}138.\allowbreak{}533) \\
NiTiGe & TiNiGe & comp. investig. & exp. (Downie 2023) & Computational only for the HH polymorph: included in DFT/Boltzmann high-throughput TE studies (Ni-based HH; 18-electron HH PF/zT screen). & Chen 2021, ACS Omega 6, 18269-18280 (10.\allowbreak{}1021/\allowbreak{}acsomega.\allowbreak{}1c02172) \\
NiTiPb & TiNiPb & comp. investig. & pred. (Bilinska 2023, predicted) & Computational only: dedicated DFT TE study of pristine HH TiXPb (X = Ni, Pd, Pt), with low predicted kappa\_L. & Khatri 2024, Computational Materials Science 244, 113250 (10.\allowbreak{}1016/\allowbreak{}j.\allowbreak{}commatsci.\allowbreak{}2024.\allowbreak{}113250) \\
NiTiSi & TiNiSi & comp. investig. & exp. (Downie 2023) & Computational only for the HH polymorph: DFT/Boltzmann TE of pristine HH NiTiSi in a Ni-based HH high-throughput study; also a 2020 conference DFT TE abstract (Bati et al., MDPI IOCC 2020; DOI not verified). & Chen 2021, ACS Omega 6, 18269-18280 (10.\allowbreak{}1021/\allowbreak{}acsomega.\allowbreak{}1c02172) \\
NiVGa & VNiGa & comp. investig. &  & Computational only: DFT/Boltzmann TE (PF, zT) of pristine HH NiVGa in Ni-based HH screens. & Chen 2021, ACS Omega 6, 18269-18280 (10.\allowbreak{}1021/\allowbreak{}acsomega.\allowbreak{}1c02172) \\
NiZrGe & ZrNiGe & comp. investig. & exp. (Saurabh 2021) & High-throughput half-Heusler transport/quality-factor screen only. No TE study of the cubic HH polymorph found; experimental TE transport exists only for orthorhombic TiNiSi-type ZrNiGe (a semimetal). & Pak 2025, J. Mater. Chem. A 13, 19455 (10.\allowbreak{}1039/\allowbreak{}d5ta00397k) \\
NiZrSi & ZrNiSi & comp. investig. & exp. (Saurabh 2022 (row notes)) & High-throughput half-Heusler transport/quality-factor screen only. No TE study of the cubic HH polymorph found; experimental TE (pristine and Sb/Hf-doped) exists only for orthorhombic TiNiSi-type ZrNiSi. & Pak 2025, J. Mater. Chem. A 13, 19455 (10.\allowbreak{}1039/\allowbreak{}d5ta00397k) \\
OsNbSb & NbOsSb & comp. investig. &  & High-throughput half-Heusler transport/quality-factor screen only. Predicted thermodynamically stable HH (DFT and CALYPSO structure search) in an inverse-design TE paper whose experimental work covered TaFeSb, not NbOsSb; no TE calculation or measurement of NbOsSb found. & Pak 2025, J. Mater. Chem. A 13, 19455 (10.\allowbreak{}1039/\allowbreak{}d5ta00397k) \\
OsTaSb & TaOsSb & comp. investig. &  & High-throughput half-Heusler transport/quality-factor screen only. Predicted thermodynamically stable HH (DFT and CALYPSO structure search) in the TaFeSb discovery paper; no TE calculation or measurement of TaOsSb found. & Pak 2025, J. Mater. Chem. A 13, 19455 (10.\allowbreak{}1039/\allowbreak{}d5ta00397k) \\
PdScAs & ScPdAs & comp. investig. & pred. (Gautier 2015, predicted) & Cubic F-43m: computational only, a DFT/deformation-potential TE study of pristine PdMX (M = Sc, Y; X = P, As, Sb) giving kappa\_L and zT for PdScAs; also an ML power-factor prediction. & Rani 2023, Materials Today Communications 36, 106461 (10.\allowbreak{}1016/\allowbreak{}j.\allowbreak{}mtcomm.\allowbreak{}2023.\allowbreak{}106461) \\
PdTiPb & TiPdPb & comp. investig. &  & Computational only: a single PBE-GGA DFT study of pristine PdTiZ (Z = Ge, Pb) including Boltzmann-transport thermoelectric properties. & Kalita 2022, Int. J. Quantum Chem. 122, e26951 (10.\allowbreak{}1002/\allowbreak{}qua.\allowbreak{}26951) \\
PdTiSn & TiPdSn & comp. investig. &  & Computational only: DFT+Boltzmann prediction of p-type TE performance of pristine TiPdSn, and a hybrid-DFT (CRYSTAL) TE study of TiMSn (M = Ni, Pd, Pt) including defects. & Kaur 2017, EPL 117, 47002 (10.\allowbreak{}1209/\allowbreak{}0295-\allowbreak{}5075/\allowbreak{}117/\allowbreak{}47002) \\
PdZrSn & ZrPdSn & comp. investig. &  & Computational only: one DFT + Boltzmann-transport study (electronic, phonon, effective mass, ZT) of pristine PdXSn (X = Zr, Hf). & Rani 2022, Molecules 27, 6567 (10.\allowbreak{}3390/\allowbreak{}molecules27196567) \\
PtNbGa & NbPtGa & comp. investig. &  & Computational only: DFT (GGA/mBJ) + Boltzmann transport assessment of pristine PtNbZ (Z = Al, Ga, In) for TE use; a separate hybrid-DFT study covers structure/electronic/optical properties only. & Abdullah 2025, RSC Adv. 15, 47036 (10.\allowbreak{}1039/\allowbreak{}d5ra06109a) \\
PtNbIn & NbPtIn & comp. investig. &  & Computational only: DFT (GGA/mBJ) + Boltzmann transport assessment of pristine PtNbZ (Z = Al, Ga, In); a separate hybrid-DFT study covers structure/electronic/optical properties only. & Abdullah 2025, RSC Adv. 15, 47036 (10.\allowbreak{}1039/\allowbreak{}d5ra06109a) \\
PtTaGa & TaPtGa & comp. investig. &  & Computational only: DFT stability and Boltzmann-transport power-factor study of pristine XTaZ (X = Pd, Pt; Z = Al, Ga, In) half-Heuslers; a separate hybrid-DFT study covers structure/electronic/optics. & Khandy 2021, Sci. Rep. 11, 20756 (10.\allowbreak{}1038/\allowbreak{}s41598-\allowbreak{}021-\allowbreak{}00314-\allowbreak{}6) \\
PtVGa & VPtGa & comp. investig. &  & High-throughput half-Heusler transport/quality-factor screen only. Computational only and not TE: a hybrid-DFT study of PtYZ (Y = V, Nb, Ta; Z = Al, Ga, In) reports structure, stability, electronic and optical properties; no transport/TE calculation or measurement for PtVGa was found. & Pak 2025, J. Mater. Chem. A 13, 19455 (10.\allowbreak{}1039/\allowbreak{}d5ta00397k) \\
RhHfSb & HfRhSb & comp. investig. &  & Computational only: an ab initio study of pristine MRhSb (M = Ti, Zr, Hf) including BoltzTraP Seebeck/TE properties alongside elastic, thermodynamic and optical properties. & Benzoudji 2019, Chin. J. Phys. 59, 434 (10.\allowbreak{}1016/\allowbreak{}j.\allowbreak{}cjph.\allowbreak{}2019.\allowbreak{}04.\allowbreak{}009) \\
RhNbGe & NbRhGe & comp. investig. & exp. (Dinges 2010, isostructural with TaRhGe (sec.)) & Computational only: DFT study of electronic, elastic, thermoelectric and optical properties of XRhZ (X = V, Nb; Z = Si, Ge), plus a DFT TE preprint on RhNbZ (Z = Ge, Sn). & Ahmed 2024, East Eur. J. Phys. (2024 no. 1), 294 (10.\allowbreak{}26565/\allowbreak{}2312-\allowbreak{}4334-\allowbreak{}2024-\allowbreak{}1-\allowbreak{}26) \\
RhNbSn & NbRhSn & comp. investig. &  & Computational only: Boltzmann transport + deformation-potential + Slack-model TE study of pristine XRhSn (X = V, Nb, Ta), plus a DFT TE preprint on RhNbZ (Z = Ge, Sn). & Yang 2023, J. Supercond. Nov. Magn. 36, 1043 (10.\allowbreak{}1007/\allowbreak{}s10948-\allowbreak{}023-\allowbreak{}06536-\allowbreak{}1) \\
RhTaGe & TaRhGe & comp. investig. & exp. (Dinges 2010) & Computational only: WIEN2k + BoltzTraP study of pristine RhTaZ (Z = Si, Ge, Sn) concluding high predicted ZT; compound itself listed as previously reported in ICSD. & Boudjeltia 2021, Mod. Phys. Lett. B 35, 2150400 (10.\allowbreak{}1142/\allowbreak{}s0217984921504005) \\
RhTaSn & TaRhSn & comp. investig. &  & Computational only: DFT+Boltzmann assessment of p-type TaRhSn, and a later Boltzmann/deformation-potential study of n-type XRhSn (X = V, Nb, Ta). & Kaur 2017, Ceram. Int. 43, 15160 (10.\allowbreak{}1016/\allowbreak{}j.\allowbreak{}ceramint.\allowbreak{}2017.\allowbreak{}08.\allowbreak{}046) \\
RhTiAs & TiRhAs & comp. investig. & exp. (Weber 2017) & High-throughput half-Heusler transport/quality-factor screen only. Computational only and not TE: TiRhAs appears in Materials Project (F-43m entry) and in a phonon/vibrational-entropy phase-stability study comparing the cubic half-Heusler and a distorted orthorhombic polymorph; no TE study found. & Pak 2025, J. Mater. Chem. A 13, 19455 (10.\allowbreak{}1039/\allowbreak{}d5ta00397k) \\
RhTiBi & TiRhBi & comp. investig. &  & Computational only: DFT + semiclassical Boltzmann study of pristine RhBiX (X = Ti, Zr, Hf) including lattice thermal conductivity and ZT. & Wei 2023, RSC Adv. 13, 11513 (10.\allowbreak{}1039/\allowbreak{}d3ra01262j) \\
RhTiSb & TiRhSb & comp. investig. &  & Computational only: an ab initio study of MRhSb (M = Ti, Zr, Hf) including BoltzTraP Seebeck coefficients; a second DFT study covers structural, mechanical and phonon properties (no TE). & Benzoudji 2019, Chin. J. Phys. 59, 434 (10.\allowbreak{}1016/\allowbreak{}j.\allowbreak{}cjph.\allowbreak{}2019.\allowbreak{}04.\allowbreak{}009) \\
RhVGe & VRhGe & comp. investig. & pred. (Gautier 2015, predicted) & Computational only: WIEN2k + BoltzTraP TE study of pristine RhVX (X = Si, Ge), and a DFT electronic/elastic/TE/optical study of XRhZ (X = V, Nb; Z = Si, Ge). & Mebed 2022, Int. J. Mod. Phys. B 37, 2350163 (10.\allowbreak{}1142/\allowbreak{}s0217979223501631) \\
RhZrBi & ZrRhBi & comp. investig. &  & Computational TE only: ab initio transport/ZT predictions for ideal ZrRhBi (also RhBiX DFT study); the compound was synthesised (multiphase, high purity, F-43m confirmed by XRD/TEM) but no measured TE data were found. & Vikram 2017, J. Mater. Chem. A 5, 6131 (10.\allowbreak{}1039/\allowbreak{}c7ta00920h) \\
Ru2ZrSi & Ru2ZrSi & comp. investig. &  & Computational only: one DFT study of L2\_1 Ru2ZrSi reporting TE transport and ZT; separately, an experimental/theoretical Ru2TiSi study proposes Zr/Hf substitution on the Ti site (alloyed, not Ru2ZrSi itself). & Benidris 2021, Solid State Commun. 339, 114466 (10.\allowbreak{}1016/\allowbreak{}j.\allowbreak{}ssc.\allowbreak{}2021.\allowbreak{}114466) \\
RuNbAs & NbRuAs & comp. investig. & pred. (Bilinska 2023, predicted) & Computational only: DFT + semiclassical Boltzmann study of pristine p-type RuVAs and RuNbAs (Seebeck, conductivity, lattice thermal conductivity, ZT). & Cherifi 2020, Phys. Status Solidi B 257, 2000271 (10.\allowbreak{}1002/\allowbreak{}pssb.\allowbreak{}202000271) \\
RuNbSb & NbRuSb & comp. investig. &  & Computational only: first-principles TE assessments of XRuSb (X = V, Nb, Ta) and of NbMSb (M = Fe, Ru, Os). & Kaur 2017, J. Phys. Chem. Solids 110, 108 (10.\allowbreak{}1016/\allowbreak{}j.\allowbreak{}jpcs.\allowbreak{}2017.\allowbreak{}05.\allowbreak{}006) \\
RuTaAs & TaRuAs & comp. investig. &  & Computational only: one DFT + Boltzmann study of TaRuAs including substitutional doping on Ru and As sites. & Nadhreen 2026, Ann. Phys. (Berlin) 538, e00096 (10.\allowbreak{}1002/\allowbreak{}andp.\allowbreak{}202500096) \\
RuTaSb & TaRuSb & comp. investig. &  & Computational only: first-principles prediction of high p-type zT for RuTaSb (low band effective mass) and follow-up DFT studies of pristine and Bi-substituted TaSbRu. & Fang 2017, Phys. Chem. Chem. Phys. 19, 4411 (10.\allowbreak{}1039/\allowbreak{}c6cp07897d) \\
RuVAs & VRuAs & comp. investig. &  & Computational only: DFT + Boltzmann TE study of pristine p-type RuVAs/RuNbAs; a further DFT study covers structural, thermodynamic, mechanical and electronic properties (no TE). & Cherifi 2020, Phys. Status Solidi B 257, 2000271 (10.\allowbreak{}1002/\allowbreak{}pssb.\allowbreak{}202000271) \\
RuVSb & VRuSb & comp. investig. &  & Computational only: first-principles TE assessment of pristine XRuSb (X = V, Nb, Ta). & Kaur 2017, J. Phys. Chem. Solids 110, 108 (10.\allowbreak{}1016/\allowbreak{}j.\allowbreak{}jpcs.\allowbreak{}2017.\allowbreak{}05.\allowbreak{}006) \\
NiNbIn & NbNiIn & underexplored &  & No dedicated TE (or other) study of HH NbNiIn found; it appears only as a computed database entry (Materials Project). &  \\
\end{longtable}
}

\begin{table}[H]\centering\scriptsize
\caption{\textbf{Parent model of each family according to each project source.} ``Scratch'': random initialisation; ``chain'': the
single sequence full cubic $\rightarrow$ full tetragonal, inverse cubic $\rightarrow$ inverse tetragonal $\rightarrow$ half cubic
$\rightarrow$ half tetragonal. Transfer script: \texttt{submit\_transfer.sh}; pipeline document: \texttt{docs/PHYSICS\_AND\_PIPELINE.md}. Only the last column rests on run records.}
\begin{tabular}{@{}l *{5}{>{\raggedright\arraybackslash}p{2.0cm}} >{\raggedright\arraybackslash}p{2.4cm}@{}}
\toprule
family & presentation (slide 20) & transfer script & pipeline doc., chain & pipeline doc., table & project overview & run records \\
\midrule
full cubic & foundation & scratch & -- & scratch & foundation & foundation, $s=0.1$ (folds 0, 1) \\
full tetragonal & full cubic & full cubic & full cubic (fold-0 SWA) & full cubic & foundation & none \\
inverse cubic & full cubic & full cubic & full cubic (fold-0 SWA) & full cubic & foundation & none \\
inverse tetragonal & inverse cubic & inverse cubic & chain & inverse cubic & foundation & none \\
half cubic & full cubic & scratch & chain (after inverse tetragonal) & scratch & foundation & none \\
half tetragonal & half cubic & half cubic & chain & half cubic & foundation & none \\
\bottomrule
\end{tabular}
\end{table}

\begin{table}[H]\centering\small
\caption{\textbf{d-DOS model and training settings}, read from the training code, the per-fold configuration files and the tensor shapes
of the deployed checkpoints (Supplementary Note~3).}
\begin{tabular}{@{}>{\raggedright}p{0.20\textwidth} >{\raggedright\arraybackslash}p{0.74\textwidth}@{}}
\toprule
item & setting \\
\midrule
encoder & ALIGNN 2026.4.1, layers unchanged; 8 ALIGNN + 8 graph-convolution layers; 512 features; batch normalisation (package default 4 + 4 layers, 256 features) \\
inputs & 97 atom features (92 CGCNN + 5 Heusler-specific); bond length: 80 radial basis functions; angle cosine: 40; embeddings 64 \\
graph & cutoff \SI{8}{\angstrom}; at most 12 neighbours; line graph of bond pairs; structures not relaxed \\
readout & mean over atoms \\
head and output & $512\rightarrow1024\rightarrow2048\rightarrow4002$; GELU; dropout 0.2 after the first two layers; softplus; $1.08\times10^{7}$ of $4.25\times10^{7}$ trainable parameters \\
loss & five weighted terms, Eq.~(\ref{eq:loss}) (loss equation of the main text) \\
optimiser & AdamW; learning rate $10^{-3}$; weight decay $5\times10^{-5}$; gradient clipping 1.0; float32 \\
schedule & reduce-on-plateau (factor 0.5; patience 25, or 15 for cubic full-Heusler folds 0--2); minimum $10^{-7}$ \\
weight averaging & SWA over the last 20\% of epochs; learning rate annealed to $5\times10^{-5}$ over 10 epochs; batch-normalisation statistics re-estimated \\
epochs, batch & 300, 128; cubic full Heusler: 150 (folds 0, 1) or 200 (folds 2--4), 64 \\
splits & five-fold per family, shuffled (seed 42); 10\% of non-test entries for validation (72/8/20\%) \\
checkpoints & best-validation checkpoint, highest validation $r$ over concatenated spectra before SWA (metrics); SWA weights (inference) \\
transfer & pretrained on all 51,257 spectra (100 epochs, SWA from epoch 85); encoder at $0.1\times$ learning rate verified for two folds (cubic full Heusler, folds 0, 1) \\
hardware & one NVIDIA H100 NVL GPU per run; PyTorch 2.2.2 (CUDA 12.1), DGL 1.1.3, Python 3.11 \\
\bottomrule
\end{tabular}

\end{table}

\begin{table}[H]\centering\scriptsize
\caption{\textbf{Provenance of the 3,625 DFT runs by selection round.} Rounds were built successively, each ordered by the
screening scores current at the time (Methods). T and P give the number of runs from the training and prediction sets, and n/a the runs
that could not be attributed. d-DOS SC: runs labelled semiconducting by their set's screen. DFT gapped: QE-PBE gap of at least
\SI{0.136}{eV}. Qualified runs pass the Seebeck, power-factor and stability criteria.}
\begin{tabular}{@{}l p{4.2cm} r rrr r rrr@{}}
\toprule
round & ordering criterion & runs & T & P & n/a & d-DOS SC & DFT gapped & qualified runs & certified \\
\midrule
1 & highest $P^{\mathrm d}_{\max}$ (early version) and largest ensemble spread & 564 & 197 & 367 & 0 & 545 & 214 & 100 & 61 \\
2 & $P^{\mathrm d}_{\max}$ order within tiers & 284 & 284 & 0 & 0 & 163 & 25 & 17 & 2 \\
3 & $P^{\mathrm d}_{\max}$ order after priority filters & 553 & 426 & 127 & 0 & 480 & 189 & 87 & 45 \\
4 & top 100 by $P^{\mathrm d}_{\max}$ among gap-filter rejects & 100 & 60 & 13 & 27 & 4 & 0 & 14 & 0 \\
5 & $P^{\mathrm d}_{\max}$ order after semiconductor and gap filters & 194 & 66 & 114 & 14 & 156 & 2 & 8 & 0 \\
6 & $S^{\mathrm d}_{\max}$ order, no filters & 272 & 189 & 82 & 1 & 136 & 119 & 71 & 47 \\
7 & semiconductor label, gap $\ge$\,0.714\,eV first, by $S^{\mathrm d}_{\max}$ & 405 & 100 & 305 & 0 & 405 & 23 & 7 & 0 \\
8 & semiconductor label in confidence tiers (0.10--1.40\,eV first) & 1,251 & 238 & 1,013 & 0 & 1,251 & 3 & 5 & 0 \\
9 & 18 or 24 valence electrons & 2 & 2 & 0 & 0 & 2 & 0 & 0 & 0 \\
\midrule
all & & 3,625 & 1,562 & 2,021 & 42 & 3,142 & 575 & 309 & 155 \\
\bottomrule
\end{tabular}

\end{table}

\begin{table}[H]\centering\small
\caption{\textbf{QE-PBE and BoltzTraP2 settings}, read from the input and output files of the calculations.}
\begin{tabular}{@{}>{\raggedright}p{0.20\textwidth} >{\raggedright\arraybackslash}p{0.74\textwidth}@{}}
\toprule
item & setting \\
\midrule
code, functional & Quantum ESPRESSO 7.1; PBE \\
pseudopotentials & SSSP 1.3.0 efficiency; PSlibrary fully relativistic PAW for SOC (1.0.0; 0.3.1 for Co and Mn, 0.2.3 for Ir) \\
cutoffs & largest SSSP value of the species: 30--90~Ry (wavefunctions), 240--1,080~Ry (density); SOC: $1.15\times$ (wavefunction) and $1.10\times$ (density) the pseudopotential values \\
smearing, SCF & Marzari--Vanderbilt, 0.01~Ry; convergence threshold $10^{-8}$~Ry; local-TF mixing, $\beta=0.3$ \\
k-meshes & $n_i=\max\bigl(6,\operatorname{round}(K/|\mathbf a_i|)\bigr)$; SCF $K=\SI{40}{\angstrom}$, shifted ($6^3$--$11^3$); NSCF $K=\SI{55}{\angstrom}$, unshifted ($9^3$--$15^3$) \\
bands & $N_\mathrm{el}/2+50$; SOC: $2.4\,N_\mathrm{el}$ \\
magnetism & collinear; starting moments 0.2--0.4~$\mu_B$ on V, Cr, Mn, Fe, Co, Ni (antiparallel retry on SCF failure); SOC runs non-magnetic \\
geometry & DXMag cell reduced to the primitive cell (spglib, tolerance $10^{-3}$); no relaxation \\
BoltzTraP2 & 24.7.2; interpolation multiplier 15; spins merged, common $\tau$; $T=300$--\SI{1200}{K} in steps of \SI{100}{K} \\
chemical potential & $|\mu-E_F|\le\SI{0.5}{eV}$ (certification); $\le\SI{1.0}{eV}$ with carrier caps $10^{20}$ and $10^{21}$~cm$^{-3}$ (ranking) \\
\bottomrule
\end{tabular}

\end{table}

\begin{table}[H]\centering\scriptsize
\caption{\textbf{Spin--orbit coupling by heavy element.} Paired scalar-relativistic and SOC runs of the same structure (analysis/soc\_paired.py).
A compound is counted under every heavy element it contains. PF ratio: median of best $\pf$ with SOC over best $\pf$ without; $\Delta E_g$: median
SOC minus scalar gap over runs gapped in both; $|S|$ and $\sigma/\tau$ ratios at the respective optima. Scalar candidates are the heavy-element
runs that are gapped and meet criteria (i)--(iii) without SOC; a run is lost when its SOC power factor falls below $\pf_0$. Lost compositions
lose every candidate run. Of the 693 SOC calculations attempted, 607 finished and 86 failed and were not retried; only finished
pairs are counted.}
\label{tab:si-soc}
\begin{tabular}{@{}l r r r r r r r r r r >{\raggedright\arraybackslash}p{3.4cm}@{}}
\toprule
 & \multicolumn{3}{c}{all paired runs} & \multicolumn{7}{c}{scalar candidates (gapped, qualified)} & \\
\cmidrule(lr){2-4}\cmidrule(lr){5-11}
element & runs & PF ratio & $\Delta E_g$ (meV) & runs & lost & PF ratio & $|S|$ ratio & $\sigma/\tau$ ratio & p, 900\,K & n, 900\,K & compositions lost \\
\midrule
Hf & 99 & 0.88 & $-$39 & 27 & 12 & 0.86 & 0.99 & 0.85 & 0.82 & 1.00 & AuHfGa, AuHfIn, IrHfSb, NiHfSn, PdHfSn, PtHfSi \\
Ta & 82 & 0.80 & $-$55 & 32 & 4 & 0.81 & 0.99 & 0.82 & 0.84 & 0.91 & NiTaGa, PtTaAl \\
W & 66 & 0.89 & $-$148 & 0 & 0 & -- & -- & -- & -- & -- & -- \\
Re & 18 & 0.16 & -- & 0 & 0 & -- & -- & -- & -- & -- & -- \\
Os & 81 & 0.88 & $-$49 & 4 & 0 & 0.88 & 1.00 & 0.89 & 0.85 & 1.08 & -- \\
Ir & 93 & 0.81 & $-$57 & 18 & 4 & 0.76 & 1.00 & 0.83 & 0.78 & 0.99 & IrHfSb, IrTiSb \\
Pt & 89 & 0.87 & $-$101 & 20 & 10 & 0.70 & 0.94 & 0.80 & 0.56 & 0.98 & PtHfSi, PtTaAl, PtTiGe, PtTiSi, PtTiSn \\
Au & 60 & 0.80 & $-$97 & 22 & 20 & 0.55 & 0.91 & 0.66 & 0.49 & 1.03 & AuHfGa, AuHfIn, AuScSi, AuTiAl, AuTiGa, AuTiIn, AuZrAl, AuZrGa, AuZrIn, AuZrTl \\
Tl & 81 & 0.77 & $-$56 & 2 & 2 & 0.55 & 0.88 & 0.71 & 0.46 & 1.11 & AuZrTl \\
Pb & 75 & 0.83 & $-$29 & 4 & 0 & 0.88 & 0.94 & 1.00 & 0.77 & 1.01 & -- \\
Bi & 65 & 0.84 & $-$32 & 12 & 0 & 0.97 & 1.01 & 0.96 & 0.76 & 0.97 & -- \\
\midrule
all & 607 & 0.87 & $-$56 & 115 & 40 & 0.82 & 0.97 & 0.86 & 0.76 & 1.00 & 20 compositions \\
\bottomrule
\end{tabular}

\end{table}

\begin{table}[H]\centering\small
\caption{\textbf{Sensitivity of qualified-run and certified counts} to the power-factor threshold, the Seebeck criterion and the
hull-distance cut. ``Threshold from corrected runs'' uses the median of mesh-corrected runs instead of the declared value. Compositions are counted over
both sets.}
\begin{tabular}{@{}l rrrr@{}}
\toprule
case & qualified runs & gapped qualified & certified entries & certified compositions \\
\midrule
declared (baseline) & 309 & 201 & 155 & 80 \\
threshold from corrected runs (1.154e+12) & 295 & 193 & 149 & 77 \\
threshold $\times$0.8 & 353 & 228 & 205 & 105 \\
threshold $\times$0.9 & 337 & 220 & 187 & 96 \\
threshold $\times$1.1 & 283 & 185 & 136 & 70 \\
threshold $\times$1.2 & 250 & 174 & 126 & 65 \\
threshold $\times$1.5 & 161 & 107 & 67 & 35 \\
Seebeck criterion $|S|$ $>$ 40 & 349 & 201 & 155 & 80 \\
Seebeck criterion $|S|$ $>$ 80 & 282 & 201 & 155 & 80 \\
Seebeck criterion $|S|$ $>$ 100 & 249 & 201 & 155 & 80 \\
hull $<$ 0.025 & 148 & 112 & 95 & 49 \\
hull $<$ 0.05 & 206 & 152 & 123 & 64 \\
hull $<$ 0.15 & 395 & 249 & 203 & 104 \\
\bottomrule
\end{tabular}

\end{table}

\begin{table}[H]\centering\scriptsize
\caption{\textbf{Development of the workflow} (Supplementary Note~6). Each row gives the first-principles observation, the change it motivated
and the evidence after it; T and P denote the training and prediction sets.}
\label{tab:si-decisions}
\begin{tabular}{@{}c c >{\raggedright}p{0.27\textwidth} >{\raggedright}p{0.21\textwidth} >{\raggedright\arraybackslash}p{0.32\textwidth}@{}}
\toprule
 & acts on & what the first-principles results showed & revision & evidence after the revision \\
\midrule
R1 & screen & The initial composite score (four d-DOS features at $E_F$) favoured metals: of its 100 top prediction-set entries, 83 carried a metal label, 7 were computed, none was certified &
rank by Cutler--Mott moments of the whole d-DOS &
top 100 by $S^{\mathrm d}_{\max}$: 83 semiconductor labels, 98 computed, 35 certified; $\Delta$ROC-AUC vs composite $=+0.068$ [0.038, 0.101] (P) \\
R2 & screen & 98 of those 100 entries lie at least \SI{0.1}{eV.atom^{-1}} above the DXMag hull &
priority filters on stability, the semiconductor label and ensemble spread &
filters order the selection rounds; the hull distance also enters certification \\
R3 & certification & most d-DOS ``semiconductors'' are QE-PBE metals: 63.1\% (T), 85.1\% (P) &
classify every run by its QE-PBE band gap &
gapped runs pass the transport criteria most often for $E_g=0.6$--\SI{1.2}{eV}: 53--61\% (T), 58--61\% (P) \\
R4 & certification & a threshold chosen during the campaign can be tuned to its results &
one power-factor threshold, the median \SI{900}{K} $\pf$ of the first 1,401 runs, then held fixed &
certified compositions vary smoothly: 105 ($0.8\times$) to 65 ($1.2\times$); 77 with the threshold from corrected runs \\
R5 & certification & 108 of the 309 qualified runs lack a QE-PBE gap (102 metals, 6 narrow-gap) and pass only with a shifted chemical potential &
require a QE-PBE gap $\ge\SI{0.136}{eV}$ &
201 gapped qualified runs remain (102 T, 99 P) \\
R6 & certification & spin--orbit coupling lowers the best $\pf$ (median ratio 0.87 over 607 runs) &
re-check compounds containing elements with atomic number $\ge72$ with SOC &
35 of 56 (T) and 40 of 59 (P) heavy gapped runs stay above the threshold; 5 Rh-based (T) unresolved, 1 (P) not run \\
R7 & ranking & among certified entries $S^{\mathrm d}_{\max}$ barely orders the DFT power factor (Spearman 0.13 T, 0.29 P) &
rank only on computed $|S|$ and $\pf$ at fixed carrier concentration &
$|S|$ ranks stable across conventions (Spearman 0.87), $\pf$ ranks not (0.19); first Pareto layer in the main-text Pareto table \\
\bottomrule
\end{tabular}

\end{table}

\begin{table*}[!htbp]
\centering
\caption{\textbf{Held-out d-DOS fidelity by family and DFT outcome per set} (complements main-text Fig.~4).
$N_\mathrm{train}$ and $N_\mathrm{pred}$ are the training- and prediction-set sizes. $\bar r$ and $\bar r_{\pm2\,\mathrm{eV}}$ are mean
per-material Pearson correlations over the full spectrum and within \SI{2}{eV} of $E_F$; RMSE is in states\,eV$^{-1}$\,atom$^{-1}$ (archive
units). BA and rec.\ are the balanced accuracy and recall of the held-out d-DOS gap flag. These metrics are upper bounds (best-validation
checkpoints; for fine-tuned models the pretraining saw the test spectra). DFT runs and certified compositions are given for the training (T) and prediction (P)
sets; PdScAs is certified in two training-set families, so the T rows sum to 76 while the set contains 75 compositions. Baselines on the
same held-out spectra: $\bar r = 0.488$ (family-mean DOS), 0.501 (nearest composition), 0.605 (mean of five nearest compositions).}
\label{tab:si-dataml}
\footnotesize
\begin{tabular}{@{}l r ccc cc r rr rr@{}}
\toprule
 & \multicolumn{6}{c}{training set: held-out ML d-DOS fidelity} & & \multicolumn{2}{c}{DFT runs} & \multicolumn{2}{c}{certified comp.} \\
\cmidrule(lr){2-7}\cmidrule(lr){9-10}\cmidrule(l){11-12}
family & $N_\mathrm{train}$ & $\bar r$ & $\bar r_{\pm2\,\mathrm{eV}}$ & RMSE & BA & rec. & $N_\mathrm{pred}$ & T & P & T & P \\
\midrule
full cubic & 9,396 & 0.898 & 0.857 & 0.312 & 0.56 & 0.13 & 4,504 & 220 & 185 & 5 & 5 \\
full tetra & 8,047 & 0.925 & 0.840 & 0.226 & 0.66 & 0.33 & 13,479 & 95 & 442 & 0 & 0 \\
half cubic & 9,396 & 0.874 & 0.865 & 0.423 & 0.58 & 0.21 & 4,914 & 614 & 438 & 70 & 74 \\
half tetra & 6,909 & 0.884 & 0.807 & 0.337 & 0.62 & 0.25 & 10,896 & 205 & 466 & 1 & 0 \\
inverse cubic & 9,072 & 0.876 & 0.826 & 0.332 & 0.52 & 0.04 & 6,702 & 294 & 120 & 0 & 0 \\
inverse tetra & 8,437 & 0.917 & 0.799 & 0.231 & 0.56 & 0.12 & 14,483 & 134 & 370 & 0 & 0 \\
\midrule
\textit{all} & 51,257 & 0.895 & 0.834 & 0.312 & 0.57 & 0.15 & 54,978 & 1,562 & 2,021 & 75 & 79 \\
\bottomrule
\end{tabular}

\end{table*}

\noindent\begin{minipage}{\textwidth}\small\captionof{table}{\textbf{Indicator triage metrics within each set.} ROC-AUC, PR-AUC and the fraction of certified entries
captured in the top 10, 20 and 30\% of each pre-DFT score, among computed entries of the stated subset. PR-AUC should be compared
with the base rate given for each subset.}\end{minipage}
{\small\begin{longtable}{@{}l ccccc@{}}
\toprule
score & ROC-AUC & PR-AUC & top 10\% & top 20\% & top 30\% \\
\midrule\endhead
\multicolumn{6}{@{}l}{\textit{prediction set, all computed}: $n$ = 2,016, positives = 79 (base rate 0.039)} \\
\quad $S^{\mathrm d}_{\max}$ & 0.938 & 0.340 & 0.71 & 0.91 & 0.99 \\
\quad $P^{\mathrm d}_{\max}$ & 0.877 & 0.137 & 0.37 & 0.80 & 0.97 \\
\quad $\xi$ & 0.852 & 0.152 & 0.53 & 0.76 & 0.89 \\
\quad $L_0^{*}$ & 0.792 & 0.117 & 0.18 & 0.38 & 0.72 \\
\quad initial composite & 0.869 & 0.168 & 0.46 & 0.76 & 0.90 \\
\quad electron count 18/24 & 0.891 & 0.157 & 0.40 & 0.80 & 1.00 \\
\quad random & 0.494 & 0.039 & 0.11 & 0.20 & 0.28 \\
\multicolumn{6}{@{}l}{\textit{prediction set, d-DOS gap 0.10--1.40 eV}: $n$ = 586, positives = 79 (base rate 0.135)} \\
\quad $S^{\mathrm d}_{\max}$ & 0.836 & 0.406 & 0.35 & 0.60 & 0.77 \\
\quad $P^{\mathrm d}_{\max}$ & 0.751 & 0.248 & 0.14 & 0.39 & 0.56 \\
\quad $\xi$ & 0.714 & 0.243 & 0.19 & 0.47 & 0.58 \\
\quad $L_0^{*}$ & 0.691 & 0.269 & 0.17 & 0.29 & 0.38 \\
\quad initial composite & 0.774 & 0.304 & 0.29 & 0.48 & 0.62 \\
\quad electron count 18/24 & 0.712 & 0.213 & 0.16 & 0.32 & 0.47 \\
\quad random & 0.469 & 0.128 & 0.11 & 0.19 & 0.25 \\
\multicolumn{6}{@{}l}{\textit{prediction set, 18 or 24 electrons}: $n$ = 502, positives = 79 (base rate 0.157)} \\
\quad $S^{\mathrm d}_{\max}$ & 0.808 & 0.407 & 0.30 & 0.54 & 0.71 \\
\quad $P^{\mathrm d}_{\max}$ & 0.688 & 0.232 & 0.10 & 0.29 & 0.46 \\
\quad $\xi$ & 0.672 & 0.240 & 0.13 & 0.38 & 0.54 \\
\quad $L_0^{*}$ & 0.585 & 0.239 & 0.15 & 0.18 & 0.32 \\
\quad initial composite & 0.736 & 0.306 & 0.24 & 0.38 & 0.57 \\
\quad electron count 18/24 & 0.500 & 0.157 & 0.10 & 0.20 & 0.30 \\
\quad random & 0.477 & 0.152 & 0.11 & 0.19 & 0.27 \\
\multicolumn{6}{@{}l}{\textit{prediction set, half cubic}: $n$ = 438, positives = 74 (base rate 0.169)} \\
\quad $S^{\mathrm d}_{\max}$ & 0.799 & 0.407 & 0.30 & 0.50 & 0.68 \\
\quad $P^{\mathrm d}_{\max}$ & 0.690 & 0.240 & 0.08 & 0.31 & 0.49 \\
\quad $\xi$ & 0.664 & 0.253 & 0.14 & 0.38 & 0.54 \\
\quad $L_0^{*}$ & 0.607 & 0.246 & 0.14 & 0.19 & 0.31 \\
\quad initial composite & 0.664 & 0.241 & 0.14 & 0.35 & 0.41 \\
\quad electron count 18/24 & 0.641 & 0.221 & 0.13 & 0.26 & 0.39 \\
\quad random & 0.469 & 0.160 & 0.09 & 0.20 & 0.28 \\
\multicolumn{6}{@{}l}{\textit{training set, all computed}: $n$ = 1,562, positives = 76 (base rate 0.049)} \\
\quad $S^{\mathrm d}_{\max}$ & 0.869 & 0.158 & 0.32 & 0.75 & 0.91 \\
\quad $P^{\mathrm d}_{\max}$ & 0.649 & 0.064 & 0.07 & 0.20 & 0.47 \\
\quad $\xi$ & 0.687 & 0.081 & 0.16 & 0.42 & 0.58 \\
\quad $L_0^{*}$ & 0.557 & 0.054 & 0.08 & 0.11 & 0.17 \\
\quad electron count 18/24 & 0.858 & 0.153 & 0.31 & 0.63 & 0.94 \\
\quad random & 0.574 & 0.075 & 0.20 & 0.33 & 0.42 \\
\multicolumn{6}{@{}l}{\textit{training set, d-DOS gap 0.10--1.40 eV}: $n$ = 661, positives = 76 (base rate 0.115)} \\
\quad $S^{\mathrm d}_{\max}$ & 0.829 & 0.350 & 0.36 & 0.61 & 0.72 \\
\quad $P^{\mathrm d}_{\max}$ & 0.550 & 0.117 & 0.04 & 0.12 & 0.23 \\
\quad $\xi$ & 0.560 & 0.128 & 0.05 & 0.21 & 0.38 \\
\quad $L_0^{*}$ & 0.522 & 0.143 & 0.09 & 0.13 & 0.24 \\
\quad electron count 18/24 & 0.754 & 0.209 & 0.18 & 0.36 & 0.54 \\
\quad random & 0.574 & 0.172 & 0.22 & 0.32 & 0.41 \\
\multicolumn{6}{@{}l}{\textit{training set, 18 or 24 electrons}: $n$ = 498, positives = 76 (base rate 0.153)} \\
\quad $S^{\mathrm d}_{\max}$ & 0.805 & 0.337 & 0.26 & 0.51 & 0.64 \\
\quad $P^{\mathrm d}_{\max}$ & 0.557 & 0.159 & 0.07 & 0.14 & 0.32 \\
\quad $\xi$ & 0.596 & 0.190 & 0.12 & 0.29 & 0.47 \\
\quad $L_0^{*}$ & 0.474 & 0.163 & 0.11 & 0.14 & 0.22 \\
\quad electron count 18/24 & 0.500 & 0.153 & 0.10 & 0.20 & 0.30 \\
\quad random & 0.567 & 0.225 & 0.18 & 0.30 & 0.40 \\
\multicolumn{6}{@{}l}{\textit{training set, half cubic}: $n$ = 614, positives = 70 (base rate 0.114)} \\
\quad $S^{\mathrm d}_{\max}$ & 0.765 & 0.202 & 0.11 & 0.34 & 0.64 \\
\quad $P^{\mathrm d}_{\max}$ & 0.618 & 0.138 & 0.07 & 0.17 & 0.41 \\
\quad $\xi$ & 0.600 & 0.148 & 0.14 & 0.30 & 0.44 \\
\quad $L_0^{*}$ & 0.565 & 0.149 & 0.10 & 0.13 & 0.22 \\
\quad electron count 18/24 & 0.732 & 0.193 & 0.17 & 0.34 & 0.51 \\
\quad random & 0.541 & 0.151 & 0.14 & 0.23 & 0.34 \\
\bottomrule
\end{longtable}
}\addtocounter{table}{-1}

\begin{table}[H]\centering\small
\caption{\textbf{Paired-bootstrap differences in ROC-AUC} (2,000 resamples over computed entries, seed 0) within each set.}
\begin{tabular}{@{}l l l r r r c@{}}
\toprule
set & subset & comparison & $n$ & pos. & $\Delta$AUC & 95\% CI \\
\midrule
prediction set & all computed & $S^{\mathrm d}_{\max}$ $-$ $P^{\mathrm d}_{\max}$ & 2,016 & 79 & +0.061 & [+0.047, +0.076] \\
prediction set & all computed & $S^{\mathrm d}_{\max}$ $-$ composite & 2,016 & 79 & +0.068 & [+0.038, +0.101] \\
prediction set & all computed & $S^{\mathrm d}_{\max}$ $-$ electron count & 2,016 & 79 & +0.047 & [+0.029, +0.063] \\
prediction set & all computed & $S^{\mathrm d}_{\max}$ $-$ $\xi$ & 2,016 & 79 & +0.085 & [+0.053, +0.124] \\
prediction set & 18 or 24 electrons & $S^{\mathrm d}_{\max}$ $-$ $P^{\mathrm d}_{\max}$ & 502 & 79 & +0.120 & [+0.081, +0.159] \\
prediction set & 18 or 24 electrons & $S^{\mathrm d}_{\max}$ $-$ composite & 502 & 79 & +0.072 & [+0.007, +0.133] \\
prediction set & 18 or 24 electrons & $S^{\mathrm d}_{\max}$ $-$ $\xi$ & 502 & 79 & +0.136 & [+0.086, +0.186] \\
training set & all computed & $S^{\mathrm d}_{\max}$ $-$ $P^{\mathrm d}_{\max}$ & 1,562 & 76 & +0.221 & [+0.171, +0.267] \\
training set & all computed & $S^{\mathrm d}_{\max}$ $-$ electron count & 1,562 & 76 & +0.011 & [$-$0.011, +0.035] \\
training set & all computed & $S^{\mathrm d}_{\max}$ $-$ $\xi$ & 1,562 & 76 & +0.183 & [+0.134, +0.238] \\
training set & 18 or 24 electrons & $S^{\mathrm d}_{\max}$ $-$ $P^{\mathrm d}_{\max}$ & 498 & 76 & +0.248 & [+0.178, +0.317] \\
training set & 18 or 24 electrons & $S^{\mathrm d}_{\max}$ $-$ $\xi$ & 498 & 76 & +0.208 & [+0.137, +0.277] \\
\bottomrule
\end{tabular}

\end{table}

\begingroup\small
\setlength{\LTcapwidth}{\textwidth}
\noindent\begin{minipage}{\textwidth}\captionof{table}{\textbf{All 80 certified compositions}, with computed $S$ (\si{\micro V.K^{-1}}) and $\pf$
(\SI{e12}{W.m^{-1}.K^{-2}.s^{-1}}) at \SI{900}{K} and $|n|\le\SI{e21}{cm^{-3}}$ for p- and n-type doping. Set: certified in the
training (T) and/or prediction (P) set. $E_g$: QE-PBE gap; $E_\mathrm{hull}$: DXMag hull distance (eV\,atom$^{-1}$); SOC: values
with spin--orbit coupling. CRTA values, not $zT$ predictions.}\end{minipage}
\begin{longtable}{@{}l l c cc c rr rr@{}}
\toprule
 & & & & & & \multicolumn{2}{c}{p-type} & \multicolumn{2}{c}{n-type} \\
composition & family & set & $E_g$ (eV) & $E_\mathrm{hull}$ & SOC & $S$ & PF/$\tau$ & $S$ & PF/$\tau$ \\
\midrule\endhead
AuHfAl & half cubic & T{+}P & 0.64 & 0.033 & yes & 222 & 0.57 & $-$129 & 0.54 \\
CoHfAs & half cubic & T{+}P & 1.29 & 0.003 & yes & 296 & 1.39 & $-$221 & 1.35 \\
CoHfBi & half cubic & T{+}P & 0.98 & 0.000 & yes & 265 & 1.19 & $-$217 & 1.15 \\
CoHfSb & half cubic & T{+}P & 1.15 & 0.016 & yes & 293 & 1.33 & $-$213 & 1.00 \\
CoNbGe & half cubic & T{+}P & 1.09 & 0.000 & no & 280 & 1.76 & $-$218 & 1.11 \\
CoNbSn & half cubic & T{+}P & 1.04 & 0.039 & no & 288 & 1.63 & $-$220 & 1.02 \\
CoTaGe & half cubic & T{+}P & 1.22 & 0.000 & yes & 282 & 1.52 & $-$203 & 0.79 \\
CoTaSi & half cubic & T{+}P & 1.27 & 0.095 & yes & 259 & 1.40 & $-$206 & 0.89 \\
CoTaSn & half cubic & T{+}P & 1.07 & 0.038 & yes & 296 & 1.39 & $-$188 & 0.71 \\
CoTiAs & half cubic & T{+}P & 1.32 & 0.031 & no & 312 & 1.48 & $-$281 & 1.07 \\
CoTiBi & half cubic & T{+}P & 0.91 & 0.000 & yes & 286 & 1.08 & $-$252 & 0.72 \\
CoTiSb & half cubic & T{+}P & 1.08 & 0.029 & no & 301 & 1.35 & $-$257 & 0.89 \\
CoVGe & half cubic & T{+}P & 0.68 & 0.002 & no & 310 & 1.43 & $-$146 & 0.48 \\
CoVSi & half cubic & T{+}P & 0.58 & 0.038 & no & 295 & 1.40 & $-$144 & 0.44 \\
CoVSn & half cubic & T{+}P & 0.66 & 0.036 & no & 320 & 1.28 & $-$172 & 0.47 \\
CoZrAs & half cubic & T{+}P & 1.23 & 0.006 & no & 274 & 1.26 & $-$237 & 1.19 \\
CoZrBi & half cubic & T{+}P & 0.99 & 0.000 & yes & 280 & 1.19 & $-$217 & 0.98 \\
CoZrSb & half cubic & T{+}P & 1.08 & 0.000 & no & 283 & 1.27 & $-$217 & 0.95 \\
Fe$_{2}$HfSi & full cubic & T{+}P & 0.33 & 0.082 & yes & 208 & 0.83 & $-$239 & 0.89 \\
Fe$_{2}$NbAl & full cubic & T{+}P & 0.30 & 0.000 & no & 212 & 1.08 & $-$221 & 0.81 \\
Fe$_{2}$TaAl & full cubic & T{+}P & 0.29 & 0.000 & yes & 187 & 0.77 & $-$219 & 0.83 \\
Fe$_{2}$TiSi & full cubic & T{+}P & 0.37 & 0.020 & no & 190 & 0.77 & $-$261 & 0.86 \\
FeNbAs & half cubic & T{+}P & 0.57 & 0.000 & no & 231 & 1.37 & $-$147 & 0.35 \\
FeNbSb & half cubic & T{+}P & 0.60 & 0.011 & no & 237 & 1.31 & $-$148 & 0.37 \\
FeTaAs & half cubic & T{+}P & 0.92 & 0.000 & yes & 231 & 1.27 & $-$102 & 0.36 \\
FeTaSb & half cubic & T{+}P & 0.93 & 0.000 & yes & 238 & 1.23 & $-$112 & 0.36 \\
FeVAs & half cubic & T{+}P & 0.36 & 0.009 & no & 246 & 1.26 & $-$136 & 0.61 \\
FeVSb & half cubic & T{+}P & 0.39 & 0.002 & no & 253 & 1.18 & $-$152 & 0.60 \\
IrNbGe & half cubic & T{+}P & 0.69 & 0.000 & yes & 239 & 1.63 & $-$88 & 0.43 \\
IrNbSn & half cubic & T{+}P & 0.62 & 0.000 & yes & 249 & 1.60 & $-$88 & 0.40 \\
IrTaGe & half cubic & T{+}P & 1.00 & 0.000 & yes & 237 & 1.40 & $-$96 & 0.50 \\
IrTaSi & half cubic & T{+}P & 0.87 & 0.080 & yes & 230 & 1.33 & $-$78 & 0.36 \\
IrTaSn & half cubic & T{+}P & 0.96 & 0.000 & yes & 247 & 1.31 & $-$92 & 0.40 \\
IrTiAs & half cubic & T{+}P & 0.67 & 0.020 & yes & 192 & 0.78 & $-$171 & 0.92 \\
IrVGe & half cubic & T{+}P & 0.36 & 0.000 & yes & 261 & 1.28 & $-$148 & 0.73 \\
NiHfGe & half cubic & T{+}P & 0.58 & 0.001 & yes & 172 & 0.80 & $-$143 & 0.83 \\
NiHfSi & half cubic & T{+}P & 0.71 & 0.090 & yes & 211 & 0.92 & $-$144 & 0.81 \\
NiNbIn & half cubic & T{+}P & 0.14 & 0.075 & no & 115 & 0.75 & $-$122 & 0.47 \\
NiScAs & half cubic & T{+}P & 0.48 & 0.043 & no & 151 & 0.67 & $-$182 & 1.12 \\
NiScBi & half cubic & T{+}P & 0.19 & 0.022 & yes & 102 & 0.38 & $-$158 & 0.88 \\
NiScSb & half cubic & T{+}P & 0.27 & 0.010 & no & 141 & 0.66 & $-$160 & 0.89 \\
NiTiGe & half cubic & T{+}P & 0.64 & 0.012 & no & 230 & 1.02 & $-$171 & 0.72 \\
NiTiPb & half cubic & T{+}P & 0.35 & 0.073 & yes & 206 & 0.70 & $-$167 & 0.62 \\
NiTiSi & half cubic & T{+}P & 0.76 & 0.092 & no & 266 & 1.08 & $-$170 & 0.73 \\
NiTiSn & half cubic & T{+}P & 0.46 & 0.043 & no & 224 & 0.93 & $-$161 & 0.64 \\
NiVGa & half cubic & T{+}P & 0.39 & 0.096 & no & 178 & 1.04 & $-$225 & 0.72 \\
NiYBi & half cubic & T{+}P & 0.21 & 0.010 & yes & 87 & 0.27 & $-$150 & 0.97 \\
NiYSb & half cubic & T{+}P & 0.30 & 0.002 & no & 141 & 0.59 & $-$153 & 0.98 \\
NiZrGe & half cubic & T{+}P & 0.69 & 0.002 & no & 216 & 1.01 & $-$146 & 0.80 \\
NiZrSi & half cubic & T{+}P & 0.82 & 0.085 & no & 283 & 1.22 & $-$148 & 0.79 \\
NiZrSn & half cubic & T{+}P & 0.51 & 0.019 & no & 196 & 0.87 & $-$138 & 0.70 \\
OsNbSb & half cubic & T{+}P & 0.26 & 0.000 & yes & 178 & 1.13 & $-$113 & 0.69 \\
OsTaSb & half cubic & T{+}P & 0.56 & 0.000 & yes & 172 & 0.97 & $-$125 & 0.88 \\
PdScAs & half cubic & T & 0.44 & 0.096 & no & 143 & 0.56 & $-$187 & 1.18 \\
PdScSb & half cubic & T{+}P & 0.28 & 0.014 & no & 137 & 0.61 & $-$161 & 0.87 \\
PdTiPb & half cubic & T{+}P & 0.37 & 0.096 & yes & 207 & 0.66 & $-$187 & 0.55 \\
PdTiSn & half cubic & T{+}P & 0.48 & 0.000 & no & 216 & 0.94 & $-$187 & 0.66 \\
PdZrSn & half cubic & T{+}P & 0.47 & 0.000 & no & 168 & 0.82 & $-$139 & 0.62 \\
PtNbGa & half cubic & T{+}P & 0.55 & 0.000 & yes & 155 & 1.04 & $-$137 & 0.35 \\
PtNbIn & half cubic & T{+}P & 0.47 & 0.075 & yes & 154 & 1.12 & $-$138 & 0.55 \\
PtScSb & half cubic & T{+}P & 0.63 & 0.017 & yes & 120 & 0.32 & $-$198 & 0.97 \\
PtTaGa & half cubic & T{+}P & 1.05 & 0.009 & yes & 163 & 0.99 & $-$120 & 0.47 \\
PtVGa & half cubic & T{+}P & 0.33 & 0.097 & yes & 186 & 1.04 & $-$137 & 0.51 \\
RhHfSb & half cubic & P & 1.10 & 0.020 & yes & 148 & 0.80 & $-$211 & 1.05 \\
RhNbGe & half cubic & T{+}P & 0.76 & 0.066 & no & 233 & 1.81 & $-$144 & 0.39 \\
RhNbSn & half cubic & T{+}P & 0.76 & 0.044 & no & 241 & 1.82 & $-$139 & 0.39 \\
RhTaGe & half cubic & P & 1.16 & 0.017 & yes & 278 & 1.68 & $-$147 & 0.34 \\
RhTaSn & half cubic & P & 1.08 & 0.044 & yes & 296 & 1.47 & $-$114 & 0.39 \\
RhTiAs & half cubic & T{+}P & 0.83 & 0.033 & no & 203 & 1.12 & $-$162 & 0.60 \\
RhTiBi & half cubic & P & 0.70 & 0.037 & yes & 210 & 0.78 & $-$258 & 0.66 \\
RhTiSb & half cubic & T{+}P & 0.75 & 0.031 & no & 214 & 1.08 & $-$175 & 0.58 \\
RhVGe & half cubic & T{+}P & 0.45 & 0.050 & no & 291 & 1.50 & $-$138 & 0.61 \\
RhZrBi & half cubic & P & 0.99 & 0.014 & yes & 167 & 0.68 & $-$230 & 1.12 \\
Ru$_{2}$ZrSi & full cubic & T{+}P & 0.20 & 0.000 & no & 88 & 0.54 & $-$192 & 1.10 \\
RuNbAs & half cubic & T{+}P & 0.45 & 0.038 & no & 187 & 1.29 & $-$110 & 0.64 \\
RuNbSb & half cubic & T{+}P & 0.45 & 0.022 & no & 200 & 1.34 & $-$103 & 0.54 \\
RuTaAs & half cubic & T{+}P & 0.43 & 0.000 & yes & 192 & 1.23 & $-$111 & 0.71 \\
RuTaSb & half cubic & T{+}P & 0.73 & 0.022 & yes & 205 & 1.25 & $-$110 & 0.60 \\
RuVAs & half cubic & T{+}P & 0.25 & 0.020 & no & 208 & 1.19 & $-$159 & 0.85 \\
RuVSb & half cubic & T{+}P & 0.26 & 0.074 & no & 227 & 1.22 & $-$164 & 0.80 \\
\bottomrule
\end{longtable}
\addtocounter{table}{-1}
\endgroup

\begin{table}[H]\centering\small
\caption{\textbf{Established half-Heusler thermoelectric compositions in the campaign.} The list is taken from the reviews and studies cited in
the main text (Zeier et al.\ 2016; Quinn and Bos 2021; Fu et al.\ 2015; Zhu et al.\ 2019) and is not exhaustive. DXMag writes half-Heusler
compositions in site order. $E_g$: QE-PBE gap of the reported entry. The certified prediction-set entries are near-duplicates of the
training-set entries. No numerical comparison with published coefficients is made. Script: \texttt{scripts/figures/known\_hh\_check.py}.}
\begin{tabular}{@{}llllrl@{}}
\toprule
composition & DXMag string & computed in & certified in & $E_g$ (eV) & outcome \\
\midrule
TiNiSn & NiTiSn & P+T & P+T & 0.46 & certified \\
ZrNiSn & NiZrSn & P+T & P+T & 0.51 & certified \\
HfNiSn & NiHfSn & P+T & - & 0.39 & SOC PF/$\tau$ 1.046 below threshold \\
TiCoSb & CoTiSb & P+T & P+T & 1.08 & certified \\
ZrCoSb & CoZrSb & P+T & P+T & 1.08 & certified \\
HfCoSb & CoHfSb & P+T & P+T & 1.15 & certified \\
VFeSb & FeVSb & P+T & P+T & 0.40 & certified \\
NbFeSb & FeNbSb & P+T & P+T & 0.60 & certified \\
TaFeSb & FeTaSb & P+T & P+T & 0.93 & certified \\
NbCoSn & CoNbSn & P+T & P+T & 1.04 & certified \\
ZrCoBi & CoZrBi & P+T & P+T & 1.00 & certified \\
\bottomrule
\end{tabular}

\end{table}

\begin{table}[H]\centering\small
\caption{\textbf{Earlier best-over-all-conditions Pareto set}, as presented in the project's internal ranking: each composition at its best
$\pf$ over 300--\SI{1200}{K} with re-optimised doping ($|\mu-E_F|\le\SI{0.5}{eV}$). $S$ in \si{\micro V.K^{-1}}, $\pf$ in
\SI{e12}{W.m^{-1}.K^{-2}.s^{-1}}. All optima lie at \SI{1200}{K}; the main text therefore ranks at \SI{900}{K} and fixed carrier concentration.}
\begin{tabular}{@{}l l c r r r@{}}
\toprule
composition & family & set & $S$ & PF/$\tau$ & combined rank \\
\midrule
CoNbGe & half cubic & T{+}P & 287 & 2.11 & 1 \\
RhNbGe & half cubic & T{+}P & 211 & 2.64 & 2 \\
RhTaSn & half cubic & P & 299 & 1.72 & 4 \\
IrTaGe & half cubic & T{+}P & 311 & 1.36 & 30 \\
CoTaSi & half cubic & T{+}P & 319 & 1.36 & 30 \\
CoHfAs & half cubic & T{+}P & 335 & 1.35 & 30 \\
CoTiAs & half cubic & T{+}P & 359 & 1.33 & 34 \\
\bottomrule
\end{tabular}

\end{table}

\section*{Supplementary Figures}

\begin{figure}[H]\centering
\includegraphics[width=\textwidth]{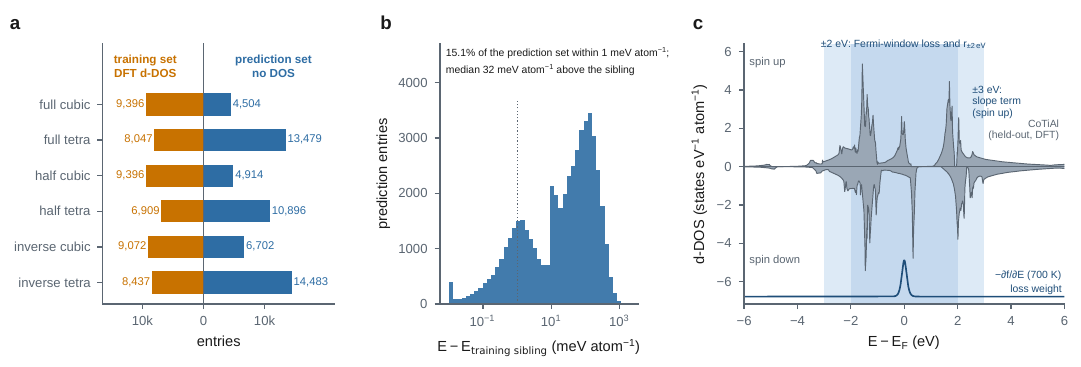}
\caption{\textbf{The two sets and the learning target.} \textbf{a},~Entries per structural family in the training set (DFT d-DOS
available) and the prediction set (no DOS). \textbf{b},~Energy of each prediction-set entry above the training entry of the same family and
composition, for the 51,103 prediction entries that have one; values below 0.01~meV\,atom$^{-1}$ are drawn in the first bin. No prediction
entry lies below its training sibling; the dotted line marks 1~meV\,atom$^{-1}$. \textbf{c},~The learning target for one held-out training
entry (CoTiAl, a gapped half-Heusler): the d-projected DOS per spin channel on the 2,001-point grid, spin down drawn negative. Shaded bands
mark the $\pm2$~eV window of the $\mathcal L_{\pm2}$ loss term and of the $r_{\pm2\,\mathrm{eV}}$ metric and the $\pm3$~eV window of the slope term;
the curve at the bottom is the 700~K Fermi-window weight of the loss (Eq.~(\ref{eq:si-loss})). Values carry the archive's factor-of-two
normalisation (Note~1). (Detailed version of the target shown in main-text Figs.~3 and~4.)}
\end{figure}

\begin{figure}[H]\centering
\includegraphics[width=0.5\textwidth]{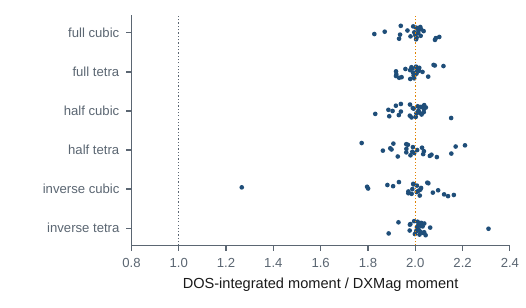}
\caption{\textbf{Normalisation of the DOS archive.} Moment from the archive's spin-resolved total DOS, per formula unit of the stored
structure, divided by the DXMag magnetisation, for 25 magnetic entries per family (Supplementary Note~1). Dotted lines mark ratios of 1 and 2.}
\end{figure}

\begin{figure}[H]\centering
\includegraphics[width=\textwidth]{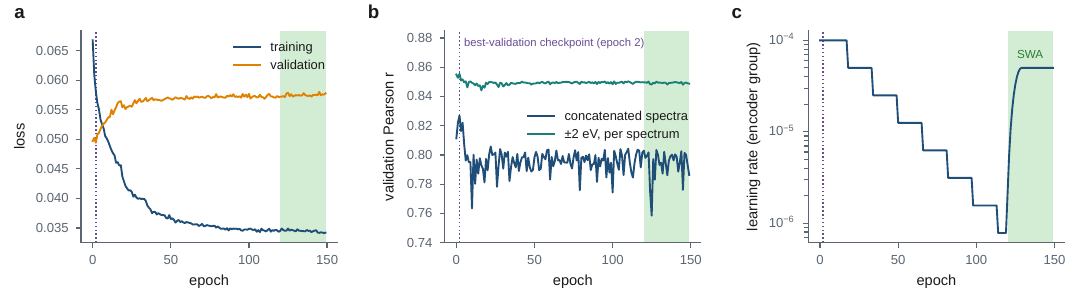}
\caption{\textbf{Training record of cubic full-Heusler fold~1}, the only fine-tuning run whose log survives (TensorBoard scalars).
\textbf{a},~Training and validation loss. \textbf{b},~Validation Pearson correlation over the concatenated spectra (the selection metric)
and the mean per-spectrum correlation within $\pm2$~eV. \textbf{c},~Learning rate of the encoder group, halved by the plateau schedule and
set to $5\times10^{-5}$ by the SWA schedule. The dotted line marks the best-validation checkpoint (epoch~2), the green band the SWA window (epochs 120--149).}
\end{figure}

\begin{figure}[H]\centering
\includegraphics[width=\textwidth]{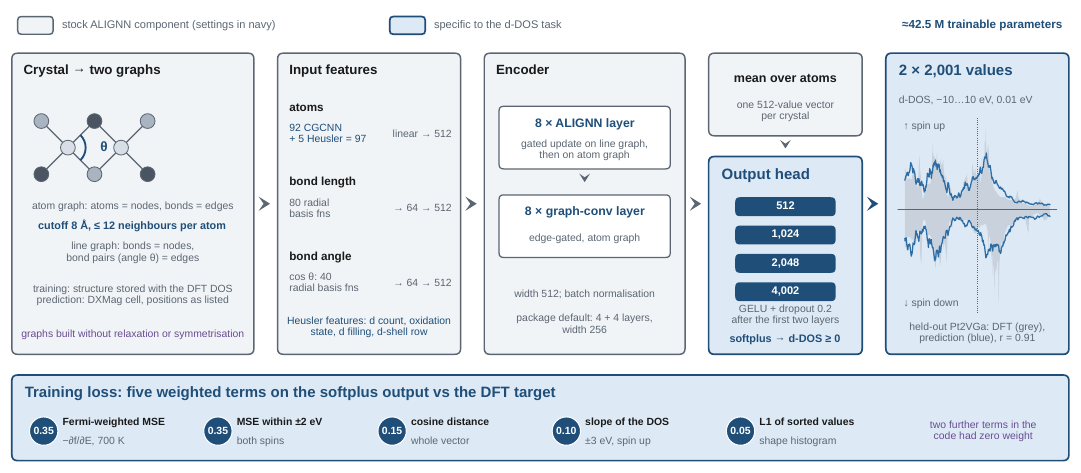}
\caption{\textbf{The d-DOS model: a stock ALIGNN encoder with a task-specific head, output and loss.} Grey cards are components of the
ALIGNN package used unchanged, with the settings of this work in navy; navy cards are specific to the d-DOS task. Graphs are built from the
structure stored with each training spectrum or from the listed DXMag cell for prediction-set entries, without relaxation. The five
Heusler atom features are the d-electron count, the largest common oxidation state, the distance of the d count from a half-filled or
filled shell, the principal quantum number of the d shell, and a second copy of the d-electron count. The output spectrum is a held-out
entry with the median Pearson $r$ (grey, DFT; blue, prediction from the best-validation checkpoint). The bottom row gives the five loss
terms and their weights (Eq.~(\ref{eq:si-loss}); Supplementary Note~3).}
\end{figure}

\begin{figure}[H]\centering
\includegraphics[width=\textwidth]{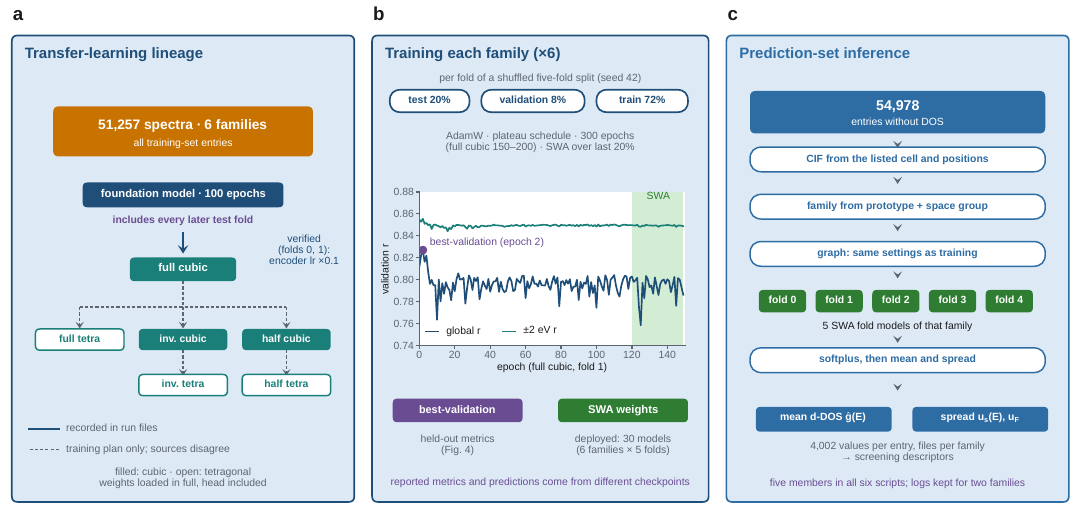}
\caption{\textbf{How the 30 deployed models were obtained and applied.} \textbf{a},~Transfer-learning lineage. A foundation model trained on
all 51,257 training-set spectra initialises the family models. The solid arrow, foundation to cubic full Heusler with the encoder at
$0.1\times$ the head's learning rate, is recorded in the run files of folds 0 and 1. The dashed tree is the lineage of the training plan;
four project sources give different parents and no checkpoint records it (Table~S4). \textbf{b},~Training of each family:
the data partition of one fold, and the validation correlation of cubic full-Heusler fold~1 over its 150 epochs (TensorBoard record), with
the best-validation checkpoint (orange) at epoch~2 and the SWA window (green). The best-validation checkpoint gives the held-out metrics;
the SWA weights give the predictions. \textbf{c},~Inference for the prediction set. Each entry is routed by family to that family's five SWA fold models;
the mean of their softplus outputs is the predicted d-DOS, and their spread $u_s(E)$, through its Fermi-window mean $u_F$, is the
uncertainty used by the reliability filter. Five members
are specified in the screening scripts of all six families; run summaries confirming them survive for two families.}
\end{figure}

\begin{figure}[H]\centering
\includegraphics[width=0.85\textwidth]{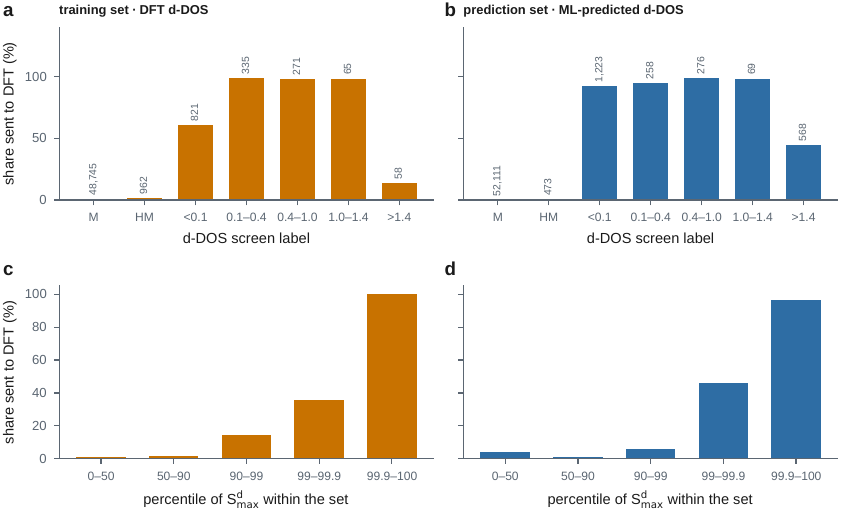}
\caption{\textbf{Selection of entries for DFT.} Share of each set's screened entries that were computed, by d-DOS screen label
(\textbf{a,b}; the screened population per label is printed above the bars) and by percentile of the d-DOS Seebeck indicator
$S^{\mathrm d}_{\max}$ within the set (\textbf{c,d}). Selection followed the scores through nine successive selection rounds (Table~S6).}
\end{figure}

\begin{figure}[H]\centering
\includegraphics[width=0.5\textwidth]{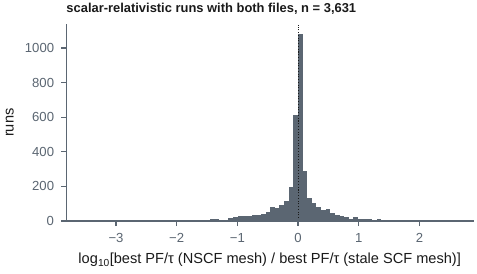}
\caption{\textbf{Effect of the mesh correction} on the best $\pf$ of scalar-relativistic runs, comparing interpolation of the
non-self-consistent eigenvalues with the earlier, erroneous interpolation of the shifted self-consistent mesh.}
\end{figure}

\begin{figure}[H]\centering
\includegraphics[width=\textwidth]{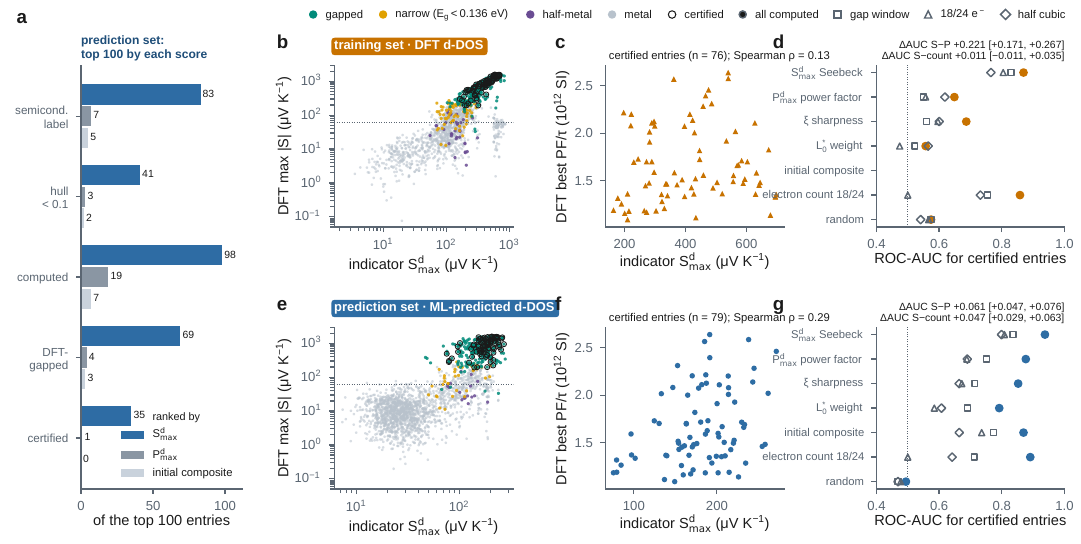}
\caption{\textbf{Workflow revisions acting on the screen (R1, R2 and R7 of Table~S10).} \textbf{a},~Composition of the 100 highest-ranked
prediction-set entries by the initial composite score (Supplementary Note~5), the d-DOS power-factor indicator $P^{\mathrm d}_{\max}$ and the
d-DOS Seebeck indicator $S^{\mathrm d}_{\max}$. \textbf{b,e},~$S^{\mathrm d}_{\max}$ of each set against the DFT intrinsic maximum $|S|$
(300--\SI{1200}{K}), coloured by QE-PBE class; open circles mark certified entries, and the dotted line is the Seebeck criterion.
Horizontal axes are set-specific. \textbf{c,f},~$S^{\mathrm d}_{\max}$ against the DFT best power factor among certified entries, with
Spearman's $\rho$. \textbf{d,g},~ROC-AUC with which each pre-DFT score ranks certified entries among computed entries: overall (filled) and
within three subsets (open). Paired-bootstrap differences are printed with 95\% intervals. The initial composite score exists only for the prediction set.
Scores are compared only within a set; ``18/24 e$^-$'' is the subset obeying the electron-count rule.
(More detailed version of main-text Figs.~5 and~7; panel~a documents the initial composite score.)}
\end{figure}

\begin{figure}[H]\centering
\includegraphics[width=\textwidth]{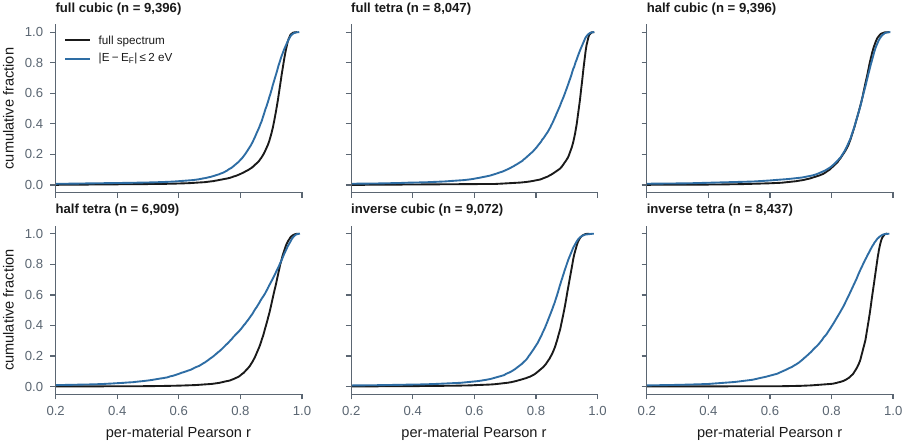}
\caption{\textbf{Held-out spectral fidelity by family} (training set, five-fold cross-validation, best-validation
checkpoints). Cumulative distributions of per-material Pearson $r$ over the full spectrum and within \SI{2}{eV} of $E_F$.}
\end{figure}

\begin{figure}[H]\centering
\includegraphics[width=\textwidth]{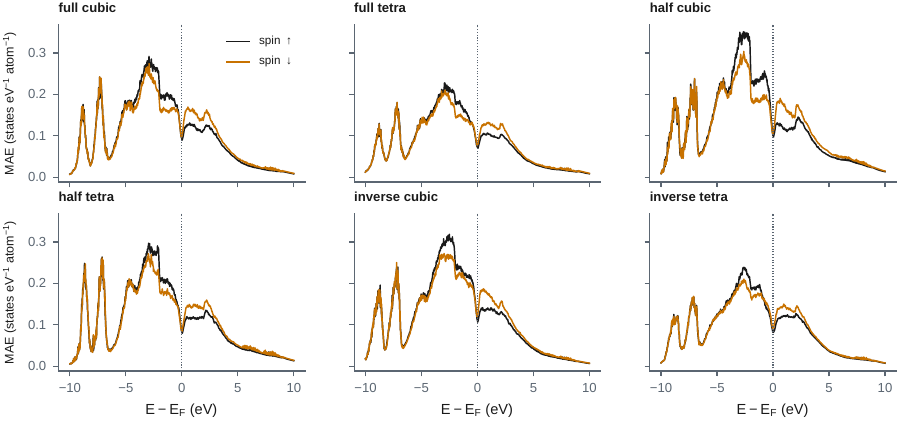}
\caption{\textbf{Energy-resolved mean absolute error} of the predicted d-DOS, averaged over the five folds of each family. The
absolute error is small near $E_F$ partly because the d-DOS itself is small there. It should not be read as higher accuracy at
the Fermi level (compare main-text Fig.~4e,f).}
\end{figure}

\begin{figure}[H]\centering
\includegraphics[width=\textwidth]{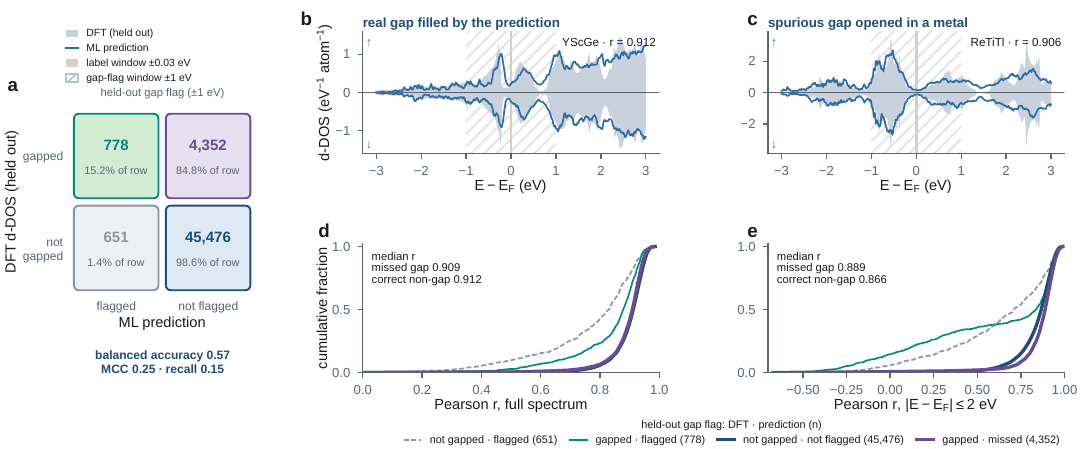}
\caption{\textbf{Held-out d-DOS fidelity in detail} (training set, best-validation checkpoints; the population panels of main-text Fig.~4
in full). \textbf{a},~Confusion matrix of the held-out gap flag. \textbf{b},~A real gap filled by the prediction (YScGe). \textbf{c},~A
spurious gap opened in a metal (ReTiTl); hatching, the $\pm\SI{1}{eV}$ flag window; tan strip, the label window. \textbf{d,e},~Cumulative
distributions of the per-spectrum Pearson $r$ over the full spectrum and within \SI{2}{eV} of $E_F$ for the four cells of a.}
\end{figure}

\begin{figure}[H]\centering
\includegraphics[width=\textwidth]{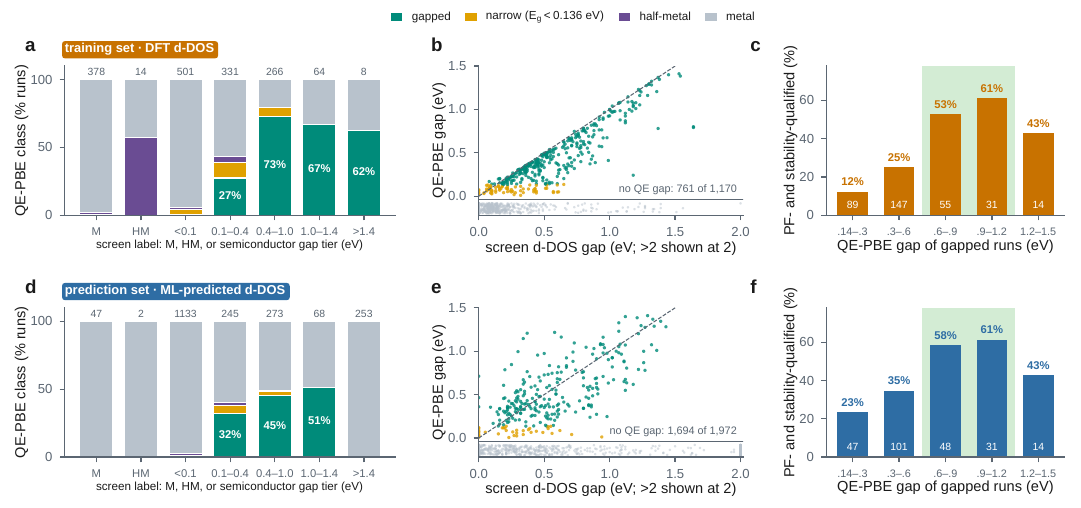}
\caption{\textbf{The screen's labels tested by QE-PBE, per set.} Top row, training set (screen on DFT d-DOS); bottom row, prediction set
(screen on ML-predicted d-DOS). \textbf{a,d},~QE-PBE electronic class of the computed runs in each screen label: M, metal; HM, half-metal;
otherwise semiconductor by d-DOS gap tier in eV. White numbers give the gapped share and the numbers above the bars give the runs per label.
\textbf{b,e},~d-DOS gap assigned by the screen against the QE-PBE gap for runs labelled semiconducting; points below the line have no QE-PBE
gap, and the dashed line is parity. \textbf{c,f},~Share of gapped runs that are qualified runs (meeting the Seebeck criterion, above the power-factor
threshold and stable), by QE-PBE gap, with the number of runs per bin; the green band marks 0.6--\SI{1.2}{eV}. All denominators are within a set. (Detailed version of main-text Fig.~5a,d with gap parity and qualified-run rates.)}
\end{figure}

\begin{figure}[H]\centering
\includegraphics[width=0.5\textwidth]{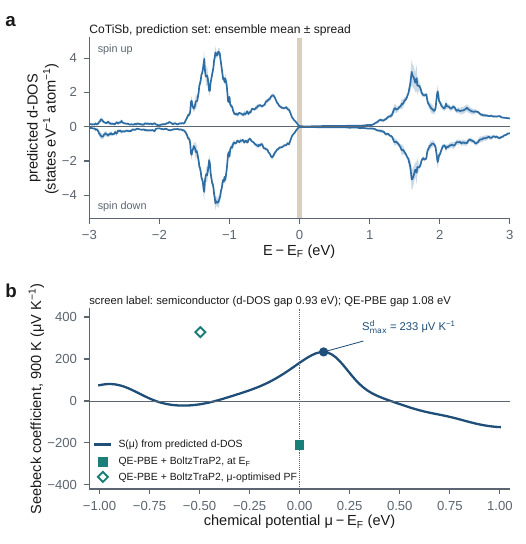}
\caption{\textbf{The indicator for a prediction-set entry} (CoTiSb, certified). \textbf{a},~Predicted d-DOS, ensemble mean $\pm$ spread.
\textbf{b},~Cutler--Mott Seebeck coefficient at \SI{900}{K} from the predicted d-DOS, with its maximum $S^{\mathrm d}_{\max}$, and the QE-PBE +
BoltzTraP2 Seebeck coefficient of the same entry at $E_F$ and at its power-factor optimum.}
\end{figure}

\begin{figure}[H]\centering
\includegraphics[width=0.5\textwidth]{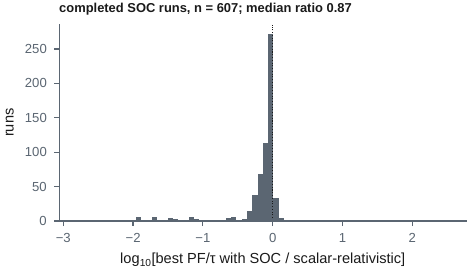}
\caption{\textbf{Effect of spin--orbit coupling} on the best $\pf$ of completed SOC runs. The SOC runs use fully relativistic PSlibrary
potentials, so the ratio also reflects the change of pseudopotential family.}
\end{figure}

\begin{figure}[H]\centering
\includegraphics[width=\textwidth]{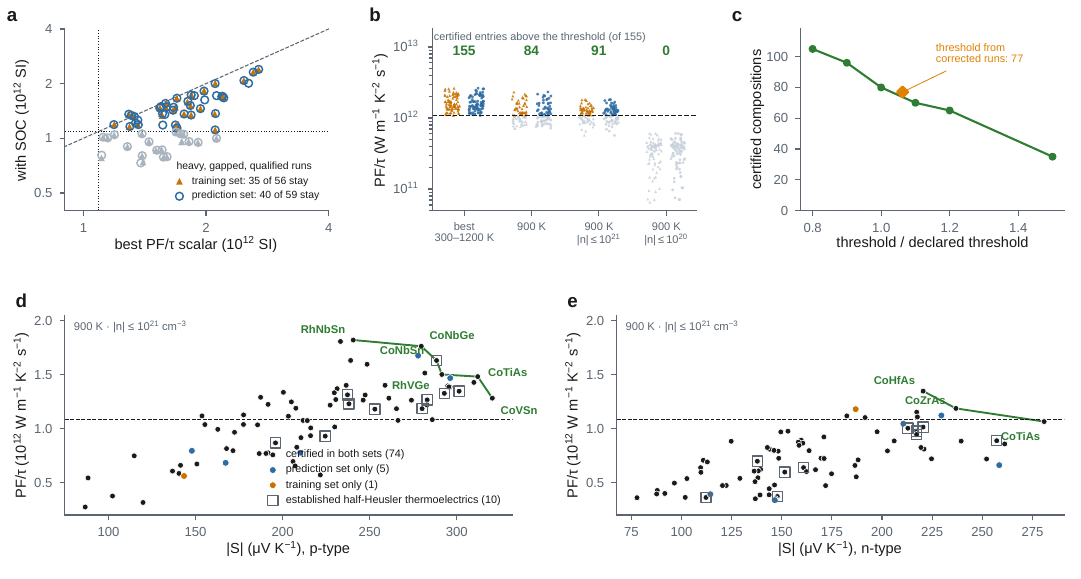}
\caption{\textbf{What survives, and how robustly.}
\textbf{a},~Best $\pf$ of heavy ($Z\ge72$) gapped qualified runs with a completed SOC run, scalar-relativistic against SOC. Symbols are
coloured when the SOC value stays above the power-factor threshold (dotted lines) and grey otherwise; triangles mark the training set and open circles the prediction set. The
SOC runs use fully relativistic PSlibrary potentials, so the comparison also reflects a change of pseudopotential family.
\textbf{b},~$\pf$ of the 155 certified entries under four conventions, counting how many stay above the declared power-factor threshold (dashed): best value
over 300--\SI{1200}{K} with $|\mu-E_F|\le\SI{0.5}{eV}$ (the certification basis); \SI{900}{K} in the same window; and \SI{900}{K} with the
carrier concentration capped at $\SI{e21}{cm^{-3}}$ or $\SI{e20}{cm^{-3}}$. Horizontal jitter is for display only.
\textbf{c},~Certified compositions against the power-factor threshold relative to its declared value; the diamond marks the threshold
re-derived from mesh-corrected runs. \textbf{d,e},~Computed $|S|$ and $\pf$ of the 80 certified compositions at \SI{900}{K} and $|n|\le\SI{e21}{cm^{-3}}$, for p-type
(\textbf{d}) and n-type (\textbf{e}) doping, coloured by the set(s) in which each composition is certified; the green line marks the first
Pareto layer, and open squares mark the established half-Heusler thermoelectrics of Table~S15. SOC values are used for compounds with $Z\ge72$. (Complements main-text Figs.~10 and~8d; panel~b gives the transport conventions referred to in the main text.)}
\end{figure}

\end{document}